\documentclass[reprint,pra,superscriptaddress]{revtex4-1}

\usepackage{amsmath}
\usepackage{amssymb}
\usepackage{enumitem}
\usepackage{graphicx}
\usepackage{dcolumn}
\usepackage{multirow}
\usepackage{color}
\newcommand{\Hamiltonian}{\mathcal{H}}

\mathchardef\mhyphen="2D

\newcommand{\units}[1]{\ensuremath{\mathrm{#1}}}

\newcommand{\bra}[1]{\ensuremath{\langle#1|}}
\newcommand{\ket}[1]{\ensuremath{|{#1}\rangle}}
\newcommand{\braket}[2]{ \langle #1 | #2 \rangle }

\newcommand{\red}[1]{\textcolor{black}{#1}}

\begin{document}

\title{High-Fidelity Entangling Gates for Quantum-Dot Hybrid Qubits Based on Exchange Interactions}

\author{Yuan-Chi Yang}
\email[]{yjc1989@gmail.com}
\affiliation{Department of Physics, University of Wisconsin-Madison, Madison, Wisconsin, 53706, USA}

\author{S. N. Coppersmith}
\email[]{snc@physics.wisc.edu}
\affiliation{Department of Physics, University of Wisconsin-Madison, Madison, Wisconsin, 53706, USA}
\affiliation{School of Physics, The University of New South Wales, Sydney NSW 2052
Australia}

\author{Mark Friesen}
\email[]{friesen@physics.wisc.edu}
\affiliation{Department of Physics, University of Wisconsin-Madison, Madison, Wisconsin, 53706, USA}

\date{\today}

\begin{abstract}
Quantum dot hybrid qubits exploit an extended charge-noise sweet spot that suppresses dephasing and has enabled the experimental achievement of high-fidelity single-qubit gates.
However, current proposals for two-qubit gates require tuning the qubits away from their sweet spots.
Here, we propose a two-hybrid-qubit coupling scheme, based on exchange interactions, that allows the qubits to remain at their sweet spots at all times.
The interaction is controlled via the inter-qubit tunnel coupling.
By simulating such gates in the presence of realistic quasistatic and $1\!/\!f$ charge noise, we show that our scheme should enable controlled-$Z$ gates of length $\sim$5~ns, and Z-CNOT gates of length $\sim$7~ns, both with fidelities $>$99.9\%.
\end{abstract}

\maketitle

\section{Introduction}
Electrically-gated quantum dot systems are promising platforms for quantum information processing~\cite{PhysRevA.57.120,Morton:2011,RevModPhys.85.961}. 
The qubits defined in these systems are typically formed of small numbers of electrons confined inside single, double, or triple quantum 
dots,
%dots~\cite{PhysRevA.57.120,PhysRevLett.105.246804, PhysRevLett.110.146804, PhysRevLett.116.110402, Wu19082014, Srinivasa2014, Scarlino2015, Kawakami18102016, Shi2012,Koh2012, KimShiSimmonsEtAl2014, KimWardSimmonsEtAl2015, Koh03122013,Kim2015,Shi2013,ncomms15923}.
which can be manipulated electrically or magnetically, via DC pulses or microwave driving.
High-fidelity gate operations have been demonstrated in several quantum-dot spin-based architectures.
For example, resonantly driven single-qubit gates have been realized in single-electron-spin~\cite{PioroLadriere2008,Veldhorst2014,Kawakami2014,Yoneda2018}, singlet-triplet~\cite{Shulman2014}, hybrid~\cite{Kim2015,Thorgrimsson2017}, and exchange-only qubits~\cite{Medford2013,Landig:2018}.
Entangling gates have also been demonstrated in single-spin~\cite{Veldhorst:2015,Zajaceaao5965,Watson2018,Hendrickx:Preprint} and singlet-triplet qubits~\cite{Nichol2017}.

\red{Optimal working points or ``sweet spots," where qubits are protected from dephasing caused by electrostatic fluctuations, are well known in superconducting systems~\cite{Vion2002}.
More recently, sweet spots have also been found in spin qubits~\cite{Kim2015,KimShiSimmonsEtAl2014,Cao2016,PhysRevLett.116.116801,Schoenfield:2017,Thorgrimsson2017,Croot:Preprint,PhysRevLett.116.110402}.}
For example, in hybrid qubits, an extended sweet spot emerges when the double dot is strongly biased~\cite{Shi2012,Koh2012,Wong2016}, or detuned, enabling high-fidelity single-qubit gates via resonant driving~\cite{KimWardSimmonsEtAl2015}.
Entangling gates between hybrid qubits have not yet been demonstrated.
However, several two-qubit gate proposals require tuning the qubits away from their sweet spots~\cite{Koh2012,Shi2012,PhysRevB.91.035430, 2015arXiv150703425M,Ferraro:2015,Michielis:2015,2018arXiv181203177F}, exposing them to the effects of charge noise, and ultimately limiting their gate fidelities.

Here, we propose and investigate a method for performing controlled-Z (CZ)
gates between a pair of exchange-coupled hybrid qubits, by modulating the inter-qubit tunnel couplings.
The gates are implemented by applying DC pulses to the tunnel barrier between the qubits, while the qubits remain near their individual sweet spots.
Such fast tunnel-coupling control has been demonstrated in several recent experiments~\cite{PhysRevLett.115.096801,PhysRevLett.116.116801,PhysRevLett.116.110402,Zajaceaao5965}.
By applying an adiabatic ramp to suppress the leakage, we obtain an optimal gate fidelity $>$99.9\%, with a gate time of around $30 \,\units{ns}$, even in the presence of a realistic level of quasistatic charge noise.
We also consider faster entangling gates with nonadiabatic ramps.
By characterizing the oscillations in the fidelity patterns caused by leakage, over a range of control parameters, we identify operating regimes with fidelities $>$99.9\%, even for gate times as short as $5 \, \units{ns}$.
We further consider a Z-CNOT gate sequence formed by combining CZ and high-fidelity single-qubit gates~\cite{PhysRevA.95.062321}, obtaining intrinsic fidelities $>$99.9\% in the absence of charge noise.
In the presence of realistic $1\!/\!f$ charge noise, we can still obtain Z-CNOT gate fidelities of order 99.9\%.

The paper is organized as follows.
In Sec.~\ref{Sec:DQHQTwoQubitModel}, we introduce our model for a pair of exchange-coupled double-quantum-dot hybrid qubits.
In Sec.~\ref{Sec:DQHQTwoQubitCZGates}, we describe our proposals for CZ and Z-CNOT gates and characterize their performance. 
In Sec.~\ref{Sec:DQHQTwoQubitDiscussion} we discuss methods to further improve the gate fidelity.
We finally conclude in Sec.~\ref{Sec:DQHQTwoQubitConclusion}.

\begin{figure*}[t]
\includegraphics[width=6in]{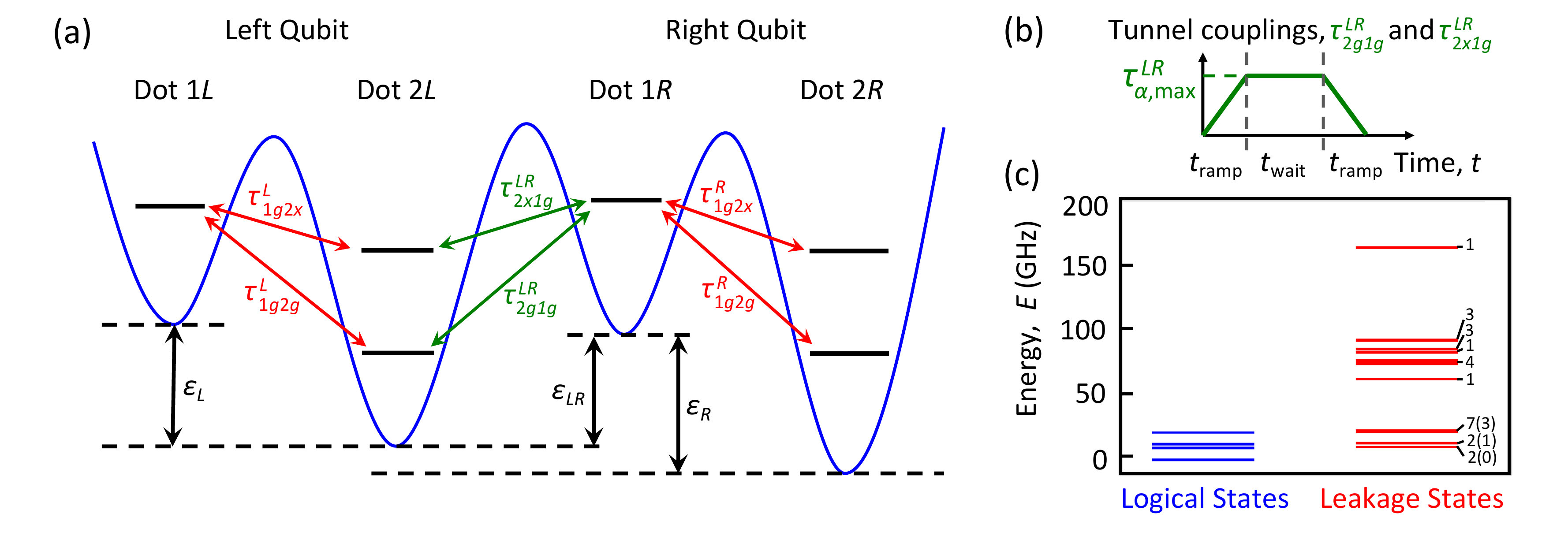}
\caption[Exchange-coupled double quantum dot hybrid qubits]{
\label{Fig:DQHQTwoQubitFig1}
Exchange-coupled quantum-dot hybrid qubits. 
(a) Schematic of the two-qubit system, comprising four dots in a linear array; the left ($L$) and right ($R$) double dots define the left and right qubits. 
The full system contains six electrons, with three electrons per qubit.
For each qubit, we assume the two lowest-energy states have the same (1,2) charge occupation, but different spin configurations.
We also assume that each double dot has one low-energy charge excitation, corresponding to a (2,1) leakage state.
The tunnel couplings between these different charge configurations are designated $\tau_{1g2g}^q$ and $\tau_{1g2x}^q$, where $q=L,R$, and $g,x$ refer to the single-electron ground and excited energy levels.
The qubit detuning parameters are designated $\varepsilon_q$.
We also assume a tunable tunnel coupling between the second and third dots, which mediates an exchange interaction via the four-dot charge configuration (1,1,2,2).
The corresponding tunnel couplings are designated $\tau_{2g1g}^{L\! R}$ and $\tau_{2x1g}^{L\! R}$, and the detuning parameter between the second and third dots is $\varepsilon_{L\! R}$.
(b) Tunnel coupling control pulse for a CZ gate:
the tunnel coupling is ramped linearly from zero to its peak value, $\tau_{\alpha,\text{max}}^{L\! R}$ ($\alpha=2g1g$ or $2x1g$), 
held constant for a waiting period, then ramped back down to zero.
(c) Typical energy levels for the four logical states (blue) and 24 leakage states (red) considered here. 
Many of the leakage levels are degenerate or nearly degenerate, with degeneracy factors indicated on the right.
The most dangerous leakage states occur in the low-energy manifold, and have the same (1,2,1,2) charge configuration as the logical states.
However, not all these states couple to the logical states via second-order tunneling processes, as discussed in Appendix~\ref{Sec:DQHQTwoQubit_SubTwoDQHQ}; the number of tunnel-coupled leakage states are indicated in parentheses.
The system parameters used for this calculation are $\{\varepsilon_L,E_{ST,L},\Delta_{1L},\Delta_{2L},\varepsilon_R,E_{ST,R},\Delta_{1R},\Delta_{2R},\varepsilon_{L\! R},\mathcal{G} \}/h = \{90,12,8.4,8.4,70,9,6.3,6.3,-80 ,20\}\,\units{GHz}$, as consistent with recent experiments~\cite{npjqi201632,Thorgrimsson2017}; we also take $\tau^{L\! R}_{2g1g}=\tau^{L\! R}_{2x1g} = 0$.
Note that we choose $\Delta_{1q}=\Delta_{2q}$ to suppress single-qubit dephasing in the far-detuned regime~\cite{Wong2016,PhysRevA.95.062321, DQHQNoise}.
}
\end{figure*}

\section{Model \label{Sec:DQHQTwoQubitModel}}
The quantum dot hybrid qubit is composed of three electrons in a double quantum dot, with total spin quantum numbers $S=\frac{1}{2}$ and $S_z = -\frac{1}{2}$~\cite{Shi2012,Koh2012}.
For example, the qubit can be formed in the left two dots depicted in Fig.~\ref{Fig:DQHQTwoQubitFig1}(a).
The detuning parameter, $\varepsilon_L$, is defined as the energy bias between these dots, while the tunneling couplings between the single-electron levels indicated in the figure, $\tau^{L}_{1g2g}$ and $\tau^{L}_{1g2x}$, are rigorously defined in Appendix~\ref{Sec:DQHQTwoQubit_SubModel}, and $g$ and $x$ refer to the single-electron ground and excited energy levels, respectively.
To suppress decoherence caused by fluctuations of the detuning parameter, $\delta \varepsilon_L$, we operate the qubit in the far-detuned regime, defined as $\varepsilon_L \gg \tau^{L}_{1g 2\alpha}$ ($\alpha=g,x$).
The logical basis states are defined by their spin configurations, $\ket{0}_L = \ket{\cdot S}_L = \ket{\downarrow\! S}_L$ and $\ket{1}_L = \ket{\cdot T}_L = \sqrt{\frac{2}{3}}\ket{\uparrow\! T_-}_L - \frac{1}{\sqrt{3}}\ket{\downarrow\! T_0}_L$.
Here, the protection against charge noise arises from the fact that both states have the same (1,2) charge configuration, with one electron in dot~$1L$ and two electrons in dot~$2L$~\cite{Wong2016, Thorgrimsson2017}.

We define $E^L_{i\alpha}$ as the single-electron energy level of dot $iL$ ($i=1,2$) in its ground ($\alpha=g$) or excited ($\alpha=x$) state.
Assuming that $E^L_1 \equiv E^L_{1x}-E^L_{1g} \gg E^L_2 \equiv E^L_{2x}-E^L_{2g}$, which can be achieved in silicon dots by choosing an appropriate filled shell~\cite{Harvey-Collard:2017}, we may limit our analysis to these two basis states and the low-energy leakage state, $\ket{S \cdot }_L = \ket{S\! \downarrow }_L$.
Projecting the system Hamiltonian onto this three-state basis~\cite{Shi2012}, as described in Appendix~\ref{Sec:DQHQTwoQubit_SubDQHQ}, we obtain the effective Hubbard Hamiltonian, 
\begin{eqnarray}
\label{Eq:DQHQTwoQubitHamiltonianLQ}
\Hamiltonian_{LQ} = 
\begin{pmatrix}
-\varepsilon_L/2 & 0 & \Delta_{1L} \\
0 & -\varepsilon_L/2 + E_{ST,L} & -\Delta_{2L} \\
\Delta_{1L} & -\Delta_{2L} & \varepsilon_L/2 \\
\end{pmatrix},
\end{eqnarray}\\
where $\Delta_{1L} = -\tau^{L}_{1g2g}$, $\Delta_{2L} = -\sqrt{3/2}\,\tau^{L}_{1g2x}$, and $E_{ST,L}$ is the singlet-triplet splitting of the two-electron configuration of dot $2L$.

We now consider a two-qubit system, including the double dot $R$, as depicted in Fig.~\ref{Fig:DQHQTwoQubitFig1}(a),
making analogous definitions and assumptions as for qubit $L$.
In addition to the intra-qubit tunnel couplings, we now also include inter-qubit tunnel couplings, $\tau^{L\! R}_{2\alpha 1g}$ ($\alpha=g,x$).
In this arrangement, single-qubit gates are performed when the latter are turned off, while entangling gates are realized when they are turned on.
To model the full system, we extend the Hubbard-like model of Eq.~(\ref{Eq:DQHQTwoQubitHamiltonianLQ}) to include the four logical states, $\{\ket{00},\,\ket{01},\,\ket{10},\,\ket{11}\}$, where $\ket{i\, j} \equiv \ket{i}_L \otimes \ket{j}_R$, and any states connected to them by tunnel couplings, up to second order in the tunneling processes shown in Fig.~\ref{Fig:DQHQTwoQubitFig1}(a), $\mathcal{O}[\tau^2]$. %, $\{\tau^{L}_{1g2g},\tau^{L}_{1g2x},\tau^{R}_{1g2g},\tau^{R}_{1g2x},\tau^{L\! R}_{1g2g},\tau^{L\! R}_{1g2x}\}$. 
These states comprise the charge  configurations $(1,2,1,2)$, $(1,2,2,1)$, $(2,1,1,2)$, $(2,1,2,1)$, and $(1,1,2,2)$, with a total of $28$ basis states.
A full description of the model is presented in Appendix~\ref{Sec:DQHQTwoQubit_SubModel},
yielding the typical set of energy levels shown in Fig.~\ref{Fig:DQHQTwoQubitFig1}(c). 

Although tunnel couplings and exchange interactions enable strong and fast entangling gates, as investigated here, they also induce new leakage channels that can reduce the overall gate fidelity~\cite{DiVincenzo2000}.
In our system, the most dangerous leakage states are found in the lowest-energy (1,2,1,2) charge manifold of Fig.~\ref{Fig:DQHQTwoQubitFig1}(c), which contains the four logical states and eleven other states with similar energies.
The states in the higher energy manifolds present a weaker threat from leakage; however they also generate Coulomb interactions, due to their different charge configurations.
As explained in Appendix~\ref{Sec:DQHQTwoQubit_SubTwoDQHQ}, this generates a new term in the Hamiltonian, given by $\Hamiltonian_{C,\text{eff}} = \frac{\mathcal{G}}{4} (\hat n_{1L}-\hat n_{2L})(\hat n_{1R}-\hat n_{2R})$, where $\hat n_{iq}$ is the charge occupation of dot $i$ in qubit $q$~\cite{2015arXiv150703425M, npjqi201632,2018arXiv181203177F}.
Although the Coulomb interaction provides an alternative scheme for entangling hybrid qubits~\cite{Koh2012,2015arXiv150703425M,PhysRevB.99.195403,2018arXiv181203177F}, it also causes leakage, and dephasing due to charge noise~\cite{2018arXiv181203177F}.
Mitigating these effects requires biasing the qubits into the large-detuning regime and performing one and two-qubit gate operations as fast as possible, highlighting the importance of strong driving for high-fidelity gates~\cite{PhysRevA.95.062321}.
We note that high-fidelity single-qubit gates in hybrid qubits have been theoretically investigated elsewhere~\cite{Wong2016,PhysRevA.95.062321,DQHQNoise}, and will not be discussed in detail here.

\section{Entangling Gates \label{Sec:DQHQTwoQubitCZGates}}
In this work, we propose to implement a CZ gate by modulating the tunnel couplings between the middle two dots, $\tau^{L\! R}_{2g1g}$ and $\tau^{L\! R}_{2x1g}$.
In the logical subspace, expressed in its adiabatic basis, the effective coupling caused by this modulation (up to ${\cal O}[\tau^2]$) has the form $Z\!Z$, which generates the desired operation, $U_\text{CZ}=\text{diag}[1,1,1,-1]$.
The full unitary evolution also includes single-qubit $Z$ rotations, which can be removed later, if desired. 
The rotation angles of these incidental gates depends on details of the pulse sequence, similar to the situation in Ref.~\cite{PhysRevA.95.062321}.

Here, we consider the simple pulsing scheme shown in Fig.~\ref{Fig:DQHQTwoQubitFig1}(b), consisting of an initial linear ramp-up period $t_\text{ramp}$, to turn the tunnel coupling on, a waiting period $t_\text{wait}$, and a final ramp-down period $t_\text{ramp}$, to turn the coupling off. 
% (A similar ramping scheme has been used for single-spin qubits~\cite{PhysRevB.83.121403,Watson2018}.) 
For hybrid qubits, the entire operation can be performed in the large-detuning regime, defined as $\varepsilon_L,\,\varepsilon_R,\,\varepsilon_{L\! R} \gg t_{\alpha i \beta j}^q$, to ensure the best protection from charge noise.
In this case, however, the low-energy manifold of leakage states in Fig.~\ref{Fig:DQHQTwoQubitFig1}(c) is close in energy to the logical states.
Indeed, three of the four logical states are nearly degenerate with leakages states (although they do not necessarily couple to these states at ${\cal O}[\tau^2]$, as indicated in the figure), which increases the probability of leakage.
An adiabatic ramp sequence may be employed, to suppress the leakage.
However, in this case, charge noise can still be a problem due to the long gate time.
High-fidelity gate pulses therefore require optimization.
Below, we show that, in the presence of $1\!/\!f$ charge noise, optimal results are achieved by employing fast, nonadiabatic ramps in which leakage occurs but is reversed by the end of the gate operation.

\begin{figure*}[t]
\includegraphics[width=7in]{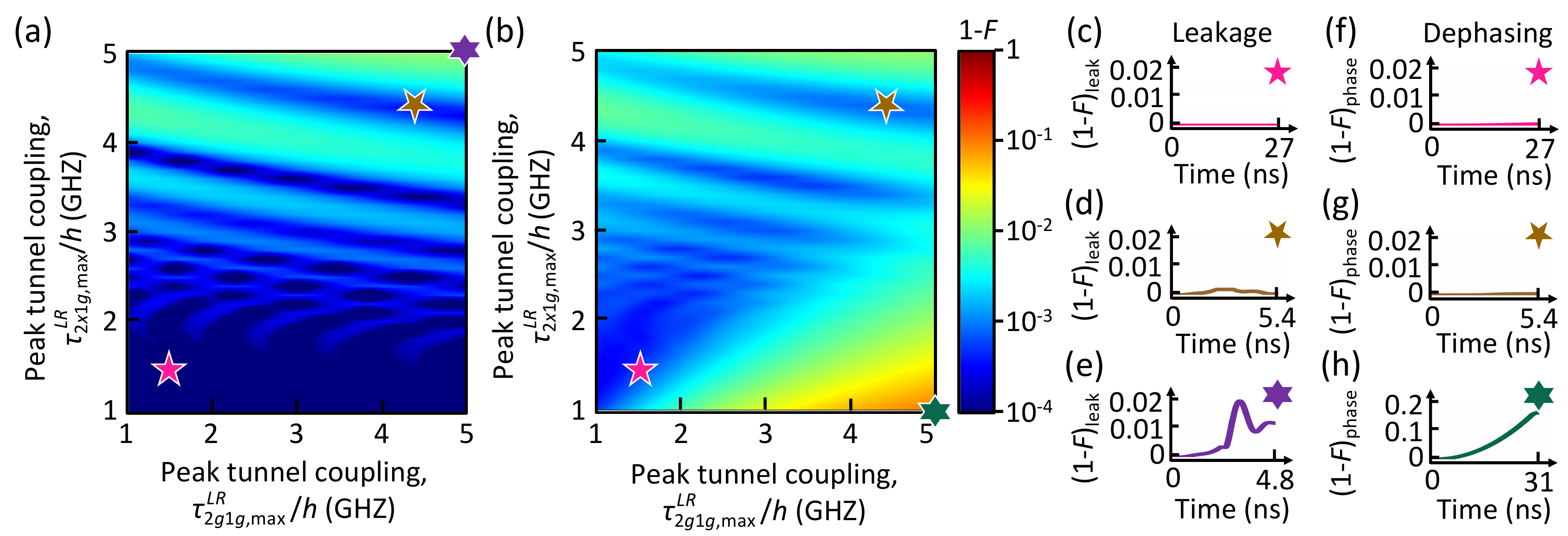}
\caption[Controlled-Z Fidelity versus maximum tunnel couplings]{
\label{Fig:DQHQTwoQubitFig2}
CZ gate fidelities.
In (a) and (b), gate infidelities are plotted as a function of pulse peak heights, $\tau^{L\! R}_{2g1g,\text{max}}$ and $\tau^{L\! R}_{2x1g,\text{max}}$, in (a) the absence, or (b) the presence of quasistatic charge noise on the detuning parameters.
Here, the simulation parameters are the same as in Fig.~\ref{Fig:DQHQTwoQubitFig1}(c) and $t_\text{ramp} = 2.25 \,\units{ns}$.
(a) In the absence of charge noise, the fidelity is dominated by leakage, as discussed in the main text. 
The leakage is enhanced in the nonadiabatic regime (e.g., the  brown or purple stars at $\{\tau^{L\! R}_{2g1g,\text{max}},\tau^{L\! R}_{2x1g,\text{max}}\}/h = \{4.2,4.4\}\,\units{GHz}$ and $\{\tau^{L\! R}_{2g1g,\text{max}},\tau^{L\! R}_{2x1g,\text{max}}\}/h = \{5.0,5.0\}\,\units{GHz}$, respectively). 
Fidelity is enhanced in the adiabatic regime (e.g., the pink star at $\{\tau^{L\! R}_{2g1g,\text{max}},\tau^{L\! R}_{2x1g,\text{max}}\}/h = \{1.5,1.5\}\,\units{GHz}$) because slower gates suppress the leakage, and charge noise is not included in this panel.
(b) In the presence of quasistatic charge noise, dephasing generally reduces the fidelity.
Here we assume detuning fluctuations with standard deviations $\sigma_{\varepsilon_L}=\sigma_{\varepsilon_R}=\sigma_{\varepsilon_{L\! R}} = 4.14 \,\units{\mu eV}$ (=1~GHz).
Dephasing is not as strong along the line $\tau^{L\! R}_{2g1g,\text{max}}\approx\tau^{L\! R}_{2x1g,\text{max}}$ [e.g., the pink and brown stars, which are the same as in (a)], where the energy dispersion depends relatively weakly on the detunings, but it is strongly enhanced near the bottom-right portion of the plot (e.g., the green star at $\{\tau^{L\! R}_{2g1g,\text{max}},\tau^{L\! R}_{2x1g,\text{max}}\}/h = \{1,5\}\,\units{GHz}$), where the opposite is true.
Note that the fidelity at the brown star is $>$99.9\%, even for short gate times, $2t_\text{ramp}+t_\text{wait}\approx 5$~ns. 
(c)-(e) Leakage contributions to the infidelity (defined in Appendix~\ref{Sec:DQHQTwoQubit_SubContributions}) vs.\ gate time, at the starred tunings shown in (a).
These calculations do not include charge noise.
As shown in Appendix~\ref{sec:Leakage-path}, the observed fringes can be attributed to phase acquired in the leakage states.
(f)-(h) Dephasing contributions to the infidelity vs.\ gate time, at the starred tunings in (b).
These calculations include charge noise, as described in Eq.~(\ref{eq:Fphase}). 
To get a sense for the errors that arise in a given evolution, we consider here a single realization of quasistatic noise ($N=1$) with $\delta \varepsilon_L = \delta \varepsilon_R = \delta \varepsilon_{LR} = 4.14 \, \units{\mu eV}$.
}
\end{figure*}

\subsection{Absence of noise}
To characterize errors in gate operations, we perform numerical simulations of CZ gates under realistic operating conditions.
We first explore the effects of leakage on the gate fidelity by performing simulations in the absence of charge noise for a range of peak tunnel couplings, $\tau^{L\! R}_{2g1g,\text{max}}$ and $\tau^{L\! R}_{2x1g,\text{max}}$, but a fixed value of $t_\text{ramp}$.
In this work, we choose $t_\text{ramp}$=2.25~ns because it is considered to be fast.
(In this procedure, $\tau^{L\! R}_{2g1g,\text{max}}$ and $\tau^{L\! R}_{2x1g,\text{max}}$, rather than $t_\text{ramp}$, determine whether the gate is adiabatic.)
$\varepsilon_{L\! R}$ is chosen to be comparable to $\varepsilon_{L}$ and $\varepsilon_{R}$, so that the energy bias between dots~$1R$ and $2L$ is relatively large, which helps to suppress decoherence caused by $\delta \varepsilon_{L\! R}$.
The unitary evolution generated by this operation causes leakage, which reduces the gate fidelity; however since the evolution is coherent, we refer to the resulting fidelity measure as ``intrinsic."
Since the ideal evolution produces a CZ gate, combined with single-qubit gates, $t_\text{wait}$ is determined by attaining the desired Makhlin invariants~\cite{Makhlin2002}.
Details of the simulations and calculations are given in Appendix~\ref{Sec:DQHQTwoQubit_SubSimulationMethod}.

All loss of fidelity in the absence of charge noise can be attributed to nonadiabaticity.
To further characterize such errors, we introduce the following classifications.
(i) ``Qubit-transition" errors arise from non-$Z\!Z$-type couplings induced while ramping, in the adiabatic basis, acting only within the logical subspace.
(ii) ``Leakage" errors correspond to transfer of probability density outside the logical subspace.
(iii) ``Phase" errors correspond to the incorrect calibration of $Z\!Z$-type couplings (e.g. when transitions occur into, then out of, the leakage subspace), or due to conventional dephasing processes.
%effective $Z\!Z$-type couplings that arise from transitions into, then out of, the leakage subspace.
Precise definitions of these contributions to the infidelity are given in Appendix~\ref{Sec:DQHQTwoQubit_SubContributions}.

The main results of our CZ-gate fidelity simulations are shown in Fig.~\ref{Fig:DQHQTwoQubitFig2}(a), while a breakdown of the qubit-transition, leakage, and phase contributions is provided in Appendix~\ref{Sec:DQHQTwoQubit_SubSimulationResults}.
This breakdown clearly shows that the intrinsic infidelity is dominated by leakage, which can be understood from the following arguments.
(i) Qubit-transition errors are caused by nonadiabatic processes within the logical subspace.
However, the relatively large energy splittings between the logical states, compared to several nearly-degenerate couplings between logical and leakage states shown in Fig.~\ref{Fig:DQHQTwoQubitFig1}(c), suppress qubit-transition errors with respect to leakage errors.
(ii) The method we use to determine $t_\text{wait}$ effectively reduces phase errors to zero, as discussed in Appendix~\ref{sec:twait}.

The lower portion of Fig.~\ref{Fig:DQHQTwoQubitFig2}(a) corresponds to the adiabatic regime, where leakage is negligible but gate times are long.
In the absence of charge noise, we obtain very high fidelities here.
However, when charge noise is included in the simulations, as described below, dephasing can suppress the fidelity.
The upper portion of Fig.~\ref{Fig:DQHQTwoQubitFig2}(a) corresponds to the nonadiabatic regime, where we observe alternating fringes of low and high fidelity, reminiscent of a diffraction pattern or coherent oscillations.
For a high-fidelity fringe [e.g., the brown star in Fig.~\ref{Fig:DQHQTwoQubitFig2}(a)], the leakage error (defined above) initially increases, but is eventually suppressed at the end of the gate evolution, as shown in Fig.~\ref{Fig:DQHQTwoQubitFig2}(d).
In a low-fidelity fringe [e.g., the purple star], the leakage suppression at the end of the gate is incomplete, as shown in Fig.~\ref{Fig:DQHQTwoQubitFig2}(e).

\subsection{Quasistatic detuning noise}
We now include charge noise in our CZ gate simulations by adding independent fluctuations to the three detuning parameters, $\varepsilon_L \!\rightarrow\! \varepsilon_L+\delta \varepsilon_L$, $\varepsilon_R \!\rightarrow\!\varepsilon_R+\delta \varepsilon_R$, and $\varepsilon_{L\! R} \!\rightarrow\!\varepsilon_{L\! R}+\delta \varepsilon_{L\! R}$.
We first consider quasistatic noise, with $\delta \varepsilon_L$, $\delta \varepsilon_R$, and $\delta \varepsilon_{L\!R}$ drawn from Gaussian distributions with standard deviations $\sigma_{\varepsilon_L}=\sigma_{\varepsilon_R}=\sigma_{\varepsilon_{L\! R}} = 4.14 \,\units{\mu eV}$ (=1~GHz) that are consistent with recent experiments~\cite{PhysRevLett.105.246804, Wu19082014,Shi2013,Thorgrimsson2017}. 
We then average the results of many simulations and compute the fidelity as before.
(For details, see Appendix~\ref{Sec:DQHQTwoQubit_SubSimulationMethod}.)

Results of these calculations are presented in Fig.~\ref{Fig:DQHQTwoQubitFig2}(b), alongside results obtained in the absence of noise [Fig.~\ref{Fig:DQHQTwoQubitFig2}(a)].
Here, any new suppression of the fidelity can be attributed to charge-noise-induced dephasing.
In general, we see that the dephasing is reduced when the two tunnel couplings, $\tau^{L\! R}_{2g1g}$ and $\tau^{L\! R}_{2x1g}$, are approximately equal, which is reminiscent of the condition for a flat single-qubit energy dispersion, $\Delta_{1q} \approx \Delta_{2q}$ ($q = L,\,R$)~\cite{Wong2016}.
In the adiabatic regime, we note that, although dephasing can strongly suppress the fidelity when the gate is slow [lower-right portion of \ref{Fig:DQHQTwoQubitFig2}(b)], there is still a wide region with fidelities $>$99.9\%, where the dependence on $\tau^{L\! R}_{2g1g}$ and $\tau^{L\! R}_{2x1g}$ is weak [e.g., the pink star in Fig.~\ref{Fig:DQHQTwoQubitFig2}(b)].
In the nonadiabatic regime (upper portion of the plot), the faster gates generally overcome dephasing caused by quasistatic noise.
The leakage-induced fringes in Fig.~\ref{Fig:DQHQTwoQubitFig2}(a) are therefore directly reflected in Fig.~\ref{Fig:DQHQTwoQubitFig2}(b), with corresponding fidelities $>$99.9\% [e.g., the brown star].
%However, if the detuning fluctuations are not quasistatic, then we expect the nonadiabatic gates to perform better than the adiabatic gates, because dephasing effects on fidelity typically decrease as gate speed increases.
%This expectation is born out by the results in Sec.~\ref{Sec:OneOverFDetuningNoise} below.

Since many well known quantum algorithms utilize CNOT gates, rather than CZ gates, we also provide a fidelity estimate for the former.
In particular, we consider the Zero-CNOT (Z-CNOT) gate, defined as $\text{Z-CNOT}= \sigma_x \otimes \ket{0}    \langle 0| + I_2 \otimes \ket{1}  \langle 1|$, where $\sigma_x$ is a Pauli matrix and $I_2$ is the $2\times 2$ identity matrix. 
The specific gate sequence is  constructed from CZ as follows:
\begin{equation} \label{eq:Z-CNOT}
\text{Z-CNOT} = Y_L(-\pi/2) \, Z_L\!\left(-\pi\right) Z_R\!\left(-\pi\right) \, \text{CZ} \,\, Y_L(\pi/2),
\end{equation}
where $D_q(\theta)$ is a rotation of angle $\theta$ about axis $D=Y,Z$ on qubit $q = L,\,R$. 
We note that $Z$ rotations can be performed virtually here, by adjusting the phase of the AC drive~\cite{2001Natur.414..883V,Watson2018}.
Following the procedure of Ref.~\cite{DQHQNoise} to optimize strong-driving protocols for single-qubit gates in the presence of quasistatic noise, we obtain single-qubit gate fidelities $>$99.996\% (much higher than CZ gates), using the parameters described in Appendix~\ref{sec:ZCNOT}.
The resulting Z-CNOT fidelities are essentially identical to those of CZ gates.
%Clearly, if low-fidelity single-qubit gates are employed, the resulting Z-CNOT fidelities would also be suppressed.

\begin{figure}[t]
\includegraphics[width=2.7in]{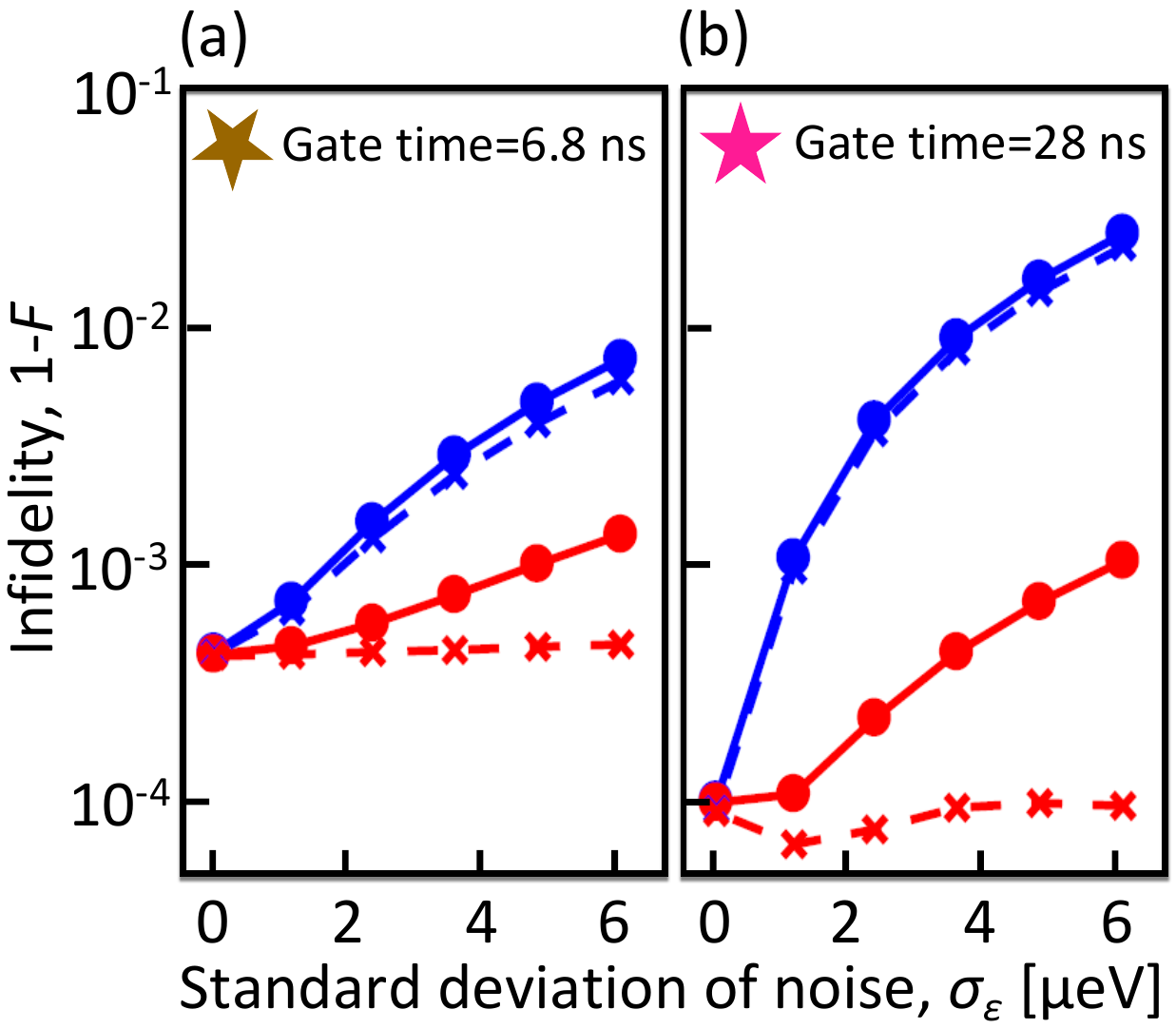}
\caption[Controlled-NOT fidelity versus noise strength]{
The total infidelity ($\bullet$) and the leakage contribution to the infidelity ($\times$) of a Z-CNOT gate, as a function of the noise standard deviation, calculated at the parameter values marked with brown and pink stars in Fig.~\ref{Fig:DQHQTwoQubitFig2}.
Here, we consider both quasistatic (red) and $1\!/\!f$ noise (blue); the lines are guides to the eye.
All single-qubit gates are performed using strong driving with smoothed rectangular pulse envelopes of amplitude $\{A_{\varepsilon},A_\Delta\}/h = \{27, 3.1\} \,\units{GHz}$, as described in Refs.~\cite{PhysRevA.95.062321,DQHQNoise}.
(a) Typical nonadiabatic gate behavior, for pulse peak heights $\{\tau^{L\! R}_{2g1g,\text{max}},\tau^{L\! R}_{2x1g,\text{max}}\}/h = \{4.2,4.4\}\,\units{GHz}$, yielding gate times of $6.8\,\units{ns}$.
The resulting gate fidelities can be $>$99.9\% for either type of noise; however, $1\!/\!f$ noise yields lower fidelities than quasistatic noise because it causes more leakage.
(b) Typical adiabatic gate behavior, for pulse peak heights $\{\tau^{L\! R}_{2g1g,\text{max}},\tau^{L\! R}_{2x1g,\text{max}}\}/h = \{1.5,1.5\}\,\units{GHz}$, yielding gate times of $28\,\units{ns}$. 
For the range of noise strengths considered here, the fidelity of the adiabatic gate is more strongly suppressed for $1\!/\!f$ noise, due to noise-induced leakage during the longer gates.
\label{Fig:DQHQTwoQubitFig3}
}
\end{figure}

\subsection{$1\!/\!f$ detuning noise \label{Sec:OneOverFDetuningNoise}}
Finally, we consider the effect of $1\!/\!f$ detuning noise on the two-qubit gate fidelity.
Since these simulations are numerically more intensive (see Appendix~\ref{sec:sims-noise}), we focus only on the two high-fidelity working points indicated with pink and brown stars in Figs.~\ref{Fig:DQHQTwoQubitFig2}(a) and \ref{Fig:DQHQTwoQubitFig2}(b).
The resulting Z-CNOT infidelities are plotted in Fig.~\ref{Fig:DQHQTwoQubitFig3} for these two tunings.
Generally, we find that $1\!/\!f$ noise suppresses the fidelity more than quasistatic noise, due to the resonant excitation of leakage states by the high-frequency components of the $1\!/\!f$ noise spectrum.
The predominance of leakage is revealed by comparing the total infidelity (solid blue lines) to the leakage contribution (dashed blue lines), which essentially overlap.
Despite the enhanced leakage, gate fidelities $>$99.9\% can still be achieved when the standard deviation of the detuning fluctuations satisfies $\sigma_\varepsilon\lesssim 2$~$\mu$eV (4.2~$\mu$eV) for $1\!/\!f$ (quasistatic) noise in the nonadiabatic regime [Fig.~\ref{Fig:DQHQTwoQubitFig3}(a)], or $\sigma_\varepsilon\lesssim 1$~$\mu$eV (6~$\mu$eV) in the adiabatic regime [Fig.~\ref{Fig:DQHQTwoQubitFig3}(b)].
Comparing Figs.~\ref{Fig:DQHQTwoQubitFig3}(a) and \ref{Fig:DQHQTwoQubitFig3}(b), we also see that, for typical $1\!/\!f$ noise levels, fast gates generally achieve higher fidelities than slow gates, in contrast to the results for quasistatic noise.

\section{Discussion\label{Sec:DQHQTwoQubitDiscussion}}
We have shown that high-fidelity two-qubit gates can be achieved in quantum-dot hybrid qubits, even in the presence of substantial detuning noise, by modulating the inter-qubit tunnel couplings and therefore the exchange interactions.
Moreover, the gates can be implemented with both qubits operated to the large-detuning regime, where their single-qubit dephasing rates are suppressed.

In this work, we have optimized only a subset of system parameters.
We now comment on alternative schemes for improving the gate fidelity.
First, we note that leakage can be suppressed by moving the nearly-degenerate leakage energy levels further away from the logical levels.
This can be accomplished by increasing $E_{ST,q}$ or $\Delta_{iq}$ ($i=1,2$; $q=L,R$), or both.
Splitting these energy levels has the additional benefit of suppressing the undesired effects of strong driving.
Leakage can also be suppressed by replacing the linear ramp, considered in this work [Fig.~\ref{Fig:DQHQTwoQubitFig1}(b)], with specially shaped ramps~\cite{Motzoi2009,Gambetta2011}, including ramps that provide shortcuts to adiabaticity~\cite{TORRONTEGUI2013117}.
Such techniques improve the intrinsic fidelity, and also provide opportunities for faster gates, which suppress noise-induced errors, particularly for the case of $1\!/\!f$ noise.

\section{Conclusion\label{Sec:DQHQTwoQubitConclusion}}
We have proposed and analyzed a two-qubit entangling gate for quantum-dot hybrid qubits based on exchange interactions mediated by tunable tunnel couplings between the qubits.
We have shown that native CZ gates are obtained by varying the tunnel couplings, either adiabatically or nonadiabatically, and we have performed simulations of gate operations in the presence of realistic levels of charge noise on the detuning parameters.

For the case of quasistatic noise with a standard deviation of $4.14\,\units{\mu eV}$ in the detuning parameters, we obtain CZ gate fidelities $>$99.9\% in both the adiabatic and nonadiabatic regimes.
In the latter case, the fidelity is found to oscillate as a function of the control parameters, so obtaining high fidelities requires choosing appropriate operating parameters.
Similarly, we find that a Z-CNOT gate with fidelity $>$99.9\% can be achieved by performing composite pulse sequences, utilizing the single-qubit gate methods proposed in Ref.~\cite{PhysRevA.95.062321,DQHQNoise}.
Finally, we perform simulations of Z-CNOT gates with $1\!/\!f$ charge noise, obtaining gate fidelities that are slightly suppressed by leakage.
However, we observe that this effect is roughly proportional to the gate time, suggesting that faster (nonadiabatic) gates are desirable. 

The exchange-based coupling scheme proposed here applies specifically to quantum-dot hybrid qubits operated in the far-detuned regime, where single-qubit dephasing is suppressed.
However, similar methods can be applied to related systems, such as singlet-triplet and exchange-only qubits, when operated at far-detuned sweet spots. 

\section*{Acknowledgments}
We are grateful to M.\ A.\ Eriksson for many enlightening discussions.
This work was supported in part by ARO (W911NF-17-1-0274) and the Vannevar Bush
Faculty Fellowship program sponsored by the Basic Research Office of the Assistant Secretary of Defense for
Research and Engineering and funded by the Office of Naval Research through grant N00014-15-1-0029. The views and conclusions contained in this document are those of the authors and should not be interpreted as representing the official policies, either expressed or implied, of the Army Research Office (ARO), or the U.S. Government. The U.S. Government is authorized to reproduce and distribute reprints for Government purposes notwithstanding any copyright notation herein.

\appendix

\section{Model \label{Sec:DQHQTwoQubit_SubModel}}
In this Appendix, we derive the Hamiltonian for the four-dot system considered in the main text.
We first evaluate a Hubbard Hamiltonian, then project it onto an appropriate Hilbert space to obtain a Hamiltonian matrix.

We derive a generalized Hubbard-like Hamiltonian for a system of six interacting electrons confined to four quantum dots, using a method similar to the one employed in Ref.~\cite{Shi2012}.
To simplify the discussion and notation, in this section we relabel the quantum dots 
$1L$, $2L$, $1R$, and $2R$ as $1$, $2$, $3$, and $4$, respectively.

The Hubbard-like model can be expressed as
\begin{eqnarray}
\label{Eq:DQHQTwoQubit_SubHubbard_Like_Model}
\hat \Hamiltonian_\text{Hubbard} &=& \hat \Hamiltonian_E + \hat \Hamiltonian_T + \hat \Hamiltonian_C, \nonumber \\
&=& \sum\limits_{i,\alpha,s} (E_{i\alpha} + \mu_i) \hat n_{i\alpha s} 
\nonumber \\
&& + \sum\limits_{i \neq j}\sum\limits_{\alpha,\beta}\sum\limits_s \tau_{i\alpha j \beta} \hat  c_{i\alpha s}^{\dagger} \hat c_{j\beta s}  
\nonumber \\
&& + \hat \Hamiltonian_{C,\text{intra}} + \hat \Hamiltonian_{C,\text{inter}},
\end{eqnarray}
where $i,j$ label the quantum dots, $\alpha,\beta$ label the single-electron orbitals in each dot, $s$ labels the electron spins, $c^{\dagger}_{i\alpha s}$ ($c_{i\alpha s}$) are the electron creation (annihilation) operators, and $\hat n_{i\alpha s} = \hat c^{\dagger}_{i\alpha s} \hat c_{i\alpha s}$ is the number operator.
The various Hamiltonian terms are identified as follows.

The $\hat \Hamiltonian_E $ term describes the intra-dot, non-interacting electron energy. 
$\mu_i$ is the electrostatic energy of an electron in dot $i$, induced by voltages applied to the top gates.
$E_{i\alpha}$ is the orbital energy, defined as
\begin{equation}
E_{i\alpha} = \int d\mathbf{r} \,\phi_{i\alpha}^*(\mathbf{r} ) \left[ \frac{\hat{\mathbf{P}}^2 }{2 m^*}+ V(\mathbf{r} ) \right]\phi_{i\alpha}(\mathbf{r} ),
\end{equation} 
where $\phi_{i\alpha}$ is a single-electron wavefunction for dot $i$ in orbital $\alpha$, $m^*$ is the effective mass of an electron in the conduction band, $\hat{\mathbf{P}}$ is the momentum operator, and $V(\mathbf{r} )$ is the quantum dot confinement potential.
For most calculations, the electron is assumed to be effectively two-dimensional, as consistent with the lowest-subband approximation, and $m^*$ is taken to be the transverse effective mass~\cite{DaviesBook}.
(Here, we consider Si, where $m^*\approx 0.2m_0$ and $m_0$ is the electron rest mass.)
For simplicity, we ignore excited states  beyond the first excited orbital.
Additionally, we assume that $E_{1x},\,E_{3x} \gg E_{2x},\,E_{4x}$, as consistent with many recent experiments~\cite{ShiSimmonsWardEtAl2014}, so that we can also ignore the first excited states in dots 1 and 3. 
Later on, we will also adopt the simplified notation for the excitation energy, $E_i\equiv E_{ix} - E_{ig} $.

The $\hat \Hamiltonian_T $ term describes the tunneling between single-electron states in different dots. Here we only consider the tunneling between nearest-neighbor dots. The tunnel couplings are given by the single-particle integrals
\begin{equation}
\tau_{i\alpha j\beta} = \int d\mathbf{r}  \,\phi_{i\alpha}^*(\mathbf{r} )\left[\frac{\hat{\mathbf{P}}^2 }{2 m^*} + V(\mathbf{r} )\right] \phi_{i\beta}(\mathbf{r} ),
\end{equation} 
where we note that $\tau_{i\alpha j\beta}^* = \tau_{j\beta i\alpha}$; for simplicity, we take $\tau_{i\alpha j\beta}$  to be real here.

\begin{widetext}

The $\hat \Hamiltonian_C$ term describes the Coulomb interactions.  It can be divided into intra-dot ($\hat \Hamiltonian_{C,\text{intra}}$) and inter-dot ($\hat \Hamiltonian_{C,i\text{nter}}$) contributions. 
The former is given by
\begin{equation}
\hat \Hamiltonian_{C,\text{intra}} = \frac{1}{2}\sum\limits_{i}\sum\limits_{\alpha,\beta}\sum\limits_{s,s^\prime} \left(C_{i\alpha i\beta} \, \hat c^{\dagger}_{i\alpha s} \hat c^{\dagger}_{i\beta s^\prime} \hat c_{i\beta s^\prime}  \hat c_{i\alpha s} + K_{i\alpha i\beta} \, \hat c^{\dagger}_{i\alpha s} \hat c^{\dagger}_{i\beta s^\prime} \hat c_{i\alpha s^\prime}  \hat c_{i\beta s}\right),
\end{equation}
which can be further separated into direct and exchange components,
\begin{eqnarray}
C_{i\alpha i\beta}  &=& \int d\mathbf{r} \, d\mathbf{r}^{\prime}\,\phi_{i\alpha}^*(\mathbf{r} )\phi_{i\beta}^*(\mathbf{r}^{\prime} )\frac{e^2}{4\pi\epsilon_r\epsilon_0|\mathbf{r}-\mathbf{r^{\prime}}|} \phi_{i\beta}(\mathbf{r} ^{\prime})\phi_{i\alpha}(\mathbf{r} ), \\
K_{i\alpha i\beta}  &=& \int d\mathbf{r} \, d\mathbf{r}^{\prime} \, \phi_{i\alpha}^*(\mathbf{r} )\phi_{i\beta}^*(\mathbf{r}^{\prime} )\frac{e^2}{4\pi\epsilon_r\epsilon_0|\mathbf{r}-\mathbf{r^{\prime}}|} \phi_{i\alpha}(\mathbf{r} ^{\prime})\phi_{i\beta}(\mathbf{r} ), 
\end{eqnarray} 
respectively, where $e$ is the electron charge, $\epsilon_0$ is the permittivity of the vacuum, and $\epsilon_r$ is the relative permittivity of the quantum well or inversion layer. 
(For low-temperature Si, $\epsilon_r=11.4$.)
The inter-dot term can be written as 
\begin{equation}
\Hamiltonian_{C,\text{inter}} = \frac{1}{2}\sum\limits_{i\neq j}\sum\limits_{k\neq l}\sum\limits_{\alpha,\beta,\gamma,\delta}\sum\limits_{s,s^\prime}  \Gamma_{i j k l}^{\alpha \beta \gamma \delta}  \, \hat c^{\dagger}_{i\alpha s} \hat c^{\dagger}_{j\beta s^\prime} \hat c_{k\gamma s^\prime} \hat c_{l\delta s},
\end{equation}
where $\Gamma_{i j k l}^{\alpha \beta \gamma \delta} $ are general two-particle integrals given by
\begin{equation}
\Gamma_{i j k l}^{\alpha \beta \gamma \delta} = \int d\mathbf{r} \, d\mathbf{r}^{\prime} \,\phi_{i\alpha}^*(\mathbf{r} )\phi_{j\beta}^*(\mathbf{r}^{\prime} )\frac{e^2}{4\pi\epsilon_r\epsilon_0|\mathbf{r}-\mathbf{r^{\prime}}|} \phi_{k\gamma}(\mathbf{r} ^{\prime})\phi_{l\delta}(\mathbf{r} ).
\end{equation} 
In the following subsections, we apply the Hubbard Hamiltonian, Eq.~(\ref{Eq:DQHQTwoQubit_SubHubbard_Like_Model}), to the problem of interest, in successive steps.
We do not explicitly evaluate the spatial integrals described above.
For such derivations, we refer the interested reader to Refs.~\cite{PhysRevB.83.161301,PhysRevB.83.235314}, as an example.

\begin{table*}
\label{table:12states}
	\caption{Basis states for the (1,2) charge configuration of the left double dot.}
\begin{tabular}{lccr}
	\hline\hline
	State label& Second-quantized expression & $S_L$ & $S_{Lz}$  \\ \hline
	$\ket{0}_L = \ket{\downarrow\! S}_L$ & $ c_{1g\downarrow}^{\dagger}c_{2g\downarrow}^{\dagger}c_{2g\uparrow}^{\dagger}\ket{\Omega}_1\ket{\Omega}_2$ & $\frac{1}{2}$ &$-\frac{1}{2}$  \\
	$\ket{v_0}_L = -\ket{\uparrow\! S}_L $ & $ -c_{1g\uparrow}^{\dagger}c_{2g\downarrow}^{\dagger}c_{2g\uparrow}^{\dagger}\ket{\Omega}_1\ket{\Omega}_2$ & $\frac{1}{2}$ & $\frac{1}{2}$  \\
	$\ket{1}_L = \sqrt{\frac{2}{3}}\ket{\uparrow\! T_-}_L - \frac{1}{\sqrt{3}}\ket{\downarrow\! T_0}_L $ & $ \left(\sqrt{\frac{2}{3}} c_{1g\uparrow}^{\dagger}c_{2g\downarrow}^{\dagger}c_{2e\downarrow}^{\dagger} - \frac{1}{\sqrt{6}} c_{1g\downarrow}^{\dagger}c_{2g\downarrow}^{\dagger}c_{2e\uparrow}^{\dagger}- \frac{1}{\sqrt{6}} c_{1g\downarrow}^{\dagger}c_{2g\uparrow}^{\dagger}c_{2e\downarrow}^{\dagger}\right)\ket{\Omega}_1\ket{\Omega}_2$ 
	\hspace{.1in}
	& $\frac{1}{2}$ & $-\frac{1}{2}$ \\ 
	$\ket{v_1}_L = \sqrt{\frac{2}{3}}\ket{\downarrow\! T_+}_L - \frac{1}{\sqrt{3}}\ket{\uparrow\! T_0}_L $ & $ \left(\sqrt{\frac{2}{3}} c_{1g\downarrow}^{\dagger}c_{2g\uparrow}^{\dagger}c_{2e\uparrow}^{\dagger} - \frac{1}{\sqrt{6}} c_{1g\uparrow}^{\dagger}c_{2g\downarrow}^{\dagger}c_{2e\uparrow}^{\dagger}- \frac{1}{\sqrt{6}} c_{1g\uparrow}^{\dagger}c_{2g\uparrow}^{\dagger}c_{2e\downarrow}^{\dagger}\right)\ket{\Omega}_1\ket{\Omega}_2$ & $\frac{1}{2}$ & $\frac{1}{2}$ \\
	$\ket{2}_L =\ket{\downarrow\! T_-}_L $ & $ c_{1g\downarrow}^{\dagger}c_{2g\downarrow}^{\dagger}c_{2e\downarrow}^{\dagger}\ket{\Omega}_1\ket{\Omega}_2$ & $\frac{3}{2}$ & $-\frac{3}{2}$ \\ 
	$\ket{3}_L =\frac{1}{\sqrt{3}}\ket{\uparrow\! T_-}_L +\sqrt{\frac{2}{3}}\ket{\downarrow\! T_0}_L$ & $\frac{1}{\sqrt{3}}\left(c_{1g\uparrow}^{\dagger}c_{2g\downarrow}^{\dagger}c_{2e\downarrow}^{\dagger}+c_{1g\downarrow}^{\dagger}c_{2g\downarrow}^{\dagger}c_{2e\uparrow}^{\dagger}+c_{1g\downarrow}^{\dagger}c_{2g\uparrow}^{\dagger}c_{2e\downarrow}^{\dagger}\right)\ket{\Omega}_1\ket{\Omega}_2$ & $\frac{3}{2}$ & $-\frac{1}{2}$ \\ 
	$\ket{v_3}_L =\frac{1}{\sqrt{3}}\ket{\downarrow\! T_+}_L +\sqrt{\frac{2}{3}}\ket{\uparrow\! T_0}_L $ 
	\hspace{.1in} 
	& $ \frac{1}{\sqrt{3}}\left(c_{1g\downarrow}^{\dagger}c_{2g\uparrow}^{\dagger}c_{2e\uparrow}^{\dagger}+c_{1g\uparrow}^{\dagger}c_{2g\downarrow}^{\dagger}c_{2e\uparrow}^{\dagger}+c_{1g\uparrow}^{\dagger}c_{2g\uparrow}^{\dagger}c_{2e\downarrow}^{\dagger}\right)\ket{\Omega}_1\ket{\Omega}_2$ & $\frac{3}{2}$ & $\frac{1}{2}$  \\ 
	$\ket{v_2}_L =\ket{\uparrow\! T_+}_L $ & $ c_{1g\uparrow}^{\dagger}c_{2g\uparrow}^{\dagger}c_{2e\uparrow}^{\dagger}\ket{\Omega}_1\ket{\Omega}_2$ & $\frac{3}{2}$ & $\frac{3}{2}$ \\ 
	\hline \hline 
	\end{tabular}
\end{table*}

\end{widetext}

\subsection{Single Quantum Dot \label{Sec:DQHQTwoQubit_SubSQD}}
We first consider a single quantum dot containing two electrons.
In this case, the tunneling ($\Hamiltonian_T$) and inter-dot Coulomb ($\Hamiltonian_{C,\text{inter}} $) contributions are both zero.
The energies of the single-electron states are simply $E_{i\alpha} + \mu_{i}$.
Considering only the two lowest-energy single-electron levels in dot $i$, the two-electron states can be defined as eigenstates of the total spin:
\begin{gather}
\ket{S}_i = c_{ig\downarrow}^{\dagger}c_{ig\uparrow}^{\dagger}\ket{\Omega}_i, \label{eq:singlet} \\
\ket{T_-}_i = c_{ig\downarrow}^{\dagger}c_{ix\downarrow}^{\dagger}\ket{\Omega}_i, \label{eq:triplet-} \\
\ket{T_0}_i = \frac{1}{\sqrt{2}}(c_{ig\downarrow}^{\dagger}c_{ix\uparrow}^{\dagger} + c_{ig\uparrow}^{\dagger}c_{ix\downarrow}^{\dagger})\ket{\Omega}_i, \\
\ket{T_+}_i = c_{ig\uparrow}^{\dagger}c_{ix\uparrow}^{\dagger}\ket{\Omega}_i, \label{eq:triplet+}
\end{gather}
where Eq.~(\ref{eq:singlet}) is a singlet state, Eqs.~(\ref{eq:triplet-})-(\ref{eq:triplet+}) are triplet states, and $\ket{\Omega}_i$ represents the vacuum state of dot $i$.
Evaluating Eq.~(\ref{Eq:DQHQTwoQubit_SubHubbard_Like_Model}) in this basis, we obtain the energies
\begin{gather}
E_{\ket{S}_i}  = 2 E_{ig} + 2 \mu_i + C_{igig} + K_{igig}, \\
E_{\ket{T}_i}  = E_{ig} +E_{ix} +2 \mu_i + C_{igix} - K_{igix},
\end{gather}
where the triplet states are degenerate in the absence of a magnetic field.
The resulting singlet-triplet splitting is given by $E_{ST,i} = E_{\ket{T}_i}- E_{\ket{S}_i} = E_i  + C_{igix} - K_{igix} -C_{igig} - K_{igig}$.

\begin{table}[b]
\caption{Basis states for the (2,1) charge configuration of the left double dot.}
	 \begin{tabular}{lccr}
	\hline \hline
	 State label& Second-quantized expression & $S_L$ & $S_{Lz}$  \\ \hline
	$\ket{\mathcal{L}}_L= \ket{S\! \downarrow}_L $ & $ c_{1g\downarrow}^{\dagger}c_{1g\uparrow}^{\dagger} c_{2g\downarrow}^{\dagger}\ket{\Omega}_1\ket{\Omega}_2$ & $\frac{1}{2}$ &$-\frac{1}{2}$  \\ 
	$\ket{v_\mathcal{L}}_L = -\ket{S\! \uparrow }_L $ 
	\hspace{.1in}
	& $- c_{1g\downarrow}^{\dagger}c_{1g\uparrow}^{\dagger} c_{2g\uparrow}^{\dagger}\ket{\Omega}_1\ket{\Omega}_2$ 
	\hspace{.1in}
	& $\frac{1}{2}$ & $\frac{1}{2}$  \\ \hline \hline
	\end{tabular}
\end{table}

\subsection{Quantum-dot hybrid qubit \label{Sec:DQHQTwoQubit_SubDQHQ} }
Next we consider a quantum-dot hybrid qubit formed of three electrons in a double dot~\cite{Shi2012}, and we assume a fixed, total spin of $S = \frac{1}{2}$ and $S_z = -\frac{1}{2}$.
For definiteness, we consider the left-most pair of dots in Fig.~1(a) of the main text, which form the qubit designated $L$. 
When $E_{1x}\gg E_{2x}$, the three lowest-energy basis states can be defined as 
\begin{widetext}
\begin{eqnarray}
\ket{\cdot S}_L&=&\ket{\downarrow\! S}_L =   c_{1g\downarrow}^{\dagger}c_{2g\downarrow}^{\dagger}c_{2g\uparrow}^{\dagger}\ket{\Omega}_1\ket{\Omega}_2, \\
\ket{\cdot T}_L &=& \sqrt{\frac{2}{3}}\ket{\uparrow\! T_-}_L - \frac{1}{\sqrt{3}}\ket{\downarrow\! T_0}_L = \left[ \sqrt{\frac{2}{3}} c_{1g\uparrow}^{\dagger}c_{2g\downarrow}^{\dagger}c_{2x\downarrow}^{\dagger} - \frac{1}{\sqrt{6}} c_{1g\downarrow}^{\dagger}c_{2g\downarrow}^{\dagger}c_{2x\uparrow}^{\dagger}- \frac{1}{\sqrt{6}} c_{1g\downarrow}^{\dagger}c_{2g\uparrow}^{\dagger}c_{2x\downarrow}^{\dagger}\right] \ket{\Omega}_1\ket{\Omega}_2,  \\
\ket{S \cdot }_L &=& \ket{S\! \downarrow}_L = c_{1g\downarrow}^{\dagger}c_{1g\uparrow}^{\dagger} c_{2g\downarrow}^{\dagger}\ket{\Omega}_1\ket{\Omega}_2. \label{eq:leak}
\end{eqnarray}
\end{widetext}
Evaluating Eq.~(\ref{Eq:DQHQTwoQubit_SubHubbard_Like_Model}) in this basis yields
\begin{eqnarray}
\Hamiltonian_{LQ} = 
\begin{pmatrix}
-\varepsilon_L/2 & 0 & \Delta_{1L} \\
0 & -\varepsilon_L/2 + E_{ST,L} & -\Delta_{2L} \\
\Delta_{1L} & -\Delta_{2L} & \varepsilon_L/2 \\
\end{pmatrix},
\end{eqnarray}\\
where $\varepsilon_L = \mu_1 - \mu_2$, $\Delta_{1L} = -\tau_{1g2g}$, $\Delta_{2L} = -\sqrt{\frac{3}{2}}\tau_{1g2x}$, and $E_{ST,L} = E_2  + C_{2 g 2 x} - K_{2 g2 x} - C_{2 g 2 g} - K_{2 g 2 g}$. 
We note that the inter-dot Coulomb interaction, $\Hamiltonian_{C, \text{inter}}$, can be absorbed into the definition of $\varepsilon_L$, to a very good approximation.
Typically, the qubit is operated in the large-detuning regime ($\varepsilon_L \gg \Delta_{1L},\, \Delta_{2L},\,E_{ST,L}$) where the dephasing due to charge noise is suppressed~\cite{Thorgrimsson2017}.
In this regime, the qubit is largely in the $(1,2)$ charge configuration, with eigenstates given by $\ket{0}_L \approx \ket{\cdot S}_L$ and $\ket{1}_L \approx \ket{\cdot T}_L$, while the leakage state, $\ket{\mathcal{L}}_L\approx \ket{S \cdot}_L$, is largely in the $(2,1)$ charge configuration.
To construct the solutions for the right-most pair of dots in Fig.~1(a), we simply replace the labels $1$, $2$, and $L$ by $3$, $4$, and $R$, respectively.

\subsection{Two exchange-coupled quantum-dot hybrid qubits \label{Sec:DQHQTwoQubit_SubTwoDQHQ} }
We now consider a pair of exchange-coupled double-quantum-dot hybrid qubits in a system of four quantum dots in a linear array, as depicted in Fig.~1(a) of the main text.
Similar arrangements have been considered in Refs.~\cite{2015arXiv150703425M,Ferraro:2015,Michielis:2015}.
The two qubits, $L$ and $R$, are coupled here through a tunable tunnel coupling between dots $2$ and $3$, with an energy bias denoted as $\varepsilon_{L\! R} = \mu_2 - \mu_3$.
To perform high-fidelity single-qubit operations, this tunnel barrier should be kept high so that the resulting tunneling is negligible.
Lowering the barrier height induces the inter-qubit couplings $\tau^{L\! R}_{2g1g}$ and $\tau^{L\! R}_{2x1g}$, as indicated in Fig.~1(a), and generates entangling gates such as CZ.
In this work, we assume that the logical states of both qubits have the same total spin quantum numbers, $S_i = \frac{1}{2}$ and $S_{iz} = -\frac{1}{2}$.
As a consequence of tunneling, the left and right double dots do not necessarily remain in these spin states~\cite{DiVincenzo2000}, yielding an accessible Hilbert space for each double dot that goes well beyond the set $\{\ket{\cdot S},\ket{\cdot T},\ket{S\cdot}\}$.
However, since the Hamiltonian contains no magnetic field terms, the total spin state of the six electrons remains $S_\text{tot}=1$, $S_{\text{tot},z}=-1$.

		\begin{table}[b]
	\caption{Spin classification of the basis states with $(1,2,1,2)$, $(1,2,2,1)$, $(2,1,1,2)$, or $(2,1,2,1)$ charge configurations, consistent with the requirement that $S_{\text{tot},z}=-1$.}
	\begin{tabular}{rrl}
	\hline \hline
	$S_{Lz}$ & $S_{Rz}$ & \hspace{.1in} No.\ states  \\ \hline 
	$-\frac{1}{2}$ & $-\frac{1}{2}$ & \hspace{.1in} $4 \times 4 = 16$  \\
	$ \frac{1}{2}$ & $-\frac{3}{2}$ & \hspace{.1in} $4 \times 1 = 4$  \\ 
	$-\frac{3}{2}$ & $ \frac{1}{2}$ & \hspace{.1in} $1 \times 4 = 4$  \\ \hline \hline
	\end{tabular}
	\end{table}

We now characterize the full Hilbert space used to study gate performance in the main text.
We first extend the Hilbert space to include states that are connected to the qubit charge configuration $(1,2,1,2)$ up to second order in tunneling processes.
[This excludes the charge configuration (2,2,1,1), for example, which is separated from (1,2,1,2) by three tunneling processes.]
We also assume that charge configurations with three electrons in one dot, or configurations with an empty dot, both have much higher energies that can be ignored in our calculations.
The charge configurations obeying these rules include $(1,2,1,2)$, $(1,2,2,1)$, $(2,1,1,2)$, $(2,1,2,1)$, and $(1,1,2,2)$.
We first focus on the states with $(1,2,1,2)$, $(1,2,2,1)$, $(2,1,1,2)$, and $(2,1,2,1)$ charge configurations, and discuss the states with the $(1,1,2,2)$ charge configuration later.

For the (1,2,1,2), (1,2,2,1), (2,1,1,2), and (2,1,2,1) states, let us first consider just the left-hand double dot, $L$ (dots 1 and 2), which can be in either the (1,2) or (2,1) charge configuration.
For the (1,2) case, there are $2^3=8$ possible spin states.
These may be classified according to their spin quantum numbers, as shown in Table~I.
For (2,1), there are also 8 possible spin states; however, only the two states listed in Table~II satisfy the rules described above.
(Recall the additional assumption that $E_{1L}\gg E_{2L}$, which effectively eliminates the excited orbital states of dot~1.)
Here, the notation $\ket{v_\alpha}_L$ indicates the spin-flipped version of state $\ket{\alpha}_L$, $\ket{\mathcal{L}}_L$ indicates a  (2,1) charge-excited leakage state, as in Eq.~(\ref{eq:leak}), and $\ket{2}_L$ and $\ket{3}_L$ represent additional new leakage states.
The states $\ket{0}_L$ and $\ket{1}_L$ represent the logical basis states of qubit~$L$, in the limit of large detuning.
The corresponding basis states for the right-hand double dot are obtained by replacing the indices $1$, $2$, and $L$ in this discussion by $3$, $4$, and $R$.

\begin{table}[t]
\caption{Enumeration of all two-qubit basis states involved in exchange coupling, following the rules described in the text, using the shorthand notation
$\ket{\alpha\beta}\equiv\ket{\alpha}_L\otimes\ket{\beta}_R$.}
\begin{tabular}{clc}
\hline \hline
No.\ & Label  &  Charge config.\\  \hline
1 & $\ket{0 \, 0}$ & $(1,2,1,2)$\\
2 & $\ket{0 \, 1}$ & $(1,2,1,2)$\\
3 & $\ket{0 \, \mathcal{L}}$ & $(1,2,2,1)$\\
4 & $\ket{1 \, 0}$ & $(1,2,1,2)$\\
5 & $\ket{1 \, 1}$ & $(1,2,1,2)$\\
6 & $\ket{1 \, \mathcal{L}}$ & $(1,2,2,1)$\\
7 & $\ket{\mathcal{L} \, 0}$ & $(2,1,1,2)$\\
8 & $\ket{\mathcal{L} \, 1}$ & $(2,1,1,2)$\\
9 & $\ket{\mathcal{L} \, \mathcal{L}}$ & $(2,1,2,1)$\\
10 & $\ket{0 \, 3}$ & $(1,2,1,2)$\\
11 & $\ket{1 \, 3}$ & $(1,2,1,2)$\\
12 & $\ket{\mathcal{L} \, 3}$ & $(2,1,1,2)$\\
13 & $\ket{v_0 \, 2}$ & $(1,2,1,2)$\\
14 & $\ket{v_1 \, 2}$ & $(1,2,1,2)$\\
15 & $\ket{v_\mathcal{L} \, 2}$ & $(2,1,1,2)$\\
16 & $\ket{3 \, 0}$ & $(1,2,1,2)$\\
17 & $\ket{3 \, 1}$ & $(1,2,1,2)$\\
18 & $\ket{3 \, \mathcal{L}}$ & $(1,2,2,1)$\\
19 & $\ket{2 \, v_0}$ & $(1,2,1,2)$\\
20 & $\ket{2 \, v_1}$ & $(1,2,1,2)$\\
21 & $\ket{2 \, v_\mathcal{L}}$  & $(1,2,2,1)$\\
22 & $\ket{3 \, 3}$  & $(1,2,1,2)$\\
23 & $\ket{v_3 \, 2}$ & $(1,2,1,2)$\\
24 & $\ket{2 \, v_3}$ & $(1,2,1,2)$\\
25 & $\ket{\downarrow \downarrow S S }$ & $(1,1,2,2)$\\
26 & $\ket{\downarrow \uparrow S T_- }$ & $(1,1,2,2)$\\ 
27 & $\ket{\uparrow \downarrow S T_- }$ & $(1,1,2,2)$\\
28 & $\ket{\downarrow \downarrow S T_0 }$ & $(1,1,2,2)$\\ \hline
\hline
\end{tabular}
\end{table}

We can combine the $L$ and $R$ basis states, described above, while satisfying the constraint that $S_{\text{tot},z}=-1$.
The resulting states are classified in Table~III, yielding 24 states in total, which we enumerate as states 1-24 in Table~IV.
Finally, the states with $(1,1,2,2)$ charge configurations can be constructed by enforcing the same spin constraint, while recalling the additional assumption that $E_{3L}\gg E_{4L}$, which effectively eliminates the excited orbital states of dot~3.
There are four basis states in this set, which we list as states 25-28 in Table~IV.

Projecting Eq.~(\ref{Eq:DQHQTwoQubit_SubHubbard_Like_Model}) onto these $28$ basis states, we obtain the full effective Hamiltonian for our system,  $\Hamiltonian_\text{eff}$. 
We can also perform a similar projection of the inter-dot Coulomb Hamiltonian, obtaining $\Hamiltonian_{C,\text{eff}} \approx \frac{\mathcal{G}}{4} (\hat n_2 - \hat n_1)(\hat n_4 - \hat n_3) $, as mentioned in the main text.
In the following, we first analyze the full effective Hamiltonian when the inter-qubit tunnel couplings are turned off, $\tau^{L\! R}_{2g1g} = \tau^{L\! R}_{2x1g}=0$. 
We then discuss the new terms arising from these tunnel couplings.
To make it easier to refer to the main text, we now switch back to the $L$ and $R$ double-dot labeling scheme used in the main text.

When $\tau^{L\! R}_{2g1g} = \tau^{L\! R}_{2x1g}=0$, the 28-dimensional Hilbert space decomposes into the following seven decoupled subspaces. (Here we refer to the two-qubit states enumerated in Table~IV.)
\begin{enumerate}[label={(\roman*)}]
	\item For states 1-9, $[\Hamiltonian_\text{eff}]_{1-9,1-9}$ is given by 
	\begin{equation}
	\Hamiltonian_{LQ} \otimes \Hamiltonian_{RQ}  + \frac{\mathcal{G}}{4}
	\begin{pmatrix}
	1 & 0 & 0 \\
	0 & 1 & 0 \\
	0 & 0 & -1
	\end{pmatrix}
	\otimes
	\begin{pmatrix}
	1 & 0 & 0 \\
	0 & 1 & 0 \\
	0 & 0 & -1
	\end{pmatrix} .
	\end{equation}
	\item For states 10-12, $[\Hamiltonian_\text{eff}]_{10-12,10-12}$ is given by
	\begin{equation}
	\Hamiltonian_{LQ}  - \frac{\varepsilon_R}{2}+ E_{ST,R}  + \frac{\mathcal{G}}{4}
	\begin{pmatrix}
	1 & 0 & 0 \\
	0 & 1 & 0 \\
	0 & 0 & -1
	\end{pmatrix} . \label{eq:10-12}
	\end{equation}
	\item For states 13-15, $[\Hamiltonian_\text{eff}]_{13-15,13-15}$ is the same as Eq.~(\ref{eq:10-12}). 
		\item For states 16-18, $[\Hamiltonian_\text{eff}]_{16-18,16-18}$ is given by 
	\begin{equation}
	\Hamiltonian_{RQ}  - \frac{\varepsilon_L}{2} + E_{ST,L} + \frac{\mathcal{G}}{4}
	\begin{pmatrix}
	1 & 0 & 0 \\
	0 & 1 & 0 \\
	0 & 0 & -1
	\end{pmatrix} . \label{eq:16-18}
	\end{equation}
	\item For states 19-21, $[\Hamiltonian_\text{eff}]_{19-21,19-21}$ is the same as Eq.~(\ref{eq:16-18}).
	\item For states 22-24, $[\Hamiltonian_\text{eff}]_{22-24,22-24}$ is given by 
	\begin{equation}
        ( - \frac{\varepsilon_L}{2} + E_{ST,L} - \frac{\varepsilon_R}{2}+ E_{ST,R} + \frac{\mathcal{G}}{4}) I_3,
	\end{equation}
	where $I_3$ is the $3\times 3$ identity matrix.
	\item Finally, for states 25-28, $[\Hamiltonian_\text{eff}]_{25-28,25-28}$ is given by
	\begin{equation}
	\begin{pmatrix}
	(- \frac{\varepsilon_L}{2} - \frac{\varepsilon_R}{2} - \varepsilon_{L\! R}) & 0_{1\times 3 } \\
	0_{3\times 1 }  & (- \frac{\varepsilon_L}{2}  - \frac{\varepsilon_R}{2}- \varepsilon_{L\! R}+ E_{ST,R}) I_3
	\end{pmatrix},
	\end{equation}
	where $0_{i \times j}$ is a matrix of zeros with dimension $i \times j$.
	\end{enumerate}

When $\tau^{L\! R}_{2g1g}$ and $\tau^{L\! R}_{2x1g}$ are non-zero, the different blocks are coupled through the following off-diagonal terms:
\begin{multline}
[\Hamiltonian_\text{eff}]_{1-24,25-28} = \\
\begin{pmatrix}
-\tau^{L\! R}_{2g1g} & 0 & 0 & 0 \\
0 & -\sqrt{\frac{2}{3}}\tau^{L\! R}_{2g1g} & 0 & \frac{1}{\sqrt{3}} \tau^{L\! R}_{2g1g} \\
0 & 0 & 0 & 0 \\
\frac{1}{\sqrt{6}}\tau^{L\! R}_{2x1g} & 0 & 0 & 0 \\
0 & -\frac{1}{3} \tau^{L\! R}_{2x1g} & \frac{2}{3} \tau^{L\! R}_{2x1g} & -\frac{1}{3\sqrt{2}} \tau^{L\! R}_{2x1g} \\
0 & 0 & 0 & 0 \\
0 & 0 & 0 & 0 \\
0 & 0 & 0 & 0 \\
0 & 0 & 0 & 0 \\
0 & -\frac{1}{\sqrt{3}} \tau^{L\! R}_{2g1g} & 0 & -\sqrt{\frac{2}{3}} \tau^{L\! R}_{2g1g} \\
0 & -\frac{1}{3\sqrt{2}} \tau^{L\! R}_{2x1g} & \frac{\sqrt{2}}{3} \tau^{L\! R}_{2x1g} & \frac{1}{3} \tau^{L\! R}_{2x1g} \\
0 & 0 & 0 & 0 \\
0 & 0 &\tau^{L\! R}_{2g1g} & 0 \\
0 & -\sqrt{\frac{2}{3}} \tau^{L\! R}_{2x1g} & \frac{1}{\sqrt{6}} \tau^{L\! R}_{2x1g} & 0 \\
0 & 0 & 0 & 0 \\
-\frac{1}{\sqrt{3}}\tau^{L\! R}_{2x1g} & 0 & 0 & 0 \\
0 & \frac{\sqrt{2}}{3} \tau^{L\! R}_{2x1g} & \frac{\sqrt{2}}{3}  \tau^{L\! R}_{2x1g} & \frac{1}{3} \tau^{L\! R}_{2x1g}  \\
0 & 0 & 0 & 0 \\
-\tau^{L\! R}_{2x1g} & 0 & 0 & 0 \\
0 & 0 & 0 & -\frac{1}{\sqrt{3}}\tau^{L\! R}_{2x1g} \\
0 & 0 & 0 & 0 \\
0 & \frac{1}{3} \tau^{L\! R}_{2x1g} & \frac{1}{3}  \tau_{2x1g}^{L\! R} & -\frac{\sqrt{2}}{3} \tau^{L\! R}_{2x1g}  \\
0 & -\frac{1}{\sqrt{3}} \tau^{L\! R}_{2x1g} & -\frac{1}{\sqrt{3}}  \tau^{L\! R}_{2x1g} & 0 \\
0 & 0 & 0 & \sqrt{\frac{2}{3}} \tau^{L\! R}_{2x1g} \\
\end{pmatrix}, \label{eq:interqubit}
\end{multline}
where the columns correspond to states with $(1,1,2,2)$ charge configurations, and the rows correspond to states with $(1,2,1,2)$, $(1,2,2,1)$, $(2,1,1,2)$, or $(2,1,2,1)$ charge configurations.
Similarly, we have 
\begin{equation}
[\Hamiltonian_\text{eff}]_{25-28,1-24}=[\Hamiltonian^\dagger_\text{eff}]_{1-24,25-28} . \label{eq:adjoint}
\end{equation}

\red{
The inter-qubit tunneling processes described in Eqs.~(\ref{eq:interqubit}) and (\ref{eq:adjoint}) couple the two-qubit basis states to the $(1,1,2,2)$ leakage states.
Although energy conservation does not allow for occupation of the $(1,1,2,2)$ states, their virtual occupation mediates the effective two-qubit interactions discussed in Sec.~\ref{Sec:DQHQTwoQubitCZGates} of the main text.
We can compute these interactions using a Schrieffer-Wolff transformation to eliminate the $(1,1,2,2)$ states~\cite{WinklerBook}, yielding an effective interaction of the form
\begin{equation}
\Hamiltonian_\text{2Q,eff}= \hbar\gamma_{2Q}\,\sigma_z^1\otimes\sigma_z^2 
\end{equation}
between the logical states, where the dominant contribution to the coupling strength is given by $\hbar\gamma_{2Q}= 16(\tau^{L\! R}_{2x1g})^2/[9(4\varepsilon_{LR}+4E_{ST,L}+g)]$.}

\section{Simulation method \label{Sec:DQHQTwoQubit_SubSimulationMethod}}
In this Appendix, we explore the performance of quantum gate operations by numerically solving the Schr$\ddot{\text{o}}$dinger equation, $i \hbar \frac{d}{d t} \ket{\psi(t)} = \Hamiltonian_\text{eff} \ket{\psi(t)} $.
The Hamiltonian parameters used in the simulations are given in the main text.
The initial states are taken to be adiabatic eigenstates, computed using the same tuning parameters.
We note that these eigenstates are generally superpositions of the logical and leakage basis states defined in Table~IV.
However, we may still label them as ``logical" or ``leakage" by adiabatically tuning the system parameters to the far-detuned regime, where $\varepsilon_L, \varepsilon_R \rightarrow \infty$ and $\Delta_{1L},\Delta_{2L},\Delta_{1R},\Delta_{2R},\tau^{L\! R}_{2g1g} , \tau^{L\! R}_{2x1g} \rightarrow 0$, and matching them up with the logical or leakage basis states.
For clarity, below we refer to such adiabatic logical states as $\ket{\widetilde{00}}$, $\ket{\widetilde{01}}$, $\ket{\widetilde{10}}$, and $\ket{\widetilde{11}}$.

For single qubit gates, we model the AC drive on the left qubit by replacing $\varepsilon_L$ with $\varepsilon_L + A_\varepsilon\, p(t) \cos (\omega t + \phi)$ and $\Delta_{\alpha L}$ with $\Delta_{\alpha L}+ A_\Delta\, p(t) \cos (\omega t + \phi)$, where $A_\varepsilon$ ($A_\Delta$) are the detuning (tunnel coupling) driving amplitudes, $\omega$ is the driving angular frequency, and $\phi$ is the phase.
$p(t)$ is the smoothed rectangular pulse envelope defined as
\begin{equation}
p(t) = \left\{ 
\begin{array}{lc} \vspace{0.03in}
\frac{t_g[1- \cos(\pi t/t_r) ]}{2(t_g-t_r)} &  (0\leq t \leq t_r), \\ \vspace{0.03in}
\frac{t_g}{t_g-t_r} & ( t_r< t < t_g - t_r), \\
\frac{t_g[1+ \cos(\pi [t-t_g+t_r]/t_r) ]}{2(t_g-t_r)} \hspace{0.2in} &  (t_g-t_r \leq t \leq t_g), 
\end{array}
\right. \label{eq:pramp}
\end{equation} 
where $t_g$ is the single-qubit gate time and we choose the smoothed ramp time to be $t_r = h/E_{ST,L} \approx 0.83\,\units{ns}$.
For a CZ gate, we must also ramp the inter-qubit tunnel coupling, which we model as as 
\begin{widetext}
\begin{equation}
\tau^{L\! R}_{\alpha}(t) = \left\{
\begin{array}{lc}
\tau^{L\! R}_{\alpha,\text{max}} \frac{t}{t_\text{ramp}} &  (0 \leq t \leq t_\text{ramp}), \\
\tau^{L\! R}_{\alpha,\text{max}}  &  (t_\text{ramp} \leq t \leq t_\text{ramp} + t_\text{wait}), \\
\tau^{L\! R}_{\alpha,\text{max}} \frac{(2 t_\text{ramp} + t_\text{wait} -t)}{t_\text{ramp}} &  (t_\text{ramp} + t_\text{wait} \leq \tau \leq 2 t_\text{ramp} + t_\text{wait}), \\
\end{array}
\right. \label{eq:tauramp}
\end{equation}
\end{widetext}
where $\alpha = 2g1g$ or $2x1g$, $t$ is the time, and $t_\text{ramp}$ ($t_\text{wait}$) are the ramping (waiting) times.

As discussed in the main text, we assume the dominant noise occurs in the detuning channels~\cite{PhysRevLett.110.146804, PhysRevLett.105.246804, Wu19082014,Shi2013,Thorgrimsson2017}.
We model this noise, here, by replacing $\{\varepsilon_L,\varepsilon_R, \varepsilon_{L\! R} \}$ with $\{\varepsilon_L+\delta\varepsilon_{L}( \tau),\varepsilon_R+\delta\varepsilon_{R}( \tau), \varepsilon_{L\! R}+\delta\varepsilon_{L\! R}( \tau) \}$, where the noise sequences $\{\delta\varepsilon_{L}( \tau),\delta\varepsilon_{R}( \tau), \delta\varepsilon_{L\! R}( \tau)\}$ are assumed to be independent, and are generated as described below.
We solve the Schr\"{o}dinger equation for a given noise sequence to obtain the final state $\ket{\psi(t)}$ at time $t$.
We then repeat this procedure for $N$ different noise realizations, denoted $\delta\varepsilon_q^{(n)}(t)$, where $n=1,\ldots,N$ and  $q = L,\,R,\,L\! R$. 
For quasistatic noise we take $N=216\,(=6^3)$, while for $1\!/\!f$ noise we take $N=10,000$.
%Finally, we compute the ensemble-averaged density matrix, 
%\begin{equation}
%\label{Eq:DQHQTwoQubit_SubEnsembleAverage}
%\langle\rho\rangle  = \frac{1}{N}  \sum\limits_{n=1}^N \ket{\psi^{(n)}(t)} \langle \psi^{(n)}(t)|.
%\end{equation} 
%In the following subsections, we provide additional details of the simulations.

\subsection{Determining $t_\text{wait}$}\label{sec:twait}
For a fixed set of Hamiltonian parameters and $t_\text{ramp}$, the procedure for choosing $t_\text{wait}$ is complicated by the fact that the pulse shape (e.g., $t_\text{wait}$) affects the two-qubit gate as well as the incidental single-qubit gates that multiply it, making it difficult to isolate the two-qubit component.
%A similar indeterminacy of the evolution operator was incurred in Ref.~\cite{PhysRevA.95.062321}, and we apply similar methods to treat it here.
To address this problem, we employ the method of Makhlin invariants~\cite{Makhlin2002}.
Here, any two-qubit gate is uniquely defined by a pair of invariants, $G_1$ and $G_2$, which can be computed from our simulations.
For example, for a CZ gate, the ideal invariants are given by $G_{1,\text{ideal}}=0$ and $G_{2,\text{ideal}}=1$.
We therefore define the combined invariant,
 \begin{equation}
D_\text{CZ} = |G_1| + |G_2-1| ,
\end{equation}
and choose $t_\text{wait}$ such that it minimizes $D_\text{CZ} $.

Since the Makhlin invariants are defined in the absence of leakage levels, we adopt the following procedure.
First, we project the full, simulated evolution operator $U$ onto the logical subspace.
The resulting 4D operator is no longer unitary;
since the Makhlin procedure assumes unitary operators, we need to correct this deficiency.
We therefore rescale the diagonal elements of the 4D evolution operator to have magnitude 1.
An appropriate procedure for modifying the off-diagonal elements is less clear; here, we simply set them to 0, as consistent with the ideal CZ unitary operation.

This procedure for determining $t_\text{wait}$ can be viewed as obtaining the optimal diagonal elements for a CZ unitary operator.
Since they are correctly normalized, these diagonal elements can be represented as phases (below, we do this explicitly).
There are four such phases, which we identify as follows: (1) a global phase, which we ignore, (2) phases associated with the two single-qubit $Z$ rotations, which we ignore by considering Makhlin invariants, and (3) the CZ phase.
The procedure described above determines the CZ phase correctly, to within the numerical accuracy of the simulation.
(This is the reason that the ``phase" error, defined below, is essentially zero in our simulations.)
Our protocol is imperfect in the sense that it overlooks certain types of off-diagonal errors.
However, it provides a well defined method for defining $t_\text{wait}$, and we find that the final gate fidelities can be well in excess of 99\%.

\subsection{Simulations in the absence of noise}
\begin{figure*}[t]
\includegraphics[width=7in]{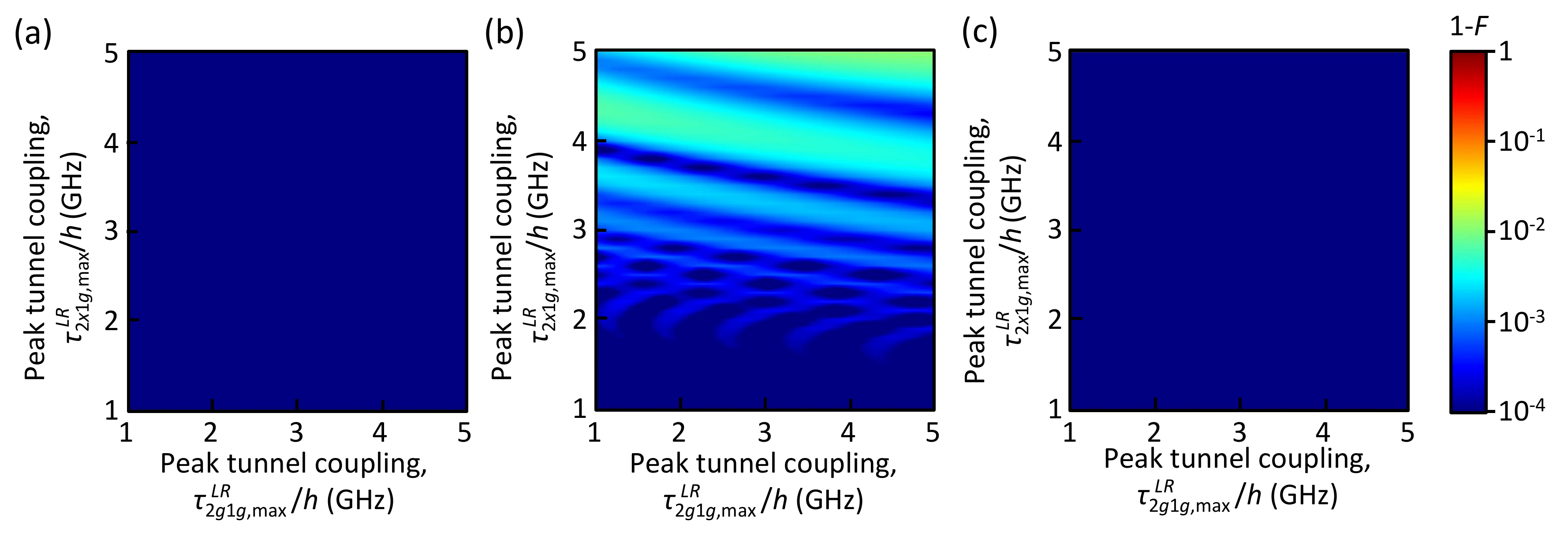}
\caption[Qubit-transition, leakage, and phase error contributions to the intrinsic infidelity of controlled-Z gate ]{
\label{Fig:DQHQTwoQubit_SubFigS1}
Contributions to the intrinsic infidelity of a CZ gate due to (a) qubit-transition, (b) leakage, or (c) phase errors, obtained at the the same simulation parameters as Fig.~2(a) of the main text. 
In the absence of noise, the dominant source of infidelity is from leakage, while qubit-transition and phase errors are both below $10^{-4}$.
The results are not surprising because energy splittings between the logical states and their dominant leakage states are much smaller than the energy splitting between the logical states.
}
\end{figure*}

In the absence of noise, it is sufficient to take $N=1$. 
We simulate and compute the fidelity of CZ gates for inter-qubit tunnel couplings in the range $1 \,\units{GHz} < \tau^{L\! R}_{2g1g,\text{max}}/h,\, \tau^{L\! R}_{2x1g,\text{max}}/h  < 5\,\units{GHz}$.
This allows us to identify the intrinsic low- and high-fidelity regimes in Fig.~1(a) of the main text.
We further classify the sources of infidelity in terms of qubit-transition, leakage, and phase errors, as described in Appendix~\ref{Sec:DQHQTwoQubit_SubContributions}, yielding the results shown in Fig.~\ref{Fig:DQHQTwoQubit_SubFigS1}.

\subsection{Simulations in the presence of noise}\label{sec:sims-noise}
In this subsection, we summarize the noise models used in our simulations.
Most of the details are published elsewhere~\cite{Kawakami18102016, TLSNoise}, but are summarized here for completeness

\textit{Quasistatic noise.} 
Here, we assume the noise sequences are time-independent (i.e., constant), and sampled from a distribution of gaussian random variables.
The three noise parameters, $\delta\varepsilon_{L}$, $\delta\varepsilon_{R}$, and $\delta\varepsilon_{L\! R}$ are also assumed to be uncorrelated.
To sample such a distribution efficiently, we replace the average with an integral, and employ Gaussian-Hermite quadrature~\cite{HermitePolynomials,AbramowitzBook}.
In practice, we find that each integral converges sufficiently when using just six sampling points.
The initial states in our simulations are determined by assuming that the quasistatic noise is already present while initializing the qubits to the detuning values $\varepsilon_L+\delta\varepsilon_{L}$, $\varepsilon_R+\delta\varepsilon_{R}$, and $\varepsilon_{L\! R}+\delta\varepsilon_{L\! R}$.

\textit{$1\!/\!f$ noise.} 
In this case, we model the detuning fluctuations as time series for which the Fourier transform of the time correlation function has a power-spectrum density given by 
\begin{equation}
\label{Eq:DQHQTwoQubit_SubSpectrum One-Over-f}
\tilde{S}(\omega) =  \left\{
  \begin{array}{lc}
   c_{\varepsilon}^2 \frac{2 \pi}{|\omega|} &  (\omega_l \leq |\omega | \leq \omega_h),\\
    0 &  (\text{otherwise}) .
  \end{array}
\right. 
\end{equation}
Here, $\omega_l/2\pi$ ($\omega_h/2\pi$) are the low (high)-frequency cutoffs, chosen to be $1\,\units{Hz}$ ($256 \,\units{GHz}$), and $c_{\varepsilon}$ is the noise amplitude.
In practice, we generate noise realizations with a low (high)-frequency cutoff of $1.2\,\units{MHz}$ ($256\,\units{GHz}$), using the method detailed in Refs.~\cite{Kawakami18102016, TLSNoise}, and we approximate the remaining low-frequency spectrum as quasistatic noise based on gaussian random variables, as described above.

\section{Process fidelity \label{Sec:DQHQTwoQubit_SubProcessFidelity}}
In this Appendix, we first derive a convenient expression for the process fidelity.
We then use it to define and characterize three different contributions to the infidelity: qubit-transition, leakage, and phase errors.

\subsection{Derivation of the process fidelity}
Following Ref.~\cite{ChuangBook}, a generic quantum process $\mathcal{E}$ acting on a $d$-dimensional Hilbert space may be expressed as
\begin{equation}
\label{Eq:DQHQTwoQubit_SubQuantumProcess}
\mathcal{E}(\rho_0) = \sum\limits_{m,n} \hat E_m \rho_0 \hat E_n ^{\dagger} \chi_{mn},
\end{equation}
where $\rho_0$ is the initial density matrix, $\mathcal{E}(\rho_0)$ is the final density matrix, $\{\hat E_m\}$ is a basis set for the vector space of $d\times d$ matrices normalized by the condition that $\text{Tr}[\hat E_m^{\dagger}\hat E_n] = d\, \delta_{mn}$, and $\chi_{mn}$ is a $d^2\times d^2$ process matrix, commonly referred to as the chi ($\chi$) matrix.
The two processes we consider here are time evolutions of a quantum gate, and time evolutions of a quantum gate averaged over many noise realizations, both defined in the logical space $H_Q = \{\ket{1} = \ket{00}, \ket{2} = \ket{01},\ket{3} = \ket{10},\ket{4} = \ket{11}\}$.
The process fidelity is defined as~\cite{PhysRevLett.102.090502}
\begin{equation}
\label{Eq:DQHQTwoQubit_SubFidelity_Chi}
F = \text{Tr} [\chi_\text{sys} \chi_\text{ideal}], 
\end{equation}
where $\chi_\text{sys}$ is the process matrix for the actual physical evolution, including strong driving effects and decoherence, and $\chi_\text{ideal}$ is the process matrix for the ideal operation.

% \subsection{Choi-Jamiolkowski formalism}
The $\chi$ matrix can be calculated using the Choi-Jamiolkowski formalism~\cite{PhysRevA.71.062310}, as follows.
First, we adopt the initial density matrix $\ket{\Phi_0} \bra{\Phi_0}$, corresponding to the special state
 \begin{equation}
 \ket{\Phi_0}  \equiv \frac{1}{2} \sum\limits_{j} \ket{j}\otimes \ket{j}, \label{eq:CJpsi}
 \end{equation}
where $\{\ket{j}\}=H_Q$.
Next, we define the basis set of $4 \times 4$ matrices used in Eq.~(\ref{Eq:DQHQTwoQubit_SubQuantumProcess}): $\{\hat E_{m(i,j)} = 2 \ket{i} \langle j|\}$, where $m(i,j)$ is an arbitrary labeling scheme that maps $(i,j)$ to $m = 1,\ldots,16$; for example, $m(i,j) = 4(i-1) + j$.
According to Eq.~(\ref{eq:CJpsi}), $\ket{\Phi_0} \bra{\Phi_0}$ must be a $16\times 16$ matrix (the same as $\chi$).
Finally, we consider a new process, $\mathcal{I}\otimes \mathcal{E}$, acting on $\ket{\Phi_0} \bra{\Phi_0}$, where $\mathcal{I}$ is the 4-dimensional identity operator.
It is then easy to show that 
\begin{widetext}
 \begin{eqnarray}
 [\mathcal{I}\otimes \mathcal{E}] (\ket{\Phi_0} \bra{\Phi_0}) 
 & = & \frac{1}{4} \sum\limits_{j,j^{\prime}} [\mathcal{I}\otimes \mathcal{E}] (\ket{j}\langle j^{\prime} | \otimes \ket{j }\langle j^{\prime} |)  \nonumber \\
  & = & \frac{1}{4} \sum\limits_{j,j^{\prime}} \ket{j}\langle j ^{\prime}| \otimes [\sum\limits_{k,l,k^{\prime},l^{\prime}}\hat E_{m(k,l)}\ket{j }\langle j^{\prime} |\hat E_{n(k^{\prime},l^{\prime})}^{\dagger} \chi_{m(k,l),n(k^{\prime},l^{\prime})}]  \nonumber \\
& = &  \sum\limits_{j,j^{\prime}} \ket{j}\langle j^{\prime}| \otimes [\sum\limits_{k,l,k^{\prime},l^{\prime}} \ket{k} \braket{l}{j }\braket{ j^{\prime} }{ l^{\prime}} \langle k^{\prime} | }{  \chi_{m(k,l),n(k^{\prime},l^{\prime})}]  \nonumber \\
& = & \sum\limits_{j,j^{\prime}} \ket{j}\langle j ^{\prime}| \otimes [\sum\limits_{k,k^{\prime}}\ket{k} \langle k^{\prime}| \chi_{m(k,j),n(k^{\prime},j^{\prime})}] \nonumber \\
& = & \sum\limits_{j,j^{\prime},k,k^{\prime}} (\ket{j} \otimes \ket{k}) (\langle j ^{\prime}| \otimes  \langle k^{\prime}| ) \chi_{m(k,j),n(k^{\prime},j^{\prime})} \nonumber \\
&=& \chi ,
\end{eqnarray}   
\end{widetext}
where $\chi$ is expressed in the bases $\{\ket{m(k,j)} = \ket{j} \otimes \ket{k}\}$ and $\{\ket{n(k^{\prime},j^{\prime})} = \ket{j^{\prime}} \otimes \ket{k^{\prime}}\}$.

For our noise-free simulations, we first compute the time-evolution operator, $U_\text{full}$, in the full 28-dimensional Hilbert space discussed in Appendix~A:
\begin{equation}
i\hbar\frac{d}{dt} U_\text{full}(t) = \mathcal{H}_\text{eff}(t) U_\text{full}(t) ,
\end{equation}
for the initial condition $U_\text{full}(t=0)=\mathcal{I}_{28 \times 28} $.
We then project $U_\text{full}$ onto the 4-dimensional logical subspace, as in Refs.~\cite{PhysRevA.97.032306,2018arXiv181203177F}, using $U_\text{sys} = P_Q U_\text{full}P_Q$, where $P_Q$ is the projection operator onto $H_Q$.
We then expand the evolution operators, $U_\text{sys}$ and $U_\text{ideal}$, in the basis $\{\hat E_m\}$, obtaining $U_\mathcal{P} = \sum\limits_m [v_{\mathcal{P}}]_m \hat E_m$ and $[\chi_\mathcal{P}]_{mn} =  [v_{\mathcal{P}}]_m [v_{\mathcal{P}}]_n^*$,  where $\mathcal{P} = \text{``ideal"}$ or $\text{``sys"}$. 
Equation~(\ref{Eq:DQHQTwoQubit_SubFidelity_Chi}) can then be written as
\begin{eqnarray}
F &=& \text{Tr} [\chi_\text{sys} \chi_\text{ideal}] \nonumber \\
&=& \sum\limits_{m,n} [v_{\text{ideal}}]_m [v_{\text{ideal}}]_n^* [v_{\text{sys}}]_n [v_{\text{sys}}]_m^* \nonumber \\
&=& \left(\sum\limits_{n}[v_{\text{ideal}}]_n^* [v_{\text{sys}}]_n \right)\left(\sum\limits_{m} [v_{\text{ideal}}]_m  [v_{\text{sys}}]_m^*\right)  \nonumber \\
&=& \frac{1}{16}\left| \text{Tr}(U_\text{ideal}^{\dagger}U_\text{sys})\right|^2. \label{eq:Fsimple}
\end{eqnarray}
(Compare to Ref.~\cite{Ghosh2010}, for example.)
Note that Eq.~(\ref{eq:Fsimple}) holds even when $U_\text{sys}$ is nonunitary.

We approximate the effect of the charge noises on the quantum process with the noise-averaged quantum process, which is given by 
\begin{equation}
\mathcal{E}_\text{sys}(\rho_0) = \frac{1}{N} \sum\limits_{n=1}^N  U_{\text{sys}}^{(n)} \rho_0 {U_{\text{sys}}^{(n)}}^{\dagger}.
\end{equation}
From Eq.~(\ref{Eq:DQHQTwoQubit_SubQuantumProcess}), the noise-averaged process matrix is then given by
\begin{equation}
\chi_\text{sys}=\frac{1}{N}\sum_n\chi^{(n)}_\text{sys} .
\end{equation}
Finally, repeating the derivation of Eq.~(\ref{eq:Fsimple}), we obtain a simple formula for the noise-averaged process fidelity:
\begin{equation}
\label{Eq:DQHQTwoQubit_SubFidelity_U}
F = \frac{1}{16N } \sum\limits_{n=1}^N |\text{Tr}(U_\text{ideal}^{\dagger}U_{\text{sys}}^{(n)})|^2.
\end{equation}

\subsection{Contributions to the process fidelity from qubit-transition, leakage, and phase errors \label{Sec:DQHQTwoQubit_SubContributions}}
The ideal unitary operator for a CZ gate in the logical basis, $H_Q$, is given by $U_\text{ideal} = \text{diag}[e^{i \phi_{\text{ideal},1}},e^{i \phi_{\text{ideal},2}},e^{i \phi_{\text{ideal},3}},e^{i \phi_{\text{ideal},4}}]$.
We may therefore express Eq.(\ref{Eq:DQHQTwoQubit_SubFidelity_U}) as
\begin{eqnarray}
\label{Eq:DQHQTwoQubit_SubFidelityContribution1}
F &=& \frac{1}{16} \left| \text{Tr}[U_\text{ideal}^{\dagger} U_{\text{sys}}] \right|^2 \nonumber \\
&=& \frac{1}{16} \left|\sum\limits_i e^{-i \, \phi_{\text{ideal},i}} [U_{\text{sys}}]_{ii}  \right|^2 \nonumber \\
&=& \frac{1}{16} \left|\sum\limits_i \left|[U_{\text{sys}}]_{ii}\right| e^{-i \,  \delta \phi_{i}} \right|^2,
\end{eqnarray} 
where $[U_{\text{sys}}]_{ii} \equiv |[U_{\text{sys}}]_{ii}|e^{i \phi_i}$, $\delta \phi_i \equiv \phi_{\text{ideal},i} - \phi_{i}$, and $\phi_i$ is the phase of the diagonal elements in the actual evolution, $U_\text{sys}$.
Since $U_{\text{full}}$ is a unitary operator, we can approximate 
\begin{eqnarray}
\label{Eq:DQHQTwoQubit_SubDiagonalError}
\left|[U_{\text{sys}}]_{ii}\right| &=& \left|[U_{\text{full}}]_{ii}\right| = \sqrt{1-\sum\limits_{j \neq i}^H \left| [U_{\text{sys}}]_{ji}\right|^2}  \\
&\approx & 1 - \frac{1}{2} \sum\limits_{j \neq i}^{H_Q} \left| [U_{\text{sys}}]_{ji}\right|^2 - \frac{1}{2} \sum\limits_{j}^{H_L} \left|[U_{\text{sys}}]_{ji}\right|^2, \nonumber
\end{eqnarray}
where we have assumed that $|[U_\text{sys}]_{ii}|\approx |[U_\text{ideal}]_{ii}|=1$.
Here, $H_L$ comprises the leakage components of the full Hilbert space, such that $H=H_Q\oplus H_L$.
Since deviations from $|[U_\text{sys}]_{ii}|=1$ represent errors for a CZ gate, we can characterize the second term in Eq.~(\ref{Eq:DQHQTwoQubit_SubDiagonalError}) as ``qubit-transition" error, because it causes unwanted transitions within the logical subspace, and we can characterize the third term as leakage error, because it causes unwanted transitions outside the logical subspace.

\begin{widetext}
Inserting Eq.~(\ref{Eq:DQHQTwoQubit_SubDiagonalError}) into Eq.~(\ref{Eq:DQHQTwoQubit_SubFidelityContribution1}) and expanding, we obtain

\begin{eqnarray}
\label{Eq:DQHQTwoQubit_SubFidelityContribution2}
F & \approx & \red{ \frac{1}{16} \left|  \sum\limits_{i}^{H_Q} \left[\left(1 - \frac{1}{2} \sum\limits_{j \neq i}^{H_Q} \left|[U_{\text{sys}}]_{ji}\right|^2 - \frac{1}{2} \sum\limits_{j}^{H_L} \left| [U_{\text{sys}}]_{ji}\right|^2\right)\left(1-i \delta \phi_i - \frac{1}{2}\delta \phi_i^2\right)\right] \right|^2 } \nonumber \\
&\approx & 1 - \frac{1}{4} \left({\sum\limits_{i}^{H_Q}\sum\limits_{j\neq i}^{H_Q}} \left|[U_{\text{sys}}]_{ji}\right|^2\right) - \frac{1}{4} \left(\sum\limits_{ i}^{H_Q}\sum\limits_{ j}^{H_L} \left|[U_{\text{sys}}]_{ji}\right|^2 \right)- \frac{1}{4} \left(\sum\limits_{i=1}^{H_Q} \delta \phi_i^2 - \frac{1}{4} \left(\sum\limits_{i=1}^{H_Q} \delta \phi_i\right)^2\right) \nonumber \\
& = & 1 - (1-F_\text{q-t}) -(1-F_\text{leak}) - (1-F_\text{phase}) ,
\end{eqnarray} 
%&=&\frac{1}{16}  \left( 4 - \frac{1}{2} \sum\limits_{i}^{H_Q} \sum\limits_{j \neq i}^{H_Q} \left|[U_{\text{sys}}]_{ji}\right|^2 - \frac{1}{2} \sum\limits_{i}^{H_Q} \sum\limits_{j}^{H_L} \left| [U_{\text{sys}}]_{ji}\right|^2 - \frac{1}{2}\sum\limits_{i}^{H_Q} \delta \phi_i^2 -i \sum\limits_{i}^{H_Q} \delta \phi_i \right) \nonumber \\
%&&\left( 4 - \frac{1}{2} \sum\limits_{i}^{H_Q} \sum\limits_{j \neq i}^{H_Q} \left|[U_{\text{sys}}]_{ji}\right|^2 - \frac{1}{2} \sum\limits_{i}^{H_Q} \sum\limits_{j}^{H_L} \left| [U_{\text{sys}}]_{ji}\right|^2 - \frac{1}{2}\sum\limits_{i}^{H_Q} \delta \phi_i^2 +i \sum\limits_{i}^{H_Q} \delta \phi_i \right) \nonumber \\
where we have also taken $\delta\phi_i$ to be small.
In the last two lines of Eq.~(\ref{Eq:DQHQTwoQubit_SubFidelityContribution2}), the terms in parentheses define the infidelity components due to qubit-transition (q-t), leakage (leak), and phase errors, respectively.
We note that the phase error is invariant under a common phase shift:
\begin{eqnarray}
(1-F_\text{phase}) &=&  \frac{1}{4} \left(\sum\limits_{i=1}^{H_Q} \delta \phi_i^2 - \frac{1}{4} \left(\sum\limits_{i=1}^{H_Q} \delta \phi_i\right) ^2\right) \nonumber \\
&\rightarrow &  \frac{1}{4} \left(\sum\limits_{i=1}^{H_Q} (\delta \phi_i + \delta \theta)^2 - \frac{1}{4} \left(\sum\limits_{i=1}^{H_Q} \delta \phi_i + \delta \theta\right) ^2\right) \nonumber \\
&= &  \frac{1}{4} \left(\left[\sum\limits_{i=1}^{H_Q}  \delta \phi_i^2 + 2\left(\sum\limits_{i=1}^{H_Q} \delta \phi_i\right)\delta \theta + 4 \delta \theta^2\right] - \frac{1}{4} \left[\left(\sum\limits_{i=1}^{H_Q} \delta \phi_i\right)^2 + 8\left(\sum\limits_{i=1}^{H_Q} \delta \phi_i\right) \delta \theta+ 16\delta \theta^2\right ]\right) \nonumber \\
&=& \frac{1}{4} \left(\sum\limits_{i=1}^{H_Q} \delta \phi_i^2 - \frac{1}{4} \left(\sum\limits_{i=1}^{H_Q} \delta \phi_i\right) ^2\right) ,  \nonumber
\end{eqnarray}
as consistent with the overall phase being being arbitrary.
Finally, we can incorporate the effects of charge noise, following the procedure used in the previous subsection, obtaining
\begin{eqnarray}
1-F_\text{q-t} & = & \frac{1}{4N} \sum\limits_{n=1}^N  \sum\limits_{i}^{H_Q}\sum\limits_{j\neq i}^{H_Q}  
|[U_{\text{sys}}^{(n)}]_{ji}|^2, \\
1-F_\text{leak} & = & \frac{1}{4N}\sum\limits_{n=1}^N \sum\limits_{i}^{H_Q}\sum\limits_{j}^{H_L}
 |[U_{\text{sys}}^{(n)}]_{ji}|^2 , \\
1-F_\text{phase} & = & \frac{1}{4N}\sum\limits_{n=1}^N \left[\sum\limits_{i=1}^{H_Q} \left( \delta\phi_i^{(n)}\right)^2 - \left(\sum\limits_{i=1}^{H_Q} \delta \phi_i^{(n)}\hspace{-.05in}\right)^{\hspace{-.05in} 2}\right] .  \hspace{.1in} \label{eq:Fphase}
\end{eqnarray}
\end{widetext}

To conclude this Appendix, we note that the Z-CNOT gate operation has only one non-zero element in each column and row of its ideal evolution matrix, $U_\text{ideal}$, which allows us to derive its $F_\text{q-t}$, $F_\text{leak}$, and $F_\text{phase}$ fidelity components in the same way as for CZ gates, yielding the results plotted in Fig.~3 of the main text and in the following Appendix.

\section{Additional simulation results\label{Sec:DQHQTwoQubit_SubSimulationResults}}
In this Appendix, we present supporting simulation results that are not included in the main text. 
We first plot the contributions to the intrinsic infidelity of a CZ gate from qubit-transition, leakage, and phase errors. 
Second, we provide a theoretical explanation for how the features observed in the leakage are analogous to coherent oscillations, and we demonstrate this numerically, for the case of CZ gates.
Finally, we plot Z-CNOT gate fidelities, simulated in the presence of quasistatic charge noise, demonstrating their similarity to CZ gates.

\subsection{Intrinsic infidelity of CZ gates}
In this subsection, we present simulation results for the fidelity of a CZ gate in the absence of charge noise, known as intrinsic fidelity, which arises due to ramping.
In Fig.~\ref{Fig:DQHQTwoQubit_SubFigS1}, we plot the separate contributions to the intrinsic infidelity, arising from qubit-transition, leakage, phase errors, as described in the previous Appendix. 
By comparing these results to the total intrinsic infidelity, shown in Fig.~2(a) of the main text, we see that the fidelity is dominated by leakage errors, while qubit-transition and phase errors both fall below $10^{-4}$.
The small size of qubit-transition errors is consistent with the relatively large energy difference between the qubit states, as described in the main text. 
The small size of phase errors can be explained by the particular method we use for choosing $t_\text{wait}$, as described above.

\begin{figure*}[t]
\includegraphics[width=5in]{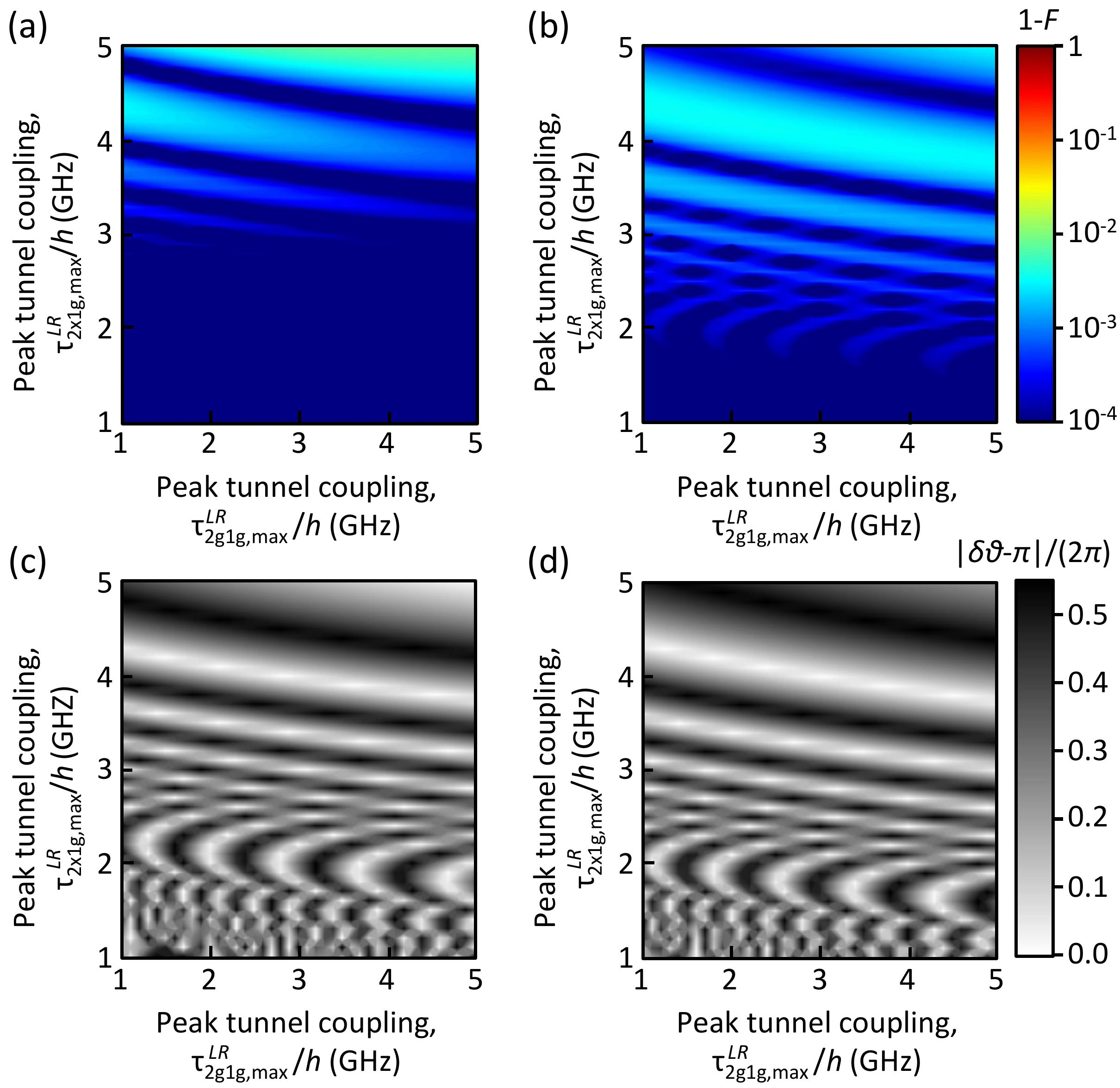}
\caption[The leakage from $\ket{10}$ and $\ket{11}$ and the corresponding phase difference between two dominant paths]{
\label{Fig:DQHQTwoQubit_SubFigS2}
Leakage contribution to the intrinsic infidelity for the initial states (a) $\ket{\widetilde{10}}$ and (b) $\ket{\widetilde{11}}$.
In Appendix~\ref{sec:Leakage-path}, we show that each of these states has a predominant leakage channel, $\ket{\widetilde{L_{10}}}$ and $\ket{\widetilde{L_{11}}}$, respectively, while $\ket{\widetilde{00}}$ and $\ket{\widetilde{01}}$ have no comparable leakage channels.
Simulation parameters here are the same as Figs.~2(a) and 4.
(c), (d) Accumulated phase differences [Eq.~(\ref{eq:LZSphase})] for evolutionary paths associated with (c) $\ket{\widetilde{10}}$ and (d) $\ket{\widetilde{11}}$, as compared to $\ket{\widetilde{L_{10}}}$ and $\ket{\widetilde{L_{11}}}$.
The fringes closely match those in panels (a) and (b), and are caused by LSZ-like effects.
Leakage errors are suppressed when the two evolutionary paths are in phase and enhanced when they are out of phase.
}
\end{figure*}

\begin{figure*}[t]
\includegraphics[width=5in]{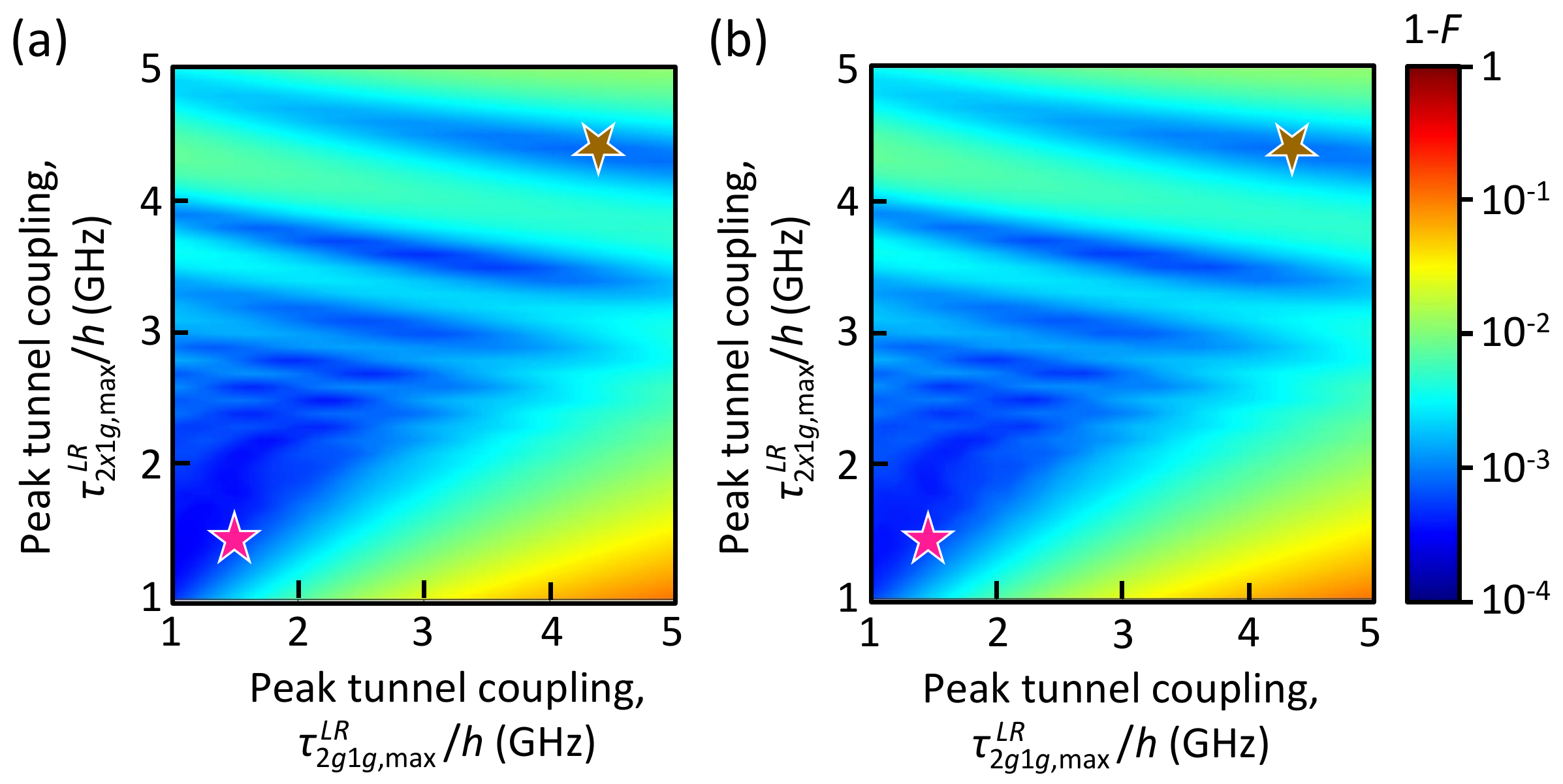}
\caption[The infidelity for controlled-Z and controlled-NOT gates ]{
\label{Fig:DQHQTwoQubit_SubFigS3}
Comparison of the infidelity for (a) Z-CNOT gates vs.\ (b) CZ gates, assuming identical simulation parameters, including quasistatic charge noise.
Panel (b) is the same as Fig.~2(b) of the main text. 
The two panels appear identical because the only difference in the gate evolutions involves single-qubit gates, whose fidelities are very high.
For the Z-CNOT gate, we find $F = 99.95\%$ (pink star, adiabatic regime), and $F = 99.91\%$ (brown star, nonadiabatic regime)
}
\end{figure*}

\subsection{Leakage}\label{sec:Leakage-path}
As demonstrated in the previous subsection, leakage is the dominant contributor to the intrinsic fidelity, and the predominant features observed in the infidelity plots resemble coherent oscillation patterns.
We can gain further insight into this behavior through the following theoretical analysis.

As noted in the main text, leakage processes occur most readily within the low-energy manifold of Fig.~1(c) of the main text. 
Even within this grouping, there are smaller sets of states that fall into highly degenerate manifolds. 
Leakage depends most strongly on the effective interactions within these sets.
As discussed in Appendix~\ref{Sec:DQHQTwoQubit_SubTwoDQHQ}, the dominant interactions are mediated by second-order tunneling processes, mediated (virtually) by the (1,1,2,2) charge states.
To understand these interactions, we perform a Schrieffer-Wolff (SW) decomposition~\cite{WinklerBook} to identify the single dominant term in each of these nearly-degenerate manifolds.
The details of these calculations are omitted for brevity, but we summarize the results here.

For a general tuning of the two-qubit device, the adiabatic eigenstates are formed of a combination of logical and leakage basis states, as discussed in Appendix~\ref{Sec:DQHQTwoQubit_SubTwoDQHQ}.
From bottom to top, the blue levels in Fig.~1(c) correspond $\ket{\widetilde{00}}$, $\ket{\widetilde{01}}$, $\ket{\widetilde{10}}$, and $\ket{\widetilde{11}}$.
It is evident from the figure that $\ket{\widetilde{00}}$ does not share a nearly-degenerate manifold with any other state. Therefore, it does not couple to any leakage levels at ${\cal O}[\tau^2]$ in the SW procedure described above.
In contrast, $\ket{\widetilde{01}}$ is nearly degenerate with two leakage levels.
However SW shows that the matrix elements coupling these states vanish at  ${\cal O}[\tau^2]$; so again, there is no effective coupling to leakage levels.
$\ket{\widetilde{10}}$ is also nearly degenerate with two leakage levels.
In this case, SW shows that the matrix element coupling one of these states vanishes, but the other does not, yielding one leakage channel at ${\cal O}[\tau^2]$.
Finally, $\ket{\widetilde{11}}$ is nearly degenerate with seven leakage levels, of which three have nonvanishing matrix elements at ${\cal O}[\tau^2]$. 
In this case, the dominant coupling is to the state lying closest in energy to $\ket{\widetilde{11}}$.
We therefore conclude that leakage occurs predominantly through these two channels, associated with the states $\ket{\widetilde{10}}$ and $\ket{\widetilde{11}}$, respectively.

The predicted leakage channels are coherent.
However, their presence can have two detrimental effects on the gate fidelity.
First, the leakage state can become occupied through non-adiabatic processes (e.g., a short ramp time or a large peak tunnel coupling, $\tau_\text{max}$).
We suppress such behavior here by applying symmetric pulses, as in Eq.~(\ref{eq:tauramp}), causing the leakage transition to reverse at the end of the pulse.
Hence, to a good approximation, the leakage state empties at the end of the sequence, analogous to a single Landau-Zener-St\"{u}ckelberg (LZS) cycle~\cite{Landau1932,Zener1932,Stuckelberg1932}.
More importantly, like LZS, the excursion through the leakage state affects the phase of the logical state, since the leakage state has a different energy, potentially causing a phase error. We mitigate this phase error by the particular method we use to choose $t_\text{wait}$, as described above.

To test this hypothesis, we compute the phase that would accumulate if the evolution passed through the leakage channel, as opposed to the logical channel.
We begin the simulation in one of the logical states, $\ket{\widetilde{10}}$ or $\ket{\widetilde{11}}$.
For example, we now consider $\ket{\widetilde{10}}_\text{init}$, where ``init" refers to the system tuning at the start of the pulse sequence.
After ramping up the tunnel coupling [Eq.~(\ref{eq:tauramp})], we project the system onto either the logical state $\ket{\widetilde{10}}_\text{peak}$, or the dominant leakage state $\ket{\widetilde{L_{10}}}_\text{peak}$, where ``peak" refers to the system tuning at the peak of the ramp sequence.
We proceed similarly through the ``wait" and ``ramp down" portions of the pulse sequence, finally projecting onto the initial state.
In this way, we can follow two evolutionary paths associated with the same pulse sequence: the leakage path vs.\ the logical path.
Finally, we compute the accumulated phase difference $\delta\vartheta$ associated with these two paths.
The explicit form of $\delta\vartheta$ is given by
\begin{widetext}
\begin{multline}
\delta \vartheta = \text{Phase}  [{}_\text{peak}\bra{\widetilde{L_{10}}}U_{\text{ramp}\uparrow}\ket{\widetilde{10}}_\text{init}]
+\text{Phase}[{}_\text{peak}\bra{\widetilde{L_{10}}}U_\text{wait}\ket{\widetilde{L_{10}}}_\text{peak}]
+\text{Phase}[{}_\text{init}\bra{\widetilde{{10}}}U_{\text{ramp}\downarrow}\ket{\widetilde{L_{10}}}_\text{peak}] \\
- \text{Phase}  [{}_\text{peak}\bra{\widetilde{{10}}}U_{\text{ramp}\uparrow}\ket{\widetilde{10}}_\text{init}]
-\text{Phase}[{}_\text{peak}\bra{\widetilde{{10}}}U_\text{wait}\ket{\widetilde{{10}}}_\text{peak}]
-\text{Phase}[{}_\text{init}\bra{\widetilde{{10}}}U_{\text{ramp}\downarrow}\ket{\widetilde{{10}}}_\text{peak}] . \label{eq:LZSphase}
\end{multline} 
\end{widetext}
We can also write an analogous version of $\delta \vartheta$ for $\ket{\widetilde{11}}$. 
These phase differences are computed and plotted in Figs.~5(c) and 5(d).
The close resemblance of these plots to the corresponding leakage plots in Figs.~5(a) and 5(b) indicates that the suppression of the fidelity due to leakage can be understood as a phase effect, analogous to LZS.
The relatively high fidelities observed in the lower portions of Figs.~5(a) and 5(b), compared to Figs.~5(c) and 5(d), are caused by the reduced occupation of the leakage states in this regime, which is not captured in the phase analysis.

\subsection{Z-CNOT gates}\label{sec:ZCNOT}
In this section, we show that Z-CNOT gate fidelities are almost identical to CZ gate fidelities.

The Z-CNOT gate sequence, defined in Eq.~(\ref{eq:Z-CNOT}) of the main text, involves AC-driven single-qubit gate operations, combined with a single CZ gate.
In this work, the AC drive is used to implement $Y$ rotations, with smooth rectangular pulse envelopes given in Eq.~(\ref{eq:pramp}), and driving amplitudes $\{A_{\varepsilon},A_\Delta\}/h = \{27, 3.1\} \,\units{GHz}$ for the detuning and tunnel-coupling, respectively.
$Z$ rotations are implemented virtually, by adjusting the phase of the rotating frame~\cite{2001Natur.414..883V,Watson2018}.
Z-CNOT gate simulations are performed in the presence of quasistatic charge noise in the same way as our previous simulations of CZ gates, assuming the same noise realization throughout the entire pulse sequence.

The results of these Z-CNOT simulations are shown in Fig.~\ref{Fig:DQHQTwoQubit_SubFigS3}(a), along with the corresponding CZ results shown in Fig.~\ref{Fig:DQHQTwoQubit_SubFigS3}(b), which were obtained at the same system tuning, and previously presented in Fig.~2(b) of the main text.
As expected, the two infidelity maps are nearly identical because the method we developed previously to implement single-qubit gates~\cite{DQHQNoise} yields very high fidelities ($>$99.996\%) in this tuning regime.
In fact, the highest Z-CNOT gate fidelities in both the adiabatic and nonadiabatic regimes are both $>$99.9\%.
For example, we obtain $F=99.95$\% when $\{\tau^{L\! R}_{2g1g, \text{max}}, \tau^{L\! R}_{2x1g, \text{max}}\}/h  = \{1.5,1.5\}\,\units{GHz}$ (pink star, adiabatic regime), and $F=99.91$\% when $\{\tau^{L\! R}_{2g1g, \text{max}}, \tau^{L\! R}_{2x1g, \text{max}}\}/h  = \{4.2,4.4\}\,\units{GHz}$ (brown star, nonadiabatic regime).

\bibliography{DQHQTwoQubit.bib}

\begin{thebibliography}{66}%
\makeatletter
\providecommand \@ifxundefined [1]{%
 \@ifx{#1\undefined}
}%
\providecommand \@ifnum [1]{%
 \ifnum #1\expandafter \@firstoftwo
 \else \expandafter \@secondoftwo
 \fi
}%
\providecommand \@ifx [1]{%
 \ifx #1\expandafter \@firstoftwo
 \else \expandafter \@secondoftwo
 \fi
}%
\providecommand \natexlab [1]{#1}%
\providecommand \enquote  [1]{``#1''}%
\providecommand \bibnamefont  [1]{#1}%
\providecommand \bibfnamefont [1]{#1}%
\providecommand \citenamefont [1]{#1}%
\providecommand \href@noop [0]{\@secondoftwo}%
\providecommand \href [0]{\begingroup \@sanitize@url \@href}%
\providecommand \@href[1]{\@@startlink{#1}\@@href}%
\providecommand \@@href[1]{\endgroup#1\@@endlink}%
\providecommand \@sanitize@url [0]{\catcode `\\12\catcode `\$12\catcode
  `\&12\catcode `\#12\catcode `\^12\catcode `\_12\catcode `\%12\relax}%
\providecommand \@@startlink[1]{}%
\providecommand \@@endlink[0]{}%
\providecommand \url  [0]{\begingroup\@sanitize@url \@url }%
\providecommand \@url [1]{\endgroup\@href {#1}{\urlprefix }}%
\providecommand \urlprefix  [0]{URL }%
\providecommand \Eprint [0]{\href }%
\providecommand \doibase [0]{http://dx.doi.org/}%
\providecommand \selectlanguage [0]{\@gobble}%
\providecommand \bibinfo  [0]{\@secondoftwo}%
\providecommand \bibfield  [0]{\@secondoftwo}%
\providecommand \translation [1]{[#1]}%
\providecommand \BibitemOpen [0]{}%
\providecommand \bibitemStop [0]{}%
\providecommand \bibitemNoStop [0]{.\EOS\space}%
\providecommand \EOS [0]{\spacefactor3000\relax}%
\providecommand \BibitemShut  [1]{\csname bibitem#1\endcsname}%
\let\auto@bib@innerbib\@empty
%</preamble>
\bibitem [{\citenamefont {Loss}\ and\ \citenamefont
  {DiVincenzo}(1998)}]{PhysRevA.57.120}%
  \BibitemOpen
  \bibfield  {author} {\bibinfo {author} {\bibfnamefont {D.}~\bibnamefont
  {Loss}}\ and\ \bibinfo {author} {\bibfnamefont {D.~P.}\ \bibnamefont
  {DiVincenzo}},\ }\href {\doibase 10.1103/PhysRevA.57.120} {\bibfield
  {journal} {\bibinfo  {journal} {Phys. Rev. A}\ }\textbf {\bibinfo {volume}
  {57}},\ \bibinfo {pages} {120} (\bibinfo {year} {1998})}\BibitemShut
  {NoStop}%
\bibitem [{\citenamefont {Morton}\ \emph {et~al.}(2011)\citenamefont {Morton},
  \citenamefont {McCamey}, \citenamefont {Eriksson},\ and\ \citenamefont
  {Lyon}}]{Morton:2011}%
  \BibitemOpen
  \bibfield  {author} {\bibinfo {author} {\bibfnamefont {J.~J.~L.}\
  \bibnamefont {Morton}}, \bibinfo {author} {\bibfnamefont {D.~R.}\
  \bibnamefont {McCamey}}, \bibinfo {author} {\bibfnamefont {M.~A.}\
  \bibnamefont {Eriksson}}, \ and\ \bibinfo {author} {\bibfnamefont {S.~A.}\
  \bibnamefont {Lyon}},\ }\href {https://doi.org/10.1038/nature10681}
  {\bibfield  {journal} {\bibinfo  {journal} {Nature}\ }\textbf {\bibinfo
  {volume} {479}},\ \bibinfo {pages} {345} (\bibinfo {year}
  {2011})}\BibitemShut {NoStop}%
\bibitem [{\citenamefont {Zwanenburg}\ \emph {et~al.}(2013)\citenamefont
  {Zwanenburg}, \citenamefont {Dzurak}, \citenamefont {Morello}, \citenamefont
  {Simmons}, \citenamefont {Hollenberg}, \citenamefont {Klimeck}, \citenamefont
  {Rogge}, \citenamefont {Coppersmith},\ and\ \citenamefont
  {Eriksson}}]{RevModPhys.85.961}%
  \BibitemOpen
  \bibfield  {author} {\bibinfo {author} {\bibfnamefont {F.~A.}\ \bibnamefont
  {Zwanenburg}}, \bibinfo {author} {\bibfnamefont {A.~S.}\ \bibnamefont
  {Dzurak}}, \bibinfo {author} {\bibfnamefont {A.}~\bibnamefont {Morello}},
  \bibinfo {author} {\bibfnamefont {M.~Y.}\ \bibnamefont {Simmons}}, \bibinfo
  {author} {\bibfnamefont {L.~C.~L.}\ \bibnamefont {Hollenberg}}, \bibinfo
  {author} {\bibfnamefont {G.}~\bibnamefont {Klimeck}}, \bibinfo {author}
  {\bibfnamefont {S.}~\bibnamefont {Rogge}}, \bibinfo {author} {\bibfnamefont
  {S.~N.}\ \bibnamefont {Coppersmith}}, \ and\ \bibinfo {author} {\bibfnamefont
  {M.~A.}\ \bibnamefont {Eriksson}},\ }\href {\doibase
  10.1103/RevModPhys.85.961} {\bibfield  {journal} {\bibinfo  {journal} {Rev.
  Mod. Phys.}\ }\textbf {\bibinfo {volume} {85}},\ \bibinfo {pages} {961}
  (\bibinfo {year} {2013})}\BibitemShut {NoStop}%
\bibitem [{\citenamefont {Pioro-Ladri\`{e}re}\ \emph
  {et~al.}(2008)\citenamefont {Pioro-Ladri\`{e}re}, \citenamefont {Obata},
  \citenamefont {Tokura}, \citenamefont {Shin}, \citenamefont {Kubo},
  \citenamefont {Yoshida}, \citenamefont {Taniyama},\ and\ \citenamefont
  {Tarucha}}]{PioroLadriere2008}%
  \BibitemOpen
  \bibfield  {author} {\bibinfo {author} {\bibfnamefont {M.}~\bibnamefont
  {Pioro-Ladri\`{e}re}}, \bibinfo {author} {\bibfnamefont {T.}~\bibnamefont
  {Obata}}, \bibinfo {author} {\bibfnamefont {Y.}~\bibnamefont {Tokura}},
  \bibinfo {author} {\bibfnamefont {Y.~S.}\ \bibnamefont {Shin}}, \bibinfo
  {author} {\bibfnamefont {T.}~\bibnamefont {Kubo}}, \bibinfo {author}
  {\bibfnamefont {K.}~\bibnamefont {Yoshida}}, \bibinfo {author} {\bibfnamefont
  {T.}~\bibnamefont {Taniyama}}, \ and\ \bibinfo {author} {\bibfnamefont
  {S.}~\bibnamefont {Tarucha}},\ }\href {http://dx.doi.org/10.1038/nphys1053}
  {\bibfield  {journal} {\bibinfo  {journal} {Nature Phys.}\ }\textbf {\bibinfo
  {volume} {4}},\ \bibinfo {pages} {776} (\bibinfo {year} {2008})}\BibitemShut
  {NoStop}%
\bibitem [{\citenamefont {Veldhorst}\ \emph {et~al.}(2014)\citenamefont
  {Veldhorst}, \citenamefont {Hwang}, \citenamefont {Yang}, \citenamefont
  {Leenstra}, \citenamefont {de~Ronde}, \citenamefont {Dehollain},
  \citenamefont {Muhonen}, \citenamefont {Hudson}, \citenamefont {Itoh},
  \citenamefont {Morello},\ and\ \citenamefont {Dzurak}}]{Veldhorst2014}%
  \BibitemOpen
  \bibfield  {author} {\bibinfo {author} {\bibfnamefont {M.}~\bibnamefont
  {Veldhorst}}, \bibinfo {author} {\bibfnamefont {J.~C.~C.}\ \bibnamefont
  {Hwang}}, \bibinfo {author} {\bibfnamefont {C.~H.}\ \bibnamefont {Yang}},
  \bibinfo {author} {\bibfnamefont {A.~W.}\ \bibnamefont {Leenstra}}, \bibinfo
  {author} {\bibfnamefont {B.}~\bibnamefont {de~Ronde}}, \bibinfo {author}
  {\bibfnamefont {J.~P.}\ \bibnamefont {Dehollain}}, \bibinfo {author}
  {\bibfnamefont {J.~T.}\ \bibnamefont {Muhonen}}, \bibinfo {author}
  {\bibfnamefont {F.~E.}\ \bibnamefont {Hudson}}, \bibinfo {author}
  {\bibfnamefont {K.~M.}\ \bibnamefont {Itoh}}, \bibinfo {author}
  {\bibfnamefont {A.}~\bibnamefont {Morello}}, \ and\ \bibinfo {author}
  {\bibfnamefont {A.~S.}\ \bibnamefont {Dzurak}},\ }\href
  {http://dx.doi.org/10.1038/nnano.2014.216} {\bibfield  {journal} {\bibinfo
  {journal} {Nature Nano.}\ }\textbf {\bibinfo {volume} {9}},\ \bibinfo {pages}
  {981} (\bibinfo {year} {2014})}\BibitemShut {NoStop}%
\bibitem [{\citenamefont {Kawakami}\ \emph {et~al.}(2014)\citenamefont
  {Kawakami}, \citenamefont {Scarlino}, \citenamefont {Ward}, \citenamefont
  {Braakman}, \citenamefont {Savage}, \citenamefont {Lagally}, \citenamefont
  {Friesen}, \citenamefont {Coppersmith}, \citenamefont {Eriksson},\ and\
  \citenamefont {Vandersypen}}]{Kawakami2014}%
  \BibitemOpen
  \bibfield  {author} {\bibinfo {author} {\bibfnamefont {E.}~\bibnamefont
  {Kawakami}}, \bibinfo {author} {\bibfnamefont {P.}~\bibnamefont {Scarlino}},
  \bibinfo {author} {\bibfnamefont {D.~R.}\ \bibnamefont {Ward}}, \bibinfo
  {author} {\bibfnamefont {F.~R.}\ \bibnamefont {Braakman}}, \bibinfo {author}
  {\bibfnamefont {D.~E.}\ \bibnamefont {Savage}}, \bibinfo {author}
  {\bibfnamefont {M.~G.}\ \bibnamefont {Lagally}}, \bibinfo {author}
  {\bibfnamefont {M.}~\bibnamefont {Friesen}}, \bibinfo {author} {\bibfnamefont
  {S.~N.}\ \bibnamefont {Coppersmith}}, \bibinfo {author} {\bibfnamefont
  {M.~A.}\ \bibnamefont {Eriksson}}, \ and\ \bibinfo {author} {\bibfnamefont
  {L.~M.~K.}\ \bibnamefont {Vandersypen}},\ }\href {\doibase
  10.1038/nnano.2014.153} {\bibfield  {journal} {\bibinfo  {journal} {Nature
  Nano.}\ }\textbf {\bibinfo {volume} {9}},\ \bibinfo {pages} {666} (\bibinfo
  {year} {2014})}\BibitemShut {NoStop}%
\bibitem [{\citenamefont {Yoneda}\ \emph {et~al.}(2018)\citenamefont {Yoneda},
  \citenamefont {Takeda}, \citenamefont {Otsuka}, \citenamefont {Nakajima},
  \citenamefont {Delbecq}, \citenamefont {Allison}, \citenamefont {Honda},
  \citenamefont {Kodera}, \citenamefont {Oda}, \citenamefont {Hoshi},
  \citenamefont {Usami}, \citenamefont {Itoh},\ and\ \citenamefont
  {Tarucha}}]{Yoneda2018}%
  \BibitemOpen
  \bibfield  {author} {\bibinfo {author} {\bibfnamefont {J.}~\bibnamefont
  {Yoneda}}, \bibinfo {author} {\bibfnamefont {K.}~\bibnamefont {Takeda}},
  \bibinfo {author} {\bibfnamefont {T.}~\bibnamefont {Otsuka}}, \bibinfo
  {author} {\bibfnamefont {T.}~\bibnamefont {Nakajima}}, \bibinfo {author}
  {\bibfnamefont {M.~R.}\ \bibnamefont {Delbecq}}, \bibinfo {author}
  {\bibfnamefont {G.}~\bibnamefont {Allison}}, \bibinfo {author} {\bibfnamefont
  {T.}~\bibnamefont {Honda}}, \bibinfo {author} {\bibfnamefont
  {T.}~\bibnamefont {Kodera}}, \bibinfo {author} {\bibfnamefont
  {S.}~\bibnamefont {Oda}}, \bibinfo {author} {\bibfnamefont {Y.}~\bibnamefont
  {Hoshi}}, \bibinfo {author} {\bibfnamefont {N.}~\bibnamefont {Usami}},
  \bibinfo {author} {\bibfnamefont {K.~M.}\ \bibnamefont {Itoh}}, \ and\
  \bibinfo {author} {\bibfnamefont {S.}~\bibnamefont {Tarucha}},\ }\href
  {https://doi.org/10.1038/s41565-017-0014-x} {\bibfield  {journal} {\bibinfo
  {journal} {Nature Nano.}\ }\textbf {\bibinfo {volume} {13}},\ \bibinfo
  {pages} {102} (\bibinfo {year} {2018})}\BibitemShut {NoStop}%
\bibitem [{\citenamefont {Shulman}\ \emph {et~al.}(2014)\citenamefont
  {Shulman}, \citenamefont {Harvey}, \citenamefont {Nichol}, \citenamefont
  {Bartlett}, \citenamefont {Doherty}, \citenamefont {Umansky},\ and\
  \citenamefont {Yacoby}}]{Shulman2014}%
  \BibitemOpen
  \bibfield  {author} {\bibinfo {author} {\bibfnamefont {M.~D.}\ \bibnamefont
  {Shulman}}, \bibinfo {author} {\bibfnamefont {S.~P.}\ \bibnamefont {Harvey}},
  \bibinfo {author} {\bibfnamefont {J.~M.}\ \bibnamefont {Nichol}}, \bibinfo
  {author} {\bibfnamefont {S.~D.}\ \bibnamefont {Bartlett}}, \bibinfo {author}
  {\bibfnamefont {A.~C.}\ \bibnamefont {Doherty}}, \bibinfo {author}
  {\bibfnamefont {V.}~\bibnamefont {Umansky}}, \ and\ \bibinfo {author}
  {\bibfnamefont {A.}~\bibnamefont {Yacoby}},\ }\href
  {http://dx.doi.org/10.1038/ncomms6156} {\bibfield  {journal} {\bibinfo
  {journal} {Nature Commun.}\ }\textbf {\bibinfo {volume} {5}},\ \bibinfo
  {pages} {5156} (\bibinfo {year} {2014})}\BibitemShut {NoStop}%
\bibitem [{\citenamefont {Kim}\ \emph {et~al.}(2015{\natexlab{a}})\citenamefont
  {Kim}, \citenamefont {Ward}, \citenamefont {Simmons}, \citenamefont {Gamble},
  \citenamefont {Blume-Kohout}, \citenamefont {Nielsen}, \citenamefont
  {Savage}, \citenamefont {Lagally}, \citenamefont {Friesen}, \citenamefont
  {Coppersmith},\ and\ \citenamefont {Eriksson}}]{Kim2015}%
  \BibitemOpen
  \bibfield  {author} {\bibinfo {author} {\bibfnamefont {D.}~\bibnamefont
  {Kim}}, \bibinfo {author} {\bibfnamefont {D.~R.}\ \bibnamefont {Ward}},
  \bibinfo {author} {\bibfnamefont {C.~B.}\ \bibnamefont {Simmons}}, \bibinfo
  {author} {\bibfnamefont {J.~K.}\ \bibnamefont {Gamble}}, \bibinfo {author}
  {\bibfnamefont {R.}~\bibnamefont {Blume-Kohout}}, \bibinfo {author}
  {\bibfnamefont {E.}~\bibnamefont {Nielsen}}, \bibinfo {author} {\bibfnamefont
  {D.~E.}\ \bibnamefont {Savage}}, \bibinfo {author} {\bibfnamefont {M.~G.}\
  \bibnamefont {Lagally}}, \bibinfo {author} {\bibfnamefont {M.}~\bibnamefont
  {Friesen}}, \bibinfo {author} {\bibfnamefont {S.~N.}\ \bibnamefont
  {Coppersmith}}, \ and\ \bibinfo {author} {\bibfnamefont {M.~A.}\ \bibnamefont
  {Eriksson}},\ }\href {\doibase 10.1038/nnano.2014.336} {\bibfield  {journal}
  {\bibinfo  {journal} {Nature Nano.}\ }\textbf {\bibinfo {volume} {10}},\
  \bibinfo {pages} {243} (\bibinfo {year} {2015}{\natexlab{a}})}\BibitemShut
  {NoStop}%
\bibitem [{\citenamefont {Thorgrimsson}\ \emph {et~al.}(2017)\citenamefont
  {Thorgrimsson}, \citenamefont {Kim}, \citenamefont {Yang}, \citenamefont
  {Smith}, \citenamefont {Simmons}, \citenamefont {Ward}, \citenamefont
  {Foote}, \citenamefont {Corrigan}, \citenamefont {Savage}, \citenamefont
  {Lagally}, \citenamefont {Friesen}, \citenamefont {Coppersmith},\ and\
  \citenamefont {Eriksson}}]{Thorgrimsson2017}%
  \BibitemOpen
  \bibfield  {author} {\bibinfo {author} {\bibfnamefont {B.}~\bibnamefont
  {Thorgrimsson}}, \bibinfo {author} {\bibfnamefont {D.}~\bibnamefont {Kim}},
  \bibinfo {author} {\bibfnamefont {Y.-C.}\ \bibnamefont {Yang}}, \bibinfo
  {author} {\bibfnamefont {L.~W.}\ \bibnamefont {Smith}}, \bibinfo {author}
  {\bibfnamefont {C.~B.}\ \bibnamefont {Simmons}}, \bibinfo {author}
  {\bibfnamefont {D.~R.}\ \bibnamefont {Ward}}, \bibinfo {author}
  {\bibfnamefont {R.~H.}\ \bibnamefont {Foote}}, \bibinfo {author}
  {\bibfnamefont {J.}~\bibnamefont {Corrigan}}, \bibinfo {author}
  {\bibfnamefont {D.~E.}\ \bibnamefont {Savage}}, \bibinfo {author}
  {\bibfnamefont {M.~G.}\ \bibnamefont {Lagally}}, \bibinfo {author}
  {\bibfnamefont {M.}~\bibnamefont {Friesen}}, \bibinfo {author} {\bibfnamefont
  {S.~N.}\ \bibnamefont {Coppersmith}}, \ and\ \bibinfo {author} {\bibfnamefont
  {M.~A.}\ \bibnamefont {Eriksson}},\ }\href {\doibase
  10.1038/s41534-017-0034-2} {\bibfield  {journal} {\bibinfo  {journal} {npj
  Quantum Inform.}\ }\textbf {\bibinfo {volume} {3}},\ \bibinfo {pages} {32}
  (\bibinfo {year} {2017})}\BibitemShut {NoStop}%
\bibitem [{\citenamefont {Medford}\ \emph {et~al.}(2013)\citenamefont
  {Medford}, \citenamefont {Beil}, \citenamefont {Taylor}, \citenamefont
  {Rashba}, \citenamefont {Lu}, \citenamefont {Gossard},\ and\ \citenamefont
  {Marcus}}]{Medford2013}%
  \BibitemOpen
  \bibfield  {author} {\bibinfo {author} {\bibfnamefont {J.}~\bibnamefont
  {Medford}}, \bibinfo {author} {\bibfnamefont {J.}~\bibnamefont {Beil}},
  \bibinfo {author} {\bibfnamefont {J.~M.}\ \bibnamefont {Taylor}}, \bibinfo
  {author} {\bibfnamefont {E.~I.}\ \bibnamefont {Rashba}}, \bibinfo {author}
  {\bibfnamefont {H.}~\bibnamefont {Lu}}, \bibinfo {author} {\bibfnamefont
  {A.~C.}\ \bibnamefont {Gossard}}, \ and\ \bibinfo {author} {\bibfnamefont
  {C.~M.}\ \bibnamefont {Marcus}},\ }\href {\doibase
  10.1103/PhysRevLett.111.050501} {\bibfield  {journal} {\bibinfo  {journal}
  {Phys. Rev. Lett.}\ }\textbf {\bibinfo {volume} {111}},\ \bibinfo {pages}
  {050501} (\bibinfo {year} {2013})}\BibitemShut {NoStop}%
\bibitem [{\citenamefont {Landig}\ \emph {et~al.}(2018)\citenamefont {Landig},
  \citenamefont {Koski}, \citenamefont {Scarlino}, \citenamefont {Mendes},
  \citenamefont {Blais}, \citenamefont {Reichl}, \citenamefont {Wegscheider},
  \citenamefont {Wallraff}, \citenamefont {Ensslin},\ and\ \citenamefont
  {Ihn}}]{Landig:2018}%
  \BibitemOpen
  \bibfield  {author} {\bibinfo {author} {\bibfnamefont {A.~J.}\ \bibnamefont
  {Landig}}, \bibinfo {author} {\bibfnamefont {J.~V.}\ \bibnamefont {Koski}},
  \bibinfo {author} {\bibfnamefont {P.}~\bibnamefont {Scarlino}}, \bibinfo
  {author} {\bibfnamefont {U.~C.}\ \bibnamefont {Mendes}}, \bibinfo {author}
  {\bibfnamefont {A.}~\bibnamefont {Blais}}, \bibinfo {author} {\bibfnamefont
  {C.}~\bibnamefont {Reichl}}, \bibinfo {author} {\bibfnamefont
  {W.}~\bibnamefont {Wegscheider}}, \bibinfo {author} {\bibfnamefont
  {A.}~\bibnamefont {Wallraff}}, \bibinfo {author} {\bibfnamefont
  {K.}~\bibnamefont {Ensslin}}, \ and\ \bibinfo {author} {\bibfnamefont
  {T.}~\bibnamefont {Ihn}},\ }\href@noop {} {\bibfield  {journal} {\bibinfo
  {journal} {Nature}\ }\textbf {\bibinfo {volume} {560}},\ \bibinfo {pages}
  {179} (\bibinfo {year} {2018})}\BibitemShut {NoStop}%
\bibitem [{\citenamefont {Veldhorst}\ \emph {et~al.}(2015)\citenamefont
  {Veldhorst}, \citenamefont {Yang}, \citenamefont {Hwang}, \citenamefont
  {Huang}, \citenamefont {Dehollain}, \citenamefont {Muhonen}, \citenamefont
  {Simmons}, \citenamefont {Laucht}, \citenamefont {Hudson}, \citenamefont
  {Itoh}, \citenamefont {Morello},\ and\ \citenamefont
  {Dzurak}}]{Veldhorst:2015}%
  \BibitemOpen
  \bibfield  {author} {\bibinfo {author} {\bibfnamefont {M.}~\bibnamefont
  {Veldhorst}}, \bibinfo {author} {\bibfnamefont {C.~H.}\ \bibnamefont {Yang}},
  \bibinfo {author} {\bibfnamefont {J.~C.~C.}\ \bibnamefont {Hwang}}, \bibinfo
  {author} {\bibfnamefont {W.}~\bibnamefont {Huang}}, \bibinfo {author}
  {\bibfnamefont {J.~P.}\ \bibnamefont {Dehollain}}, \bibinfo {author}
  {\bibfnamefont {J.~T.}\ \bibnamefont {Muhonen}}, \bibinfo {author}
  {\bibfnamefont {S.}~\bibnamefont {Simmons}}, \bibinfo {author} {\bibfnamefont
  {A.}~\bibnamefont {Laucht}}, \bibinfo {author} {\bibfnamefont {F.~E.}\
  \bibnamefont {Hudson}}, \bibinfo {author} {\bibfnamefont {K.~M.}\
  \bibnamefont {Itoh}}, \bibinfo {author} {\bibfnamefont {A.}~\bibnamefont
  {Morello}}, \ and\ \bibinfo {author} {\bibfnamefont {A.~S.}\ \bibnamefont
  {Dzurak}},\ }\href {https://doi.org/10.1038/nature15263} {\bibfield
  {journal} {\bibinfo  {journal} {Nature}\ }\textbf {\bibinfo {volume} {526}},\
  \bibinfo {pages} {410} (\bibinfo {year} {2015})}\BibitemShut {NoStop}%
\bibitem [{\citenamefont {Zajac}\ \emph {et~al.}(2018)\citenamefont {Zajac},
  \citenamefont {Sigillito}, \citenamefont {Russ}, \citenamefont {Borjans},
  \citenamefont {Taylor}, \citenamefont {Burkard},\ and\ \citenamefont
  {Petta}}]{Zajaceaao5965}%
  \BibitemOpen
  \bibfield  {author} {\bibinfo {author} {\bibfnamefont {D.~M.}\ \bibnamefont
  {Zajac}}, \bibinfo {author} {\bibfnamefont {A.~J.}\ \bibnamefont
  {Sigillito}}, \bibinfo {author} {\bibfnamefont {M.}~\bibnamefont {Russ}},
  \bibinfo {author} {\bibfnamefont {F.}~\bibnamefont {Borjans}}, \bibinfo
  {author} {\bibfnamefont {J.~M.}\ \bibnamefont {Taylor}}, \bibinfo {author}
  {\bibfnamefont {G.}~\bibnamefont {Burkard}}, \ and\ \bibinfo {author}
  {\bibfnamefont {J.~R.}\ \bibnamefont {Petta}},\ }\href
  {http://science.sciencemag.org/content/early/2017/12/06/science.aao5965}
  {\bibfield  {journal} {\bibinfo  {journal} {Science}\ }\textbf {\bibinfo
  {volume} {359}},\ \bibinfo {pages} {439} (\bibinfo {year}
  {2018})}\BibitemShut {NoStop}%
\bibitem [{\citenamefont {Watson}\ \emph {et~al.}(2018)\citenamefont {Watson},
  \citenamefont {Philips}, \citenamefont {Kawakami}, \citenamefont {Ward},
  \citenamefont {Scarlino}, \citenamefont {Veldhorst}, \citenamefont {Savage},
  \citenamefont {Lagally}, \citenamefont {Friesen}, \citenamefont
  {Coppersmith}, \citenamefont {Eriksson},\ and\ \citenamefont
  {Vandersypen}}]{Watson2018}%
  \BibitemOpen
  \bibfield  {author} {\bibinfo {author} {\bibfnamefont {T.~F.}\ \bibnamefont
  {Watson}}, \bibinfo {author} {\bibfnamefont {S.~G.~J.}\ \bibnamefont
  {Philips}}, \bibinfo {author} {\bibfnamefont {E.}~\bibnamefont {Kawakami}},
  \bibinfo {author} {\bibfnamefont {D.~R.}\ \bibnamefont {Ward}}, \bibinfo
  {author} {\bibfnamefont {P.}~\bibnamefont {Scarlino}}, \bibinfo {author}
  {\bibfnamefont {M.}~\bibnamefont {Veldhorst}}, \bibinfo {author}
  {\bibfnamefont {D.~E.}\ \bibnamefont {Savage}}, \bibinfo {author}
  {\bibfnamefont {M.~G.}\ \bibnamefont {Lagally}}, \bibinfo {author}
  {\bibfnamefont {M.}~\bibnamefont {Friesen}}, \bibinfo {author} {\bibfnamefont
  {S.~N.}\ \bibnamefont {Coppersmith}}, \bibinfo {author} {\bibfnamefont
  {M.~A.}\ \bibnamefont {Eriksson}}, \ and\ \bibinfo {author} {\bibfnamefont
  {L.~M.~K.}\ \bibnamefont {Vandersypen}},\ }\href
  {https://www.nature.com/articles/nature25766#supplementary-information}
  {\bibfield  {journal} {\bibinfo  {journal} {Nature}\ }\textbf {\bibinfo
  {volume} {555}},\ \bibinfo {pages} {633} (\bibinfo {year}
  {2018})}\BibitemShut {NoStop}%
\bibitem [{\citenamefont {Hendrickx}\ \emph {et~al.}(2020)\citenamefont
  {Hendrickx}, \citenamefont {Franke}, \citenamefont {Sammak}, \citenamefont
  {Scappucci},\ and\ \citenamefont {Veldhorst}}]{Hendrickx:Preprint}%
  \BibitemOpen
  \bibfield  {author} {\bibinfo {author} {\bibfnamefont {N.~W.}\ \bibnamefont
  {Hendrickx}}, \bibinfo {author} {\bibfnamefont {D.~P.}\ \bibnamefont
  {Franke}}, \bibinfo {author} {\bibfnamefont {A.}~\bibnamefont {Sammak}},
  \bibinfo {author} {\bibfnamefont {G.}~\bibnamefont {Scappucci}}, \ and\
  \bibinfo {author} {\bibfnamefont {M.}~\bibnamefont {Veldhorst}},\ }\href@noop
  {} {\bibfield  {journal} {\bibinfo  {journal} {Nature}\ }\textbf {\bibinfo
  {volume} {577}},\ \bibinfo {pages} {487} (\bibinfo {year}
  {2020})}\BibitemShut {NoStop}%
\bibitem [{\citenamefont {Nichol}\ \emph {et~al.}(2017)\citenamefont {Nichol},
  \citenamefont {Orona}, \citenamefont {Harvey}, \citenamefont {Fallahi},
  \citenamefont {Gardner}, \citenamefont {Manfra},\ and\ \citenamefont
  {Yacoby}}]{Nichol2017}%
  \BibitemOpen
  \bibfield  {author} {\bibinfo {author} {\bibfnamefont {J.~M.}\ \bibnamefont
  {Nichol}}, \bibinfo {author} {\bibfnamefont {L.~A.}\ \bibnamefont {Orona}},
  \bibinfo {author} {\bibfnamefont {S.~P.}\ \bibnamefont {Harvey}}, \bibinfo
  {author} {\bibfnamefont {S.}~\bibnamefont {Fallahi}}, \bibinfo {author}
  {\bibfnamefont {G.~C.}\ \bibnamefont {Gardner}}, \bibinfo {author}
  {\bibfnamefont {M.~J.}\ \bibnamefont {Manfra}}, \ and\ \bibinfo {author}
  {\bibfnamefont {A.}~\bibnamefont {Yacoby}},\ }\href {\doibase
  10.1038/s41534-016-0003-1} {\bibfield  {journal} {\bibinfo  {journal} {npj
  Quantum Inform.}\ }\textbf {\bibinfo {volume} {3}},\ \bibinfo {pages} {3}
  (\bibinfo {year} {2017})}\BibitemShut {NoStop}%
\bibitem [{\citenamefont {Vion}\ \emph {et~al.}(2002)\citenamefont {Vion},
  \citenamefont {Aassime}, \citenamefont {Cottet}, \citenamefont {Joyez},
  \citenamefont {Pothier}, \citenamefont {Urbina}, \citenamefont {Esteve},\
  and\ \citenamefont {Devoret}}]{Vion2002}%
  \BibitemOpen
  \bibfield  {author} {\bibinfo {author} {\bibfnamefont {D.}~\bibnamefont
  {Vion}}, \bibinfo {author} {\bibfnamefont {A.}~\bibnamefont {Aassime}},
  \bibinfo {author} {\bibfnamefont {A.}~\bibnamefont {Cottet}}, \bibinfo
  {author} {\bibfnamefont {P.}~\bibnamefont {Joyez}}, \bibinfo {author}
  {\bibfnamefont {H.}~\bibnamefont {Pothier}}, \bibinfo {author} {\bibfnamefont
  {C.}~\bibnamefont {Urbina}}, \bibinfo {author} {\bibfnamefont
  {D.}~\bibnamefont {Esteve}}, \ and\ \bibinfo {author} {\bibfnamefont {M.~H.}\
  \bibnamefont {Devoret}},\ }\href@noop {} {\bibfield  {journal} {\bibinfo
  {journal} {Science}\ }\textbf {\bibinfo {volume} {296}},\ \bibinfo {pages}
  {886} (\bibinfo {year} {2002})}\BibitemShut {NoStop}%
\bibitem [{\citenamefont {Kim}\ \emph {et~al.}(2014)\citenamefont {Kim},
  \citenamefont {Shi}, \citenamefont {Simmons}, \citenamefont {Ward},
  \citenamefont {Prance}, \citenamefont {Koh}, \citenamefont {Gamble},
  \citenamefont {Savage}, \citenamefont {Lagally}, \citenamefont {Friesen},
  \citenamefont {Coppersmith},\ and\ \citenamefont
  {Eriksson}}]{KimShiSimmonsEtAl2014}%
  \BibitemOpen
  \bibfield  {author} {\bibinfo {author} {\bibfnamefont {D.}~\bibnamefont
  {Kim}}, \bibinfo {author} {\bibfnamefont {Z.}~\bibnamefont {Shi}}, \bibinfo
  {author} {\bibfnamefont {C.~B.}\ \bibnamefont {Simmons}}, \bibinfo {author}
  {\bibfnamefont {D.~R.}\ \bibnamefont {Ward}}, \bibinfo {author}
  {\bibfnamefont {J.~R.}\ \bibnamefont {Prance}}, \bibinfo {author}
  {\bibfnamefont {T.~S.}\ \bibnamefont {Koh}}, \bibinfo {author} {\bibfnamefont
  {J.~K.}\ \bibnamefont {Gamble}}, \bibinfo {author} {\bibfnamefont {D.~E.}\
  \bibnamefont {Savage}}, \bibinfo {author} {\bibfnamefont {M.~G.}\
  \bibnamefont {Lagally}}, \bibinfo {author} {\bibfnamefont {M.}~\bibnamefont
  {Friesen}}, \bibinfo {author} {\bibfnamefont {S.~N.}\ \bibnamefont
  {Coppersmith}}, \ and\ \bibinfo {author} {\bibfnamefont {M.~A.}\ \bibnamefont
  {Eriksson}},\ }\href {http://dx.doi.org/10.1038/nature13407} {\bibfield
  {journal} {\bibinfo  {journal} {Nature}\ }\textbf {\bibinfo {volume} {511}},\
  \bibinfo {pages} {70} (\bibinfo {year} {2014})}\BibitemShut {NoStop}%
\bibitem [{\citenamefont {Cao}\ \emph {et~al.}(2016)\citenamefont {Cao},
  \citenamefont {Li}, \citenamefont {Yu}, \citenamefont {Wang}, \citenamefont
  {Chen}, \citenamefont {Song}, \citenamefont {Xiao}, \citenamefont {Guo},
  \citenamefont {Jiang}, \citenamefont {Hu},\ and\ \citenamefont
  {Guo}}]{Cao2016}%
  \BibitemOpen
  \bibfield  {author} {\bibinfo {author} {\bibfnamefont {G.}~\bibnamefont
  {Cao}}, \bibinfo {author} {\bibfnamefont {H.-O.}\ \bibnamefont {Li}},
  \bibinfo {author} {\bibfnamefont {G.-D.}\ \bibnamefont {Yu}}, \bibinfo
  {author} {\bibfnamefont {B.-C.}\ \bibnamefont {Wang}}, \bibinfo {author}
  {\bibfnamefont {B.-B.}\ \bibnamefont {Chen}}, \bibinfo {author}
  {\bibfnamefont {X.-X.}\ \bibnamefont {Song}}, \bibinfo {author}
  {\bibfnamefont {M.}~\bibnamefont {Xiao}}, \bibinfo {author} {\bibfnamefont
  {G.-C.}\ \bibnamefont {Guo}}, \bibinfo {author} {\bibfnamefont {H.-W.}\
  \bibnamefont {Jiang}}, \bibinfo {author} {\bibfnamefont {X.}~\bibnamefont
  {Hu}}, \ and\ \bibinfo {author} {\bibfnamefont {G.-P.}\ \bibnamefont {Guo}},\
  }\href {\doibase 10.1103/PhysRevLett.116.086801} {\bibfield  {journal}
  {\bibinfo  {journal} {Phys. Rev. Lett.}\ }\textbf {\bibinfo {volume} {116}},\
  \bibinfo {pages} {086801} (\bibinfo {year} {2016})}\BibitemShut {NoStop}%
\bibitem [{\citenamefont {Martins}\ \emph {et~al.}(2016)\citenamefont
  {Martins}, \citenamefont {Malinowski}, \citenamefont {Nissen}, \citenamefont
  {Barnes}, \citenamefont {Fallahi}, \citenamefont {Gardner}, \citenamefont
  {Manfra}, \citenamefont {Marcus},\ and\ \citenamefont
  {Kuemmeth}}]{PhysRevLett.116.116801}%
  \BibitemOpen
  \bibfield  {author} {\bibinfo {author} {\bibfnamefont {F.}~\bibnamefont
  {Martins}}, \bibinfo {author} {\bibfnamefont {F.~K.}\ \bibnamefont
  {Malinowski}}, \bibinfo {author} {\bibfnamefont {P.~D.}\ \bibnamefont
  {Nissen}}, \bibinfo {author} {\bibfnamefont {E.}~\bibnamefont {Barnes}},
  \bibinfo {author} {\bibfnamefont {S.}~\bibnamefont {Fallahi}}, \bibinfo
  {author} {\bibfnamefont {G.~C.}\ \bibnamefont {Gardner}}, \bibinfo {author}
  {\bibfnamefont {M.~J.}\ \bibnamefont {Manfra}}, \bibinfo {author}
  {\bibfnamefont {C.~M.}\ \bibnamefont {Marcus}}, \ and\ \bibinfo {author}
  {\bibfnamefont {F.}~\bibnamefont {Kuemmeth}},\ }\href {\doibase
  10.1103/PhysRevLett.116.116801} {\bibfield  {journal} {\bibinfo  {journal}
  {Phys. Rev. Lett.}\ }\textbf {\bibinfo {volume} {116}},\ \bibinfo {pages}
  {116801} (\bibinfo {year} {2016})}\BibitemShut {NoStop}%
\bibitem [{\citenamefont {Schoenfield}\ \emph {et~al.}(2017)\citenamefont
  {Schoenfield}, \citenamefont {Freeman},\ and\ \citenamefont
  {Jiang}}]{Schoenfield:2017}%
  \BibitemOpen
  \bibfield  {author} {\bibinfo {author} {\bibfnamefont {J.~S.}\ \bibnamefont
  {Schoenfield}}, \bibinfo {author} {\bibfnamefont {B.~M.}\ \bibnamefont
  {Freeman}}, \ and\ \bibinfo {author} {\bibfnamefont {H.}~\bibnamefont
  {Jiang}},\ }\href {\doibase 10.1038/s41467-017-00073-x} {\bibfield  {journal}
  {\bibinfo  {journal} {Nature Commun.}\ }\textbf {\bibinfo {volume} {8}},\
  \bibinfo {pages} {64} (\bibinfo {year} {2017})}\BibitemShut {NoStop}%
\bibitem [{\citenamefont {Croot}\ \emph {et~al.}(2020)\citenamefont {Croot},
  \citenamefont {Mi}, \citenamefont {Putz}, \citenamefont {Benito},
  \citenamefont {Borjans}, \citenamefont {Burkard},\ and\ \citenamefont
  {Petta}}]{Croot:Preprint}%
  \BibitemOpen
  \bibfield  {author} {\bibinfo {author} {\bibfnamefont {X.}~\bibnamefont
  {Croot}}, \bibinfo {author} {\bibfnamefont {X.}~\bibnamefont {Mi}}, \bibinfo
  {author} {\bibfnamefont {S.}~\bibnamefont {Putz}}, \bibinfo {author}
  {\bibfnamefont {M.}~\bibnamefont {Benito}}, \bibinfo {author} {\bibfnamefont
  {F.}~\bibnamefont {Borjans}}, \bibinfo {author} {\bibfnamefont
  {G.}~\bibnamefont {Burkard}}, \ and\ \bibinfo {author} {\bibfnamefont
  {J.~R.}\ \bibnamefont {Petta}},\ }\href@noop {} {\bibfield  {journal}
  {\bibinfo  {journal} {Phys. Rev. Research}\ }\textbf {\bibinfo {volume}
  {2}},\ \bibinfo {pages} {012006(R)} (\bibinfo {year} {2020})}\BibitemShut
  {NoStop}%
\bibitem [{\citenamefont {Reed}\ \emph {et~al.}(2016)\citenamefont {Reed},
  \citenamefont {Maune}, \citenamefont {Andrews}, \citenamefont {Borselli},
  \citenamefont {Eng}, \citenamefont {Jura}, \citenamefont {Kiselev},
  \citenamefont {Ladd}, \citenamefont {Merkel}, \citenamefont {Milosavljevic},
  \citenamefont {Pritchett}, \citenamefont {Rakher}, \citenamefont {Ross},
  \citenamefont {Schmitz}, \citenamefont {Smith}, \citenamefont {Wright},
  \citenamefont {Gyure},\ and\ \citenamefont
  {Hunter}}]{PhysRevLett.116.110402}%
  \BibitemOpen
  \bibfield  {author} {\bibinfo {author} {\bibfnamefont {M.~D.}\ \bibnamefont
  {Reed}}, \bibinfo {author} {\bibfnamefont {B.~M.}\ \bibnamefont {Maune}},
  \bibinfo {author} {\bibfnamefont {R.~W.}\ \bibnamefont {Andrews}}, \bibinfo
  {author} {\bibfnamefont {M.~G.}\ \bibnamefont {Borselli}}, \bibinfo {author}
  {\bibfnamefont {K.}~\bibnamefont {Eng}}, \bibinfo {author} {\bibfnamefont
  {M.~P.}\ \bibnamefont {Jura}}, \bibinfo {author} {\bibfnamefont {A.~A.}\
  \bibnamefont {Kiselev}}, \bibinfo {author} {\bibfnamefont {T.~D.}\
  \bibnamefont {Ladd}}, \bibinfo {author} {\bibfnamefont {S.~T.}\ \bibnamefont
  {Merkel}}, \bibinfo {author} {\bibfnamefont {I.}~\bibnamefont
  {Milosavljevic}}, \bibinfo {author} {\bibfnamefont {E.~J.}\ \bibnamefont
  {Pritchett}}, \bibinfo {author} {\bibfnamefont {M.~T.}\ \bibnamefont
  {Rakher}}, \bibinfo {author} {\bibfnamefont {R.~S.}\ \bibnamefont {Ross}},
  \bibinfo {author} {\bibfnamefont {A.~E.}\ \bibnamefont {Schmitz}}, \bibinfo
  {author} {\bibfnamefont {A.}~\bibnamefont {Smith}}, \bibinfo {author}
  {\bibfnamefont {J.~A.}\ \bibnamefont {Wright}}, \bibinfo {author}
  {\bibfnamefont {M.~F.}\ \bibnamefont {Gyure}}, \ and\ \bibinfo {author}
  {\bibfnamefont {A.~T.}\ \bibnamefont {Hunter}},\ }\href {\doibase
  10.1103/PhysRevLett.116.110402} {\bibfield  {journal} {\bibinfo  {journal}
  {Phys. Rev. Lett.}\ }\textbf {\bibinfo {volume} {116}},\ \bibinfo {pages}
  {110402} (\bibinfo {year} {2016})}\BibitemShut {NoStop}%
\bibitem [{\citenamefont {Shi}\ \emph {et~al.}(2012)\citenamefont {Shi},
  \citenamefont {Simmons}, \citenamefont {Prance}, \citenamefont {Gamble},
  \citenamefont {Koh}, \citenamefont {Shim}, \citenamefont {Hu}, \citenamefont
  {Savage}, \citenamefont {Lagally}, \citenamefont {Eriksson}, \citenamefont
  {Friesen},\ and\ \citenamefont {Coppersmith}}]{Shi2012}%
  \BibitemOpen
  \bibfield  {author} {\bibinfo {author} {\bibfnamefont {Z.}~\bibnamefont
  {Shi}}, \bibinfo {author} {\bibfnamefont {C.~B.}\ \bibnamefont {Simmons}},
  \bibinfo {author} {\bibfnamefont {J.~R.}\ \bibnamefont {Prance}}, \bibinfo
  {author} {\bibfnamefont {J.~K.}\ \bibnamefont {Gamble}}, \bibinfo {author}
  {\bibfnamefont {T.~S.}\ \bibnamefont {Koh}}, \bibinfo {author} {\bibfnamefont
  {Y.-P.}\ \bibnamefont {Shim}}, \bibinfo {author} {\bibfnamefont
  {X.}~\bibnamefont {Hu}}, \bibinfo {author} {\bibfnamefont {D.~E.}\
  \bibnamefont {Savage}}, \bibinfo {author} {\bibfnamefont {M.~G.}\
  \bibnamefont {Lagally}}, \bibinfo {author} {\bibfnamefont {M.~A.}\
  \bibnamefont {Eriksson}}, \bibinfo {author} {\bibfnamefont {M.}~\bibnamefont
  {Friesen}}, \ and\ \bibinfo {author} {\bibfnamefont {S.~N.}\ \bibnamefont
  {Coppersmith}},\ }\href
  {http://link.aps.org/doi/10.1103/PhysRevLett.108.140503} {\bibfield
  {journal} {\bibinfo  {journal} {Phys. Rev. Lett.}\ }\textbf {\bibinfo
  {volume} {108}},\ \bibinfo {pages} {140503} (\bibinfo {year}
  {2012})}\BibitemShut {NoStop}%
\bibitem [{\citenamefont {Koh}\ \emph {et~al.}(2012)\citenamefont {Koh},
  \citenamefont {Gamble}, \citenamefont {Friesen}, \citenamefont {Eriksson},\
  and\ \citenamefont {Coppersmith}}]{Koh2012}%
  \BibitemOpen
  \bibfield  {author} {\bibinfo {author} {\bibfnamefont {T.~S.}\ \bibnamefont
  {Koh}}, \bibinfo {author} {\bibfnamefont {J.~K.}\ \bibnamefont {Gamble}},
  \bibinfo {author} {\bibfnamefont {M.}~\bibnamefont {Friesen}}, \bibinfo
  {author} {\bibfnamefont {M.~A.}\ \bibnamefont {Eriksson}}, \ and\ \bibinfo
  {author} {\bibfnamefont {S.~N.}\ \bibnamefont {Coppersmith}},\ }\href
  {\doibase 10.1103/PhysRevLett.109.250503} {\bibfield  {journal} {\bibinfo
  {journal} {Phys. Rev. Lett.}\ }\textbf {\bibinfo {volume} {109}},\ \bibinfo
  {pages} {250503} (\bibinfo {year} {2012})}\BibitemShut {NoStop}%
\bibitem [{\citenamefont {Wong}(2016)}]{Wong2016}%
  \BibitemOpen
  \bibfield  {author} {\bibinfo {author} {\bibfnamefont {C.~H.}\ \bibnamefont
  {Wong}},\ }\href {\doibase 10.1103/PhysRevB.93.035409} {\bibfield  {journal}
  {\bibinfo  {journal} {Phys. Rev. B}\ }\textbf {\bibinfo {volume} {93}},\
  \bibinfo {pages} {035409} (\bibinfo {year} {2016})}\BibitemShut {NoStop}%
\bibitem [{\citenamefont {Kim}\ \emph {et~al.}(2015{\natexlab{b}})\citenamefont
  {Kim}, \citenamefont {Ward}, \citenamefont {Simmons}, \citenamefont {Savage},
  \citenamefont {Lagally}, \citenamefont {Friesen}, \citenamefont
  {Coppersmith},\ and\ \citenamefont {Eriksson}}]{KimWardSimmonsEtAl2015}%
  \BibitemOpen
  \bibfield  {author} {\bibinfo {author} {\bibfnamefont {D.}~\bibnamefont
  {Kim}}, \bibinfo {author} {\bibfnamefont {D.~R.}\ \bibnamefont {Ward}},
  \bibinfo {author} {\bibfnamefont {C.~B.}\ \bibnamefont {Simmons}}, \bibinfo
  {author} {\bibfnamefont {D.~E.}\ \bibnamefont {Savage}}, \bibinfo {author}
  {\bibfnamefont {M.~G.}\ \bibnamefont {Lagally}}, \bibinfo {author}
  {\bibfnamefont {M.}~\bibnamefont {Friesen}}, \bibinfo {author} {\bibfnamefont
  {S.~N.}\ \bibnamefont {Coppersmith}}, \ and\ \bibinfo {author} {\bibfnamefont
  {M.~A.}\ \bibnamefont {Eriksson}},\ }\href
  {http://dx.doi.org/10.1038/npjqi.2015.4} {\bibfield  {journal} {\bibinfo
  {journal} {npj Quantum Inform.}\ }\textbf {\bibinfo {volume} {1}},\ \bibinfo
  {pages} {15004} (\bibinfo {year} {2015}{\natexlab{b}})}\BibitemShut {NoStop}%
\bibitem [{\citenamefont {Mehl}(2015)}]{PhysRevB.91.035430}%
  \BibitemOpen
  \bibfield  {author} {\bibinfo {author} {\bibfnamefont {S.}~\bibnamefont
  {Mehl}},\ }\href {\doibase 10.1103/PhysRevB.91.035430} {\bibfield  {journal}
  {\bibinfo  {journal} {Phys. Rev. B}\ }\textbf {\bibinfo {volume} {91}},\
  \bibinfo {pages} {035430} (\bibinfo {year} {2015})}\BibitemShut {NoStop}%
\bibitem [{\citenamefont {{Mehl}}()}]{2015arXiv150703425M}%
  \BibitemOpen
  \bibfield  {author} {\bibinfo {author} {\bibfnamefont {S.}~\bibnamefont
  {{Mehl}}},\ }\href@noop {} {\bibfield  {journal} {\bibinfo  {journal} {ArXiv
  e-prints}\ }}\Eprint {http://arxiv.org/abs/1507.03425} {1507.03425}
  \BibitemShut {NoStop}%
\bibitem [{\citenamefont {Ferraro}\ \emph {et~al.}(2015)\citenamefont
  {Ferraro}, \citenamefont {De~Michielis}, \citenamefont {Fanciulli},\ and\
  \citenamefont {Prati}}]{Ferraro:2015}%
  \BibitemOpen
  \bibfield  {author} {\bibinfo {author} {\bibfnamefont {E.}~\bibnamefont
  {Ferraro}}, \bibinfo {author} {\bibfnamefont {M.}~\bibnamefont
  {De~Michielis}}, \bibinfo {author} {\bibfnamefont {M.}~\bibnamefont
  {Fanciulli}}, \ and\ \bibinfo {author} {\bibfnamefont {E.}~\bibnamefont
  {Prati}},\ }\href@noop {} {\bibfield  {journal} {\bibinfo  {journal} {Quantum
  Inform. Process.}\ }\textbf {\bibinfo {volume} {14}},\ \bibinfo {pages} {47}
  (\bibinfo {year} {2015})}\BibitemShut {NoStop}%
\bibitem [{\citenamefont {{De Michielis}}\ \emph {et~al.}(2015)\citenamefont
  {{De Michielis}}, \citenamefont {Ferraro}, \citenamefont {Fanciulli},\ and\
  \citenamefont {Prati}}]{Michielis:2015}%
  \BibitemOpen
  \bibfield  {author} {\bibinfo {author} {\bibfnamefont {M.}~\bibnamefont {{De
  Michielis}}}, \bibinfo {author} {\bibfnamefont {E.}~\bibnamefont {Ferraro}},
  \bibinfo {author} {\bibfnamefont {M.}~\bibnamefont {Fanciulli}}, \ and\
  \bibinfo {author} {\bibfnamefont {E.}~\bibnamefont {Prati}},\ }\href@noop {}
  {\bibfield  {journal} {\bibinfo  {journal} {J. Phys. A}\ }\textbf {\bibinfo
  {volume} {48}},\ \bibinfo {pages} {065304} (\bibinfo {year}
  {2015})}\BibitemShut {NoStop}%
\bibitem [{\citenamefont {{Frees}}\ \emph {et~al.}(2019)\citenamefont
  {{Frees}}, \citenamefont {{Mehl}}, \citenamefont {Gamble}, \citenamefont
  {{Friesen}},\ and\ \citenamefont {{Coppersmith}}}]{2018arXiv181203177F}%
  \BibitemOpen
  \bibfield  {author} {\bibinfo {author} {\bibfnamefont {A.}~\bibnamefont
  {{Frees}}}, \bibinfo {author} {\bibfnamefont {S.}~\bibnamefont {{Mehl}}},
  \bibinfo {author} {\bibfnamefont {J.~K.}\ \bibnamefont {Gamble}}, \bibinfo
  {author} {\bibfnamefont {M.}~\bibnamefont {{Friesen}}}, \ and\ \bibinfo
  {author} {\bibfnamefont {S.~N.}\ \bibnamefont {{Coppersmith}}},\ }\href@noop
  {} {\bibfield  {journal} {\bibinfo  {journal} {npj Quantum Information}\
  }\textbf {\bibinfo {volume} {5}},\ \bibinfo {pages} {73} (\bibinfo {year}
  {2019})}\BibitemShut {NoStop}%
\bibitem [{\citenamefont {Bertrand}\ \emph {et~al.}(2015)\citenamefont
  {Bertrand}, \citenamefont {Flentje}, \citenamefont {Takada}, \citenamefont
  {Yamamoto}, \citenamefont {Tarucha}, \citenamefont {Ludwig}, \citenamefont
  {Wieck}, \citenamefont {B\"auerle},\ and\ \citenamefont
  {Meunier}}]{PhysRevLett.115.096801}%
  \BibitemOpen
  \bibfield  {author} {\bibinfo {author} {\bibfnamefont {B.}~\bibnamefont
  {Bertrand}}, \bibinfo {author} {\bibfnamefont {H.}~\bibnamefont {Flentje}},
  \bibinfo {author} {\bibfnamefont {S.}~\bibnamefont {Takada}}, \bibinfo
  {author} {\bibfnamefont {M.}~\bibnamefont {Yamamoto}}, \bibinfo {author}
  {\bibfnamefont {S.}~\bibnamefont {Tarucha}}, \bibinfo {author} {\bibfnamefont
  {A.}~\bibnamefont {Ludwig}}, \bibinfo {author} {\bibfnamefont {A.~D.}\
  \bibnamefont {Wieck}}, \bibinfo {author} {\bibfnamefont {C.}~\bibnamefont
  {B\"auerle}}, \ and\ \bibinfo {author} {\bibfnamefont {T.}~\bibnamefont
  {Meunier}},\ }\href {\doibase 10.1103/PhysRevLett.115.096801} {\bibfield
  {journal} {\bibinfo  {journal} {Phys. Rev. Lett.}\ }\textbf {\bibinfo
  {volume} {115}},\ \bibinfo {pages} {096801} (\bibinfo {year}
  {2015})}\BibitemShut {NoStop}%
\bibitem [{\citenamefont {Yang}\ \emph {et~al.}(2017)\citenamefont {Yang},
  \citenamefont {Coppersmith},\ and\ \citenamefont
  {Friesen}}]{PhysRevA.95.062321}%
  \BibitemOpen
  \bibfield  {author} {\bibinfo {author} {\bibfnamefont {Y.-C.}\ \bibnamefont
  {Yang}}, \bibinfo {author} {\bibfnamefont {S.~N.}\ \bibnamefont
  {Coppersmith}}, \ and\ \bibinfo {author} {\bibfnamefont {M.}~\bibnamefont
  {Friesen}},\ }\href {\doibase 10.1103/PhysRevA.95.062321} {\bibfield
  {journal} {\bibinfo  {journal} {Phys. Rev. A}\ }\textbf {\bibinfo {volume}
  {95}},\ \bibinfo {pages} {062321} (\bibinfo {year} {2017})}\BibitemShut
  {NoStop}%
\bibitem [{\citenamefont {Ward}\ \emph {et~al.}(2016)\citenamefont {Ward},
  \citenamefont {Kim}, \citenamefont {Savage}, \citenamefont {Lagally},
  \citenamefont {Foote}, \citenamefont {Friesen}, \citenamefont {Coppersmith},\
  and\ \citenamefont {Eriksson}}]{npjqi201632}%
  \BibitemOpen
  \bibfield  {author} {\bibinfo {author} {\bibfnamefont {D.~R.}\ \bibnamefont
  {Ward}}, \bibinfo {author} {\bibfnamefont {D.}~\bibnamefont {Kim}}, \bibinfo
  {author} {\bibfnamefont {D.~E.}\ \bibnamefont {Savage}}, \bibinfo {author}
  {\bibfnamefont {M.~G.}\ \bibnamefont {Lagally}}, \bibinfo {author}
  {\bibfnamefont {R.~H.}\ \bibnamefont {Foote}}, \bibinfo {author}
  {\bibfnamefont {M.}~\bibnamefont {Friesen}}, \bibinfo {author} {\bibfnamefont
  {S.~N.}\ \bibnamefont {Coppersmith}}, \ and\ \bibinfo {author} {\bibfnamefont
  {M.~A.}\ \bibnamefont {Eriksson}},\ }\href {\doibase 10.1038/npjqi.2016.32}
  {\bibfield  {journal} {\bibinfo  {journal} {npj Quantum Inform.}\ }\textbf
  {\bibinfo {volume} {2}},\ \bibinfo {pages} {16032} (\bibinfo {year}
  {2016})}\BibitemShut {NoStop}%
\bibitem [{\citenamefont {Yang}\ \emph {et~al.}(2019)\citenamefont {Yang},
  \citenamefont {Coppersmith},\ and\ \citenamefont {Friesen}}]{DQHQNoise}%
  \BibitemOpen
  \bibfield  {author} {\bibinfo {author} {\bibfnamefont {Y.-C.}\ \bibnamefont
  {Yang}}, \bibinfo {author} {\bibfnamefont {S.~N.}\ \bibnamefont
  {Coppersmith}}, \ and\ \bibinfo {author} {\bibfnamefont {M.}~\bibnamefont
  {Friesen}},\ }\href {\doibase 10.1103/PhysRevA.100.022337} {\bibfield
  {journal} {\bibinfo  {journal} {Phys. Rev. A}\ }\textbf {\bibinfo {volume}
  {100}},\ \bibinfo {pages} {022337} (\bibinfo {year} {2019})}\BibitemShut
  {NoStop}%
\bibitem [{\citenamefont {Harvey-Collard}\ \emph {et~al.}(2017)\citenamefont
  {Harvey-Collard}, \citenamefont {Jacobson}, \citenamefont {Rudolph},
  \citenamefont {Dominguez}, \citenamefont {Ten~Eyck}, \citenamefont {Wendt},
  \citenamefont {Pluym}, \citenamefont {Gamble}, \citenamefont {Lilly},
  \citenamefont {Pioro-Ladri{\`e}re},\ and\ \citenamefont
  {Carroll}}]{Harvey-Collard:2017}%
  \BibitemOpen
  \bibfield  {author} {\bibinfo {author} {\bibfnamefont {P.}~\bibnamefont
  {Harvey-Collard}}, \bibinfo {author} {\bibfnamefont {N.~T.}\ \bibnamefont
  {Jacobson}}, \bibinfo {author} {\bibfnamefont {M.}~\bibnamefont {Rudolph}},
  \bibinfo {author} {\bibfnamefont {J.}~\bibnamefont {Dominguez}}, \bibinfo
  {author} {\bibfnamefont {G.~A.}\ \bibnamefont {Ten~Eyck}}, \bibinfo {author}
  {\bibfnamefont {J.~R.}\ \bibnamefont {Wendt}}, \bibinfo {author}
  {\bibfnamefont {T.}~\bibnamefont {Pluym}}, \bibinfo {author} {\bibfnamefont
  {J.~K.}\ \bibnamefont {Gamble}}, \bibinfo {author} {\bibfnamefont {M.~P.}\
  \bibnamefont {Lilly}}, \bibinfo {author} {\bibfnamefont {M.}~\bibnamefont
  {Pioro-Ladri{\`e}re}}, \ and\ \bibinfo {author} {\bibfnamefont {M.~S.}\
  \bibnamefont {Carroll}},\ }\href {\doibase 10.1038/s41467-017-01113-2}
  {\bibfield  {journal} {\bibinfo  {journal} {Nature Commun.}\ }\textbf
  {\bibinfo {volume} {8}},\ \bibinfo {pages} {1029} (\bibinfo {year}
  {2017})}\BibitemShut {NoStop}%
\bibitem [{\citenamefont {DiVincenzo}\ \emph {et~al.}(2000)\citenamefont
  {DiVincenzo}, \citenamefont {Bacon}, \citenamefont {Kempe}, \citenamefont
  {Burkard},\ and\ \citenamefont {Whaley}}]{DiVincenzo2000}%
  \BibitemOpen
  \bibfield  {author} {\bibinfo {author} {\bibfnamefont {D.~P.}\ \bibnamefont
  {DiVincenzo}}, \bibinfo {author} {\bibfnamefont {D.}~\bibnamefont {Bacon}},
  \bibinfo {author} {\bibfnamefont {J.}~\bibnamefont {Kempe}}, \bibinfo
  {author} {\bibfnamefont {G.}~\bibnamefont {Burkard}}, \ and\ \bibinfo
  {author} {\bibfnamefont {K.~B.}\ \bibnamefont {Whaley}},\ }\href {\doibase
  10.1038/35042541} {\bibfield  {journal} {\bibinfo  {journal} {Nature}\
  }\textbf {\bibinfo {volume} {408}},\ \bibinfo {pages} {339} (\bibinfo {year}
  {2000})}\BibitemShut {NoStop}%
\bibitem [{\citenamefont {Setser}\ and\ \citenamefont
  {Kestner}(2019)}]{PhysRevB.99.195403}%
  \BibitemOpen
  \bibfield  {author} {\bibinfo {author} {\bibfnamefont {A.~A.}\ \bibnamefont
  {Setser}}\ and\ \bibinfo {author} {\bibfnamefont {J.~P.}\ \bibnamefont
  {Kestner}},\ }\href {\doibase 10.1103/PhysRevB.99.195403} {\bibfield
  {journal} {\bibinfo  {journal} {Phys. Rev. B}\ }\textbf {\bibinfo {volume}
  {99}},\ \bibinfo {pages} {195403} (\bibinfo {year} {2019})}\BibitemShut
  {NoStop}%
\bibitem [{\citenamefont {Makhlin}(2002)}]{Makhlin2002}%
  \BibitemOpen
  \bibfield  {author} {\bibinfo {author} {\bibfnamefont {Y.}~\bibnamefont
  {Makhlin}},\ }\href {\doibase 10.1023/A:1022144002391} {\bibfield  {journal}
  {\bibinfo  {journal} {Quantum Inform. Process.}\ }\textbf {\bibinfo {volume}
  {1}},\ \bibinfo {pages} {243} (\bibinfo {year} {2002})}\BibitemShut {NoStop}%
\bibitem [{\citenamefont {Petersson}\ \emph {et~al.}(2010)\citenamefont
  {Petersson}, \citenamefont {Petta}, \citenamefont {Lu},\ and\ \citenamefont
  {Gossard}}]{PhysRevLett.105.246804}%
  \BibitemOpen
  \bibfield  {author} {\bibinfo {author} {\bibfnamefont {K.~D.}\ \bibnamefont
  {Petersson}}, \bibinfo {author} {\bibfnamefont {J.~R.}\ \bibnamefont
  {Petta}}, \bibinfo {author} {\bibfnamefont {H.}~\bibnamefont {Lu}}, \ and\
  \bibinfo {author} {\bibfnamefont {A.~C.}\ \bibnamefont {Gossard}},\ }\href
  {\doibase 10.1103/PhysRevLett.105.246804} {\bibfield  {journal} {\bibinfo
  {journal} {Phys. Rev. Lett.}\ }\textbf {\bibinfo {volume} {105}},\ \bibinfo
  {pages} {246804} (\bibinfo {year} {2010})}\BibitemShut {NoStop}%
\bibitem [{\citenamefont {Wu}\ \emph {et~al.}(2014)\citenamefont {Wu},
  \citenamefont {Ward}, \citenamefont {Prance}, \citenamefont {Kim},
  \citenamefont {Gamble}, \citenamefont {Mohr}, \citenamefont {Shi},
  \citenamefont {Savage}, \citenamefont {Lagally}, \citenamefont {Friesen},
  \citenamefont {Coppersmith},\ and\ \citenamefont {Eriksson}}]{Wu19082014}%
  \BibitemOpen
  \bibfield  {author} {\bibinfo {author} {\bibfnamefont {X.}~\bibnamefont
  {Wu}}, \bibinfo {author} {\bibfnamefont {D.~R.}\ \bibnamefont {Ward}},
  \bibinfo {author} {\bibfnamefont {J.~R.}\ \bibnamefont {Prance}}, \bibinfo
  {author} {\bibfnamefont {D.}~\bibnamefont {Kim}}, \bibinfo {author}
  {\bibfnamefont {J.~K.}\ \bibnamefont {Gamble}}, \bibinfo {author}
  {\bibfnamefont {R.~T.}\ \bibnamefont {Mohr}}, \bibinfo {author}
  {\bibfnamefont {Z.}~\bibnamefont {Shi}}, \bibinfo {author} {\bibfnamefont
  {D.~E.}\ \bibnamefont {Savage}}, \bibinfo {author} {\bibfnamefont {M.~G.}\
  \bibnamefont {Lagally}}, \bibinfo {author} {\bibfnamefont {M.}~\bibnamefont
  {Friesen}}, \bibinfo {author} {\bibfnamefont {S.~N.}\ \bibnamefont
  {Coppersmith}}, \ and\ \bibinfo {author} {\bibfnamefont {M.~A.}\ \bibnamefont
  {Eriksson}},\ }\href {\doibase 10.1073/pnas.1412230111} {\bibfield  {journal}
  {\bibinfo  {journal} {Proc. Nat. Acad. Sci.}\ }\textbf {\bibinfo {volume}
  {111}},\ \bibinfo {pages} {11938} (\bibinfo {year} {2014})}\BibitemShut
  {NoStop}%
\bibitem [{\citenamefont {Shi}\ \emph {et~al.}(2013)\citenamefont {Shi},
  \citenamefont {Simmons}, \citenamefont {Ward}, \citenamefont {Prance},
  \citenamefont {Mohr}, \citenamefont {Koh}, \citenamefont {Gamble},
  \citenamefont {Wu}, \citenamefont {Savage}, \citenamefont {Lagally},
  \citenamefont {Friesen}, \citenamefont {Coppersmith},\ and\ \citenamefont
  {Eriksson}}]{Shi2013}%
  \BibitemOpen
  \bibfield  {author} {\bibinfo {author} {\bibfnamefont {Z.}~\bibnamefont
  {Shi}}, \bibinfo {author} {\bibfnamefont {C.~B.}\ \bibnamefont {Simmons}},
  \bibinfo {author} {\bibfnamefont {D.~R.}\ \bibnamefont {Ward}}, \bibinfo
  {author} {\bibfnamefont {J.~R.}\ \bibnamefont {Prance}}, \bibinfo {author}
  {\bibfnamefont {R.~T.}\ \bibnamefont {Mohr}}, \bibinfo {author}
  {\bibfnamefont {T.~S.}\ \bibnamefont {Koh}}, \bibinfo {author} {\bibfnamefont
  {J.~K.}\ \bibnamefont {Gamble}}, \bibinfo {author} {\bibfnamefont
  {X.}~\bibnamefont {Wu}}, \bibinfo {author} {\bibfnamefont {D.~E.}\
  \bibnamefont {Savage}}, \bibinfo {author} {\bibfnamefont {M.~G.}\
  \bibnamefont {Lagally}}, \bibinfo {author} {\bibfnamefont {M.}~\bibnamefont
  {Friesen}}, \bibinfo {author} {\bibfnamefont {S.~N.}\ \bibnamefont
  {Coppersmith}}, \ and\ \bibinfo {author} {\bibfnamefont {M.~A.}\ \bibnamefont
  {Eriksson}},\ }\href {\doibase 10.1103/PhysRevB.88.075416} {\bibfield
  {journal} {\bibinfo  {journal} {Phys. Rev. B}\ }\textbf {\bibinfo {volume}
  {88}},\ \bibinfo {pages} {075416} (\bibinfo {year} {2013})}\BibitemShut
  {NoStop}%
\bibitem [{\citenamefont {{Vandersypen}}\ \emph {et~al.}(2001)\citenamefont
  {{Vandersypen}}, \citenamefont {{Steffen}}, \citenamefont {{Breyta}},
  \citenamefont {{Yannoni}}, \citenamefont {{Sherwood}},\ and\ \citenamefont
  {{Chuang}}}]{2001Natur.414..883V}%
  \BibitemOpen
  \bibfield  {author} {\bibinfo {author} {\bibfnamefont {L.~M.~K.}\
  \bibnamefont {{Vandersypen}}}, \bibinfo {author} {\bibfnamefont
  {M.}~\bibnamefont {{Steffen}}}, \bibinfo {author} {\bibfnamefont
  {G.}~\bibnamefont {{Breyta}}}, \bibinfo {author} {\bibfnamefont {C.~S.}\
  \bibnamefont {{Yannoni}}}, \bibinfo {author} {\bibfnamefont {M.~H.}\
  \bibnamefont {{Sherwood}}}, \ and\ \bibinfo {author} {\bibfnamefont {I.~L.}\
  \bibnamefont {{Chuang}}},\ }\href {\doibase 10.1038/414883a} {\bibfield
  {journal} {\bibinfo  {journal} {Nature}\ }\textbf {\bibinfo {volume} {414}},\
  \bibinfo {pages} {883} (\bibinfo {year} {2001})}\BibitemShut {NoStop}%
\bibitem [{\citenamefont {Motzoi}\ \emph {et~al.}(2009)\citenamefont {Motzoi},
  \citenamefont {Gambetta}, \citenamefont {Rebentrost},\ and\ \citenamefont
  {Wilhelm}}]{Motzoi2009}%
  \BibitemOpen
  \bibfield  {author} {\bibinfo {author} {\bibfnamefont {F.}~\bibnamefont
  {Motzoi}}, \bibinfo {author} {\bibfnamefont {J.~M.}\ \bibnamefont
  {Gambetta}}, \bibinfo {author} {\bibfnamefont {P.}~\bibnamefont
  {Rebentrost}}, \ and\ \bibinfo {author} {\bibfnamefont {F.~K.}\ \bibnamefont
  {Wilhelm}},\ }\href {\doibase 10.1103/PhysRevLett.103.110501} {\bibfield
  {journal} {\bibinfo  {journal} {Phys. Rev. Lett.}\ }\textbf {\bibinfo
  {volume} {103}},\ \bibinfo {pages} {110501} (\bibinfo {year}
  {2009})}\BibitemShut {NoStop}%
\bibitem [{\citenamefont {Gambetta}\ \emph {et~al.}(2011)\citenamefont
  {Gambetta}, \citenamefont {Motzoi}, \citenamefont {Merkel},\ and\
  \citenamefont {Wilhelm}}]{Gambetta2011}%
  \BibitemOpen
  \bibfield  {author} {\bibinfo {author} {\bibfnamefont {J.~M.}\ \bibnamefont
  {Gambetta}}, \bibinfo {author} {\bibfnamefont {F.}~\bibnamefont {Motzoi}},
  \bibinfo {author} {\bibfnamefont {S.~T.}\ \bibnamefont {Merkel}}, \ and\
  \bibinfo {author} {\bibfnamefont {F.~K.}\ \bibnamefont {Wilhelm}},\ }\href
  {\doibase 10.1103/PhysRevA.83.012308} {\bibfield  {journal} {\bibinfo
  {journal} {Phys. Rev. A}\ }\textbf {\bibinfo {volume} {83}},\ \bibinfo
  {pages} {012308} (\bibinfo {year} {2011})}\BibitemShut {NoStop}%
\bibitem [{\citenamefont {Torrontegui}\ \emph {et~al.}(2013)\citenamefont
  {Torrontegui}, \citenamefont {Ib{\'a}{\~n}ez}, \citenamefont
  {Mart{\'\i}nez-Garaot}, \citenamefont {Modugno}, \citenamefont {del Campo},
  \citenamefont {Gu{\'e}ry-Odelin}, \citenamefont {Ruschhaupt}, \citenamefont
  {Chen},\ and\ \citenamefont {Muga}}]{TORRONTEGUI2013117}%
  \BibitemOpen
  \bibfield  {author} {\bibinfo {author} {\bibfnamefont {E.}~\bibnamefont
  {Torrontegui}}, \bibinfo {author} {\bibfnamefont {S.}~\bibnamefont
  {Ib{\'a}{\~n}ez}}, \bibinfo {author} {\bibfnamefont {S.}~\bibnamefont
  {Mart{\'\i}nez-Garaot}}, \bibinfo {author} {\bibfnamefont {M.}~\bibnamefont
  {Modugno}}, \bibinfo {author} {\bibfnamefont {A.}~\bibnamefont {del Campo}},
  \bibinfo {author} {\bibfnamefont {D.}~\bibnamefont {Gu{\'e}ry-Odelin}},
  \bibinfo {author} {\bibfnamefont {A.}~\bibnamefont {Ruschhaupt}}, \bibinfo
  {author} {\bibfnamefont {X.}~\bibnamefont {Chen}}, \ and\ \bibinfo {author}
  {\bibfnamefont {J.~G.}\ \bibnamefont {Muga}},\ }in\ \href {\doibase
  https://doi.org/10.1016/B978-0-12-408090-4.00002-5} {\emph {\bibinfo
  {booktitle} {Advances in Atomic, Molecular, and Optical Physics}}},\
  Vol.~\bibinfo {volume} {62},\ \bibinfo {editor} {edited by\ \bibinfo {editor}
  {\bibfnamefont {E.}~\bibnamefont {Arimondo}}, \bibinfo {editor}
  {\bibfnamefont {P.~R.}\ \bibnamefont {Berman}}, \ and\ \bibinfo {editor}
  {\bibfnamefont {C.~C.}\ \bibnamefont {Lin}}}\ (\bibinfo  {publisher}
  {Academic Press},\ \bibinfo {year} {2013})\ pp.\ \bibinfo {pages} {117 --
  169}\BibitemShut {NoStop}%
\bibitem [{\citenamefont {Davies}(1998)}]{DaviesBook}%
  \BibitemOpen
  \bibfield  {author} {\bibinfo {author} {\bibfnamefont {J.~H.}\ \bibnamefont
  {Davies}},\ }\href@noop {} {\emph {\bibinfo {title} {The Physics of
  Low-Dimensional Semiconductors}}}\ (\bibinfo  {publisher} {Cambridge
  University Press},\ \bibinfo {address} {Cambridge, UK},\ \bibinfo {year}
  {1998})\BibitemShut {NoStop}%
\bibitem [{\citenamefont {Shi}\ \emph {et~al.}(2014)\citenamefont {Shi},
  \citenamefont {Simmons}, \citenamefont {Ward}, \citenamefont {Prance},
  \citenamefont {Wu}, \citenamefont {Koh}, \citenamefont {Gamble},
  \citenamefont {Savage}, \citenamefont {Lagally}, \citenamefont {Friesen},
  \citenamefont {Coppersmith},\ and\ \citenamefont
  {Eriksson}}]{ShiSimmonsWardEtAl2014}%
  \BibitemOpen
  \bibfield  {author} {\bibinfo {author} {\bibfnamefont {Z.}~\bibnamefont
  {Shi}}, \bibinfo {author} {\bibfnamefont {C.~B.}\ \bibnamefont {Simmons}},
  \bibinfo {author} {\bibfnamefont {D.~R.}\ \bibnamefont {Ward}}, \bibinfo
  {author} {\bibfnamefont {J.~R.}\ \bibnamefont {Prance}}, \bibinfo {author}
  {\bibfnamefont {X.}~\bibnamefont {Wu}}, \bibinfo {author} {\bibfnamefont
  {T.~S.}\ \bibnamefont {Koh}}, \bibinfo {author} {\bibfnamefont {J.~K.}\
  \bibnamefont {Gamble}}, \bibinfo {author} {\bibfnamefont {D.~E.}\
  \bibnamefont {Savage}}, \bibinfo {author} {\bibfnamefont {M.~G.}\
  \bibnamefont {Lagally}}, \bibinfo {author} {\bibfnamefont {M.}~\bibnamefont
  {Friesen}}, \bibinfo {author} {\bibfnamefont {S.~N.}\ \bibnamefont
  {Coppersmith}}, \ and\ \bibinfo {author} {\bibfnamefont {M.~A.}\ \bibnamefont
  {Eriksson}},\ }\href {http://dx.doi.org/10.1038/ncomms4020} {\bibfield
  {journal} {\bibinfo  {journal} {Nature Commun.}\ }\textbf {\bibinfo {volume}
  {5}},\ \bibinfo {pages} {3020} (\bibinfo {year} {2014})}\BibitemShut
  {NoStop}%
\bibitem [{\citenamefont {Yang}\ \emph {et~al.}(2011)\citenamefont {Yang},
  \citenamefont {Wang},\ and\ \citenamefont {Das~Sarma}}]{PhysRevB.83.161301}%
  \BibitemOpen
  \bibfield  {author} {\bibinfo {author} {\bibfnamefont {S.}~\bibnamefont
  {Yang}}, \bibinfo {author} {\bibfnamefont {X.}~\bibnamefont {Wang}}, \ and\
  \bibinfo {author} {\bibfnamefont {S.}~\bibnamefont {Das~Sarma}},\ }\href
  {\doibase 10.1103/PhysRevB.83.161301} {\bibfield  {journal} {\bibinfo
  {journal} {Phys. Rev. B}\ }\textbf {\bibinfo {volume} {83}},\ \bibinfo
  {pages} {161301} (\bibinfo {year} {2011})}\BibitemShut {NoStop}%
\bibitem [{\citenamefont {Das~Sarma}\ \emph {et~al.}(2011)\citenamefont
  {Das~Sarma}, \citenamefont {Wang},\ and\ \citenamefont
  {Yang}}]{PhysRevB.83.235314}%
  \BibitemOpen
  \bibfield  {author} {\bibinfo {author} {\bibfnamefont {S.}~\bibnamefont
  {Das~Sarma}}, \bibinfo {author} {\bibfnamefont {X.}~\bibnamefont {Wang}}, \
  and\ \bibinfo {author} {\bibfnamefont {S.}~\bibnamefont {Yang}},\ }\href
  {\doibase 10.1103/PhysRevB.83.235314} {\bibfield  {journal} {\bibinfo
  {journal} {Phys. Rev. B}\ }\textbf {\bibinfo {volume} {83}},\ \bibinfo
  {pages} {235314} (\bibinfo {year} {2011})}\BibitemShut {NoStop}%
\bibitem [{\citenamefont {Winkler}(2003)}]{WinklerBook}%
  \BibitemOpen
  \bibfield  {author} {\bibinfo {author} {\bibfnamefont {R.}~\bibnamefont
  {Winkler}},\ }\href@noop {} {\emph {\bibinfo {title} {Spin-Orbit Coupling
  Effects in Two-Dimensional Electron and Hole System}}},\ \bibinfo {series}
  {Springer Tracts in Modern Physics}, Vol.\ \bibinfo {volume} {191}\ (\bibinfo
   {publisher} {Springer, Berlin},\ \bibinfo {year} {2003})\ pp.\ \bibinfo
  {pages} {201--205}\BibitemShut {NoStop}%
\bibitem [{\citenamefont {Dial}\ \emph {et~al.}(2013)\citenamefont {Dial},
  \citenamefont {Shulman}, \citenamefont {Harvey}, \citenamefont {Bluhm},
  \citenamefont {Umansky},\ and\ \citenamefont
  {Yacoby}}]{PhysRevLett.110.146804}%
  \BibitemOpen
  \bibfield  {author} {\bibinfo {author} {\bibfnamefont {O.~E.}\ \bibnamefont
  {Dial}}, \bibinfo {author} {\bibfnamefont {M.~D.}\ \bibnamefont {Shulman}},
  \bibinfo {author} {\bibfnamefont {S.~P.}\ \bibnamefont {Harvey}}, \bibinfo
  {author} {\bibfnamefont {H.}~\bibnamefont {Bluhm}}, \bibinfo {author}
  {\bibfnamefont {V.}~\bibnamefont {Umansky}}, \ and\ \bibinfo {author}
  {\bibfnamefont {A.}~\bibnamefont {Yacoby}},\ }\href {\doibase
  10.1103/PhysRevLett.110.146804} {\bibfield  {journal} {\bibinfo  {journal}
  {Phys. Rev. Lett.}\ }\textbf {\bibinfo {volume} {110}},\ \bibinfo {pages}
  {146804} (\bibinfo {year} {2013})}\BibitemShut {NoStop}%
\bibitem [{\citenamefont {Kawakami}\ \emph {et~al.}(2016)\citenamefont
  {Kawakami}, \citenamefont {Jullien}, \citenamefont {Scarlino}, \citenamefont
  {Ward}, \citenamefont {Savage}, \citenamefont {Lagally}, \citenamefont
  {Dobrovitski}, \citenamefont {Friesen}, \citenamefont {Coppersmith},
  \citenamefont {Eriksson},\ and\ \citenamefont
  {Vandersypen}}]{Kawakami18102016}%
  \BibitemOpen
  \bibfield  {author} {\bibinfo {author} {\bibfnamefont {E.}~\bibnamefont
  {Kawakami}}, \bibinfo {author} {\bibfnamefont {T.}~\bibnamefont {Jullien}},
  \bibinfo {author} {\bibfnamefont {P.}~\bibnamefont {Scarlino}}, \bibinfo
  {author} {\bibfnamefont {D.~R.}\ \bibnamefont {Ward}}, \bibinfo {author}
  {\bibfnamefont {D.~E.}\ \bibnamefont {Savage}}, \bibinfo {author}
  {\bibfnamefont {M.~G.}\ \bibnamefont {Lagally}}, \bibinfo {author}
  {\bibfnamefont {V.~V.}\ \bibnamefont {Dobrovitski}}, \bibinfo {author}
  {\bibfnamefont {M.}~\bibnamefont {Friesen}}, \bibinfo {author} {\bibfnamefont
  {S.~N.}\ \bibnamefont {Coppersmith}}, \bibinfo {author} {\bibfnamefont
  {M.~A.}\ \bibnamefont {Eriksson}}, \ and\ \bibinfo {author} {\bibfnamefont
  {L.~M.~K.}\ \bibnamefont {Vandersypen}},\ }\href {\doibase
  10.1073/pnas.1603251113} {\bibfield  {journal} {\bibinfo  {journal} {Proc.
  Nat. Acad. Sci.}\ }\textbf {\bibinfo {volume} {113}},\ \bibinfo {pages}
  {11738} (\bibinfo {year} {2016})}\BibitemShut {NoStop}%
\bibitem [{\citenamefont {{Yang}}\ \emph {et~al.}(2019)\citenamefont {{Yang}},
  \citenamefont {{Coppersmith}},\ and\ \citenamefont {{Friesen}}}]{TLSNoise}%
  \BibitemOpen
  \bibfield  {author} {\bibinfo {author} {\bibfnamefont {Y.-C.}\ \bibnamefont
  {{Yang}}}, \bibinfo {author} {\bibfnamefont {S.~N.}\ \bibnamefont
  {{Coppersmith}}}, \ and\ \bibinfo {author} {\bibfnamefont {M.}~\bibnamefont
  {{Friesen}}},\ }\href {\doibase 10.1038/s41534-019-0127-1} {\bibfield
  {journal} {\bibinfo  {journal} {npj Quantum Inform.}\ }\textbf {\bibinfo
  {volume} {5}},\ \bibinfo {eid} {12} (\bibinfo {year} {2019})}\BibitemShut
  {NoStop}%
\bibitem [{\citenamefont {Salzer}\ \emph {et~al.}(1952)\citenamefont {Salzer},
  \citenamefont {Zucker},\ and\ \citenamefont {Capuano}}]{HermitePolynomials}%
  \BibitemOpen
  \bibfield  {author} {\bibinfo {author} {\bibfnamefont {H.~E.}\ \bibnamefont
  {Salzer}}, \bibinfo {author} {\bibfnamefont {R.}~\bibnamefont {Zucker}}, \
  and\ \bibinfo {author} {\bibfnamefont {R.}~\bibnamefont {Capuano}},\
  }\href@noop {} {\bibfield  {journal} {\bibinfo  {journal} {J. Res. Nat. Bur.
  Stand.}\ }\textbf {\bibinfo {volume} {48}},\ \bibinfo {pages} {2294}
  (\bibinfo {year} {1952})}\BibitemShut {NoStop}%
\bibitem [{\citenamefont {Abramowitz}\ and\ \citenamefont
  {Stegun}(1972)}]{AbramowitzBook}%
  \BibitemOpen
  \bibinfo {editor} {\bibfnamefont {M.}~\bibnamefont {Abramowitz}}\ and\
  \bibinfo {editor} {\bibfnamefont {I.~A.}\ \bibnamefont {Stegun}},\ eds.,\
  \href@noop {} {\emph {\bibinfo {title} {Handbook of Mathematical
  Functions}}},\ \bibinfo {edition} {10th}\ ed.\ (\bibinfo  {publisher} {Dover,
  New York},\ \bibinfo {year} {1972})\BibitemShut {NoStop}%
\bibitem [{\citenamefont {Nielsen}\ and\ \citenamefont
  {Chuang}(2010)}]{ChuangBook}%
  \BibitemOpen
  \bibfield  {author} {\bibinfo {author} {\bibfnamefont {M.~A.}\ \bibnamefont
  {Nielsen}}\ and\ \bibinfo {author} {\bibfnamefont {I.~L.}\ \bibnamefont
  {Chuang}},\ }\href@noop {} {\emph {\bibinfo {title} {Quantum Computation and
  Quantum Information}}},\ \bibinfo {edition} {{10th Anniversary}}\ ed.\
  (\bibinfo  {publisher} {Cambridge University Press, Cambridge},\ \bibinfo
  {year} {2010})\BibitemShut {NoStop}%
\bibitem [{\citenamefont {Chow}\ \emph {et~al.}(2009)\citenamefont {Chow},
  \citenamefont {Gambetta}, \citenamefont {Tornberg}, \citenamefont {Koch},
  \citenamefont {Bishop}, \citenamefont {Houck}, \citenamefont {Johnson},
  \citenamefont {Frunzio}, \citenamefont {Girvin},\ and\ \citenamefont
  {Schoelkopf}}]{PhysRevLett.102.090502}%
  \BibitemOpen
  \bibfield  {author} {\bibinfo {author} {\bibfnamefont {J.~M.}\ \bibnamefont
  {Chow}}, \bibinfo {author} {\bibfnamefont {J.~M.}\ \bibnamefont {Gambetta}},
  \bibinfo {author} {\bibfnamefont {L.}~\bibnamefont {Tornberg}}, \bibinfo
  {author} {\bibfnamefont {J.}~\bibnamefont {Koch}}, \bibinfo {author}
  {\bibfnamefont {L.~S.}\ \bibnamefont {Bishop}}, \bibinfo {author}
  {\bibfnamefont {A.~A.}\ \bibnamefont {Houck}}, \bibinfo {author}
  {\bibfnamefont {B.~R.}\ \bibnamefont {Johnson}}, \bibinfo {author}
  {\bibfnamefont {L.}~\bibnamefont {Frunzio}}, \bibinfo {author} {\bibfnamefont
  {S.~M.}\ \bibnamefont {Girvin}}, \ and\ \bibinfo {author} {\bibfnamefont
  {R.~J.}\ \bibnamefont {Schoelkopf}},\ }\href {\doibase
  10.1103/PhysRevLett.102.090502} {\bibfield  {journal} {\bibinfo  {journal}
  {Phys. Rev. Lett.}\ }\textbf {\bibinfo {volume} {102}},\ \bibinfo {pages}
  {090502} (\bibinfo {year} {2009})}\BibitemShut {NoStop}%
\bibitem [{\citenamefont {Gilchrist}\ \emph {et~al.}(2005)\citenamefont
  {Gilchrist}, \citenamefont {Langford},\ and\ \citenamefont
  {Nielsen}}]{PhysRevA.71.062310}%
  \BibitemOpen
  \bibfield  {author} {\bibinfo {author} {\bibfnamefont {A.}~\bibnamefont
  {Gilchrist}}, \bibinfo {author} {\bibfnamefont {N.~K.}\ \bibnamefont
  {Langford}}, \ and\ \bibinfo {author} {\bibfnamefont {M.~A.}\ \bibnamefont
  {Nielsen}},\ }\href {\doibase 10.1103/PhysRevA.71.062310} {\bibfield
  {journal} {\bibinfo  {journal} {Phys. Rev. A}\ }\textbf {\bibinfo {volume}
  {71}},\ \bibinfo {pages} {062310} (\bibinfo {year} {2005})}\BibitemShut
  {NoStop}%
\bibitem [{\citenamefont {Wood}\ and\ \citenamefont
  {Gambetta}(2018)}]{PhysRevA.97.032306}%
  \BibitemOpen
  \bibfield  {author} {\bibinfo {author} {\bibfnamefont {C.~J.}\ \bibnamefont
  {Wood}}\ and\ \bibinfo {author} {\bibfnamefont {J.~M.}\ \bibnamefont
  {Gambetta}},\ }\href {\doibase 10.1103/PhysRevA.97.032306} {\bibfield
  {journal} {\bibinfo  {journal} {Phys. Rev. A}\ }\textbf {\bibinfo {volume}
  {97}},\ \bibinfo {pages} {032306} (\bibinfo {year} {2018})}\BibitemShut
  {NoStop}%
\bibitem [{\citenamefont {Ghosh}\ and\ \citenamefont
  {Geller}(2010)}]{Ghosh2010}%
  \BibitemOpen
  \bibfield  {author} {\bibinfo {author} {\bibfnamefont {J.}~\bibnamefont
  {Ghosh}}\ and\ \bibinfo {author} {\bibfnamefont {M.~R.}\ \bibnamefont
  {Geller}},\ }\href {\doibase 10.1103/PhysRevA.81.052340} {\bibfield
  {journal} {\bibinfo  {journal} {Phys. Rev. A}\ }\textbf {\bibinfo {volume}
  {81}},\ \bibinfo {pages} {052340} (\bibinfo {year} {2010})}\BibitemShut
  {NoStop}%
\bibitem [{\citenamefont {Landau}(1932)}]{Landau1932}%
  \BibitemOpen
  \bibfield  {author} {\bibinfo {author} {\bibfnamefont {L.}~\bibnamefont
  {Landau}},\ }\href@noop {} {\bibfield  {journal} {\bibinfo  {journal} {Phys.
  Z. Sowjetunion}\ }\textbf {\bibinfo {volume} {2}},\ \bibinfo {pages} {46}
  (\bibinfo {year} {1932})}\BibitemShut {NoStop}%
\bibitem [{\citenamefont {Zener}(1932)}]{Zener1932}%
  \BibitemOpen
  \bibfield  {author} {\bibinfo {author} {\bibfnamefont {C.}~\bibnamefont
  {Zener}},\ }\href@noop {} {\bibfield  {journal} {\bibinfo  {journal} {Proc.
  R. Soc. Lond. A}\ }\textbf {\bibinfo {volume} {137}},\ \bibinfo {pages} {696}
  (\bibinfo {year} {1932})}\BibitemShut {NoStop}%
\bibitem [{\citenamefont {St\"{u}ckelberg}(1932)}]{Stuckelberg1932}%
  \BibitemOpen
  \bibfield  {author} {\bibinfo {author} {\bibfnamefont {E.~C.~G.}\
  \bibnamefont {St\"{u}ckelberg}},\ }\href@noop {} {\bibfield  {journal}
  {\bibinfo  {journal} {Helv. Phys. Acta}\ }\textbf {\bibinfo {volume} {5}},\
  \bibinfo {pages} {369} (\bibinfo {year} {1932})}\BibitemShut {NoStop}%
\end{thebibliography}%

\end{document}